\documentclass[12pt,a4paper,final]{article}
\usepackage{cite}
\pdfoutput=1
\usepackage[]{footmisc}
\usepackage{comment}
\usepackage{graphicx}
\usepackage[dvipsnames]{xcolor}
\usepackage[centertags]{amsmath}
\usepackage{amsfonts}
\usepackage{amssymb} 
\usepackage{amsthm}
\usepackage{slashed}
\usepackage{enumerate,enumitem}
\usepackage{amsbsy}
\usepackage{mathrsfs}
\usepackage{url}
\usepackage{bbm}
\usepackage{color}
\usepackage{framed}
\usepackage{subfig}
\usepackage{caption}
\usepackage{float}
\usepackage{epstopdf}
\usepackage[colorlinks=true,
            linkcolor=RoyalBlue,
            urlcolor=PineGreen,
            citecolor=BrickRed]{hyperref}
\hypersetup{linktocpage}
\def\({\left(}
\def\){\right)}
\def\[{\left[}
\def\]{\right]}
\newcommand{\T}{\mathbf{T}}
\newcommand{\Q}{\mathbf{Q}}
\newcommand{\clambda}{\hat{l}}

\newcommand{\bra}{\langle}
\newcommand{\ket}{\rangle}

\def\one{{\rm 1\kern -.9mm l}}                  
\newcommand{\beq}{\begin{equation}}
\newcommand{\eeq}{\end{equation}}
\newcommand{\beqa}{\begin{eqnarray}}
\newcommand{\eeqa}{\end{eqnarray}}

\newcommand{\bz}{\bar{z}}

\newcommand{\pp}{{\mathbbmss{p}}}

\newcommand{\CC}{\mathbb{C}}
\newcommand{\mi}{{\mathbbm i}}
\newcommand\blank[1]{}
\newcommand{\fract}[2]{{\textstyle\frac{#1}{#2}}}

\newcommand{\ri}{\right}

\newcommand{\lf}{\left}

\newcommand{\CS}{{\cal S}}

\newcommand\ZZ{{\mathbb Z}}
\newcommand\RR{{\mathbb R}}

\newcommand\TT{{\mathbb T}}

\newcommand{\bbalpha}{\alpha\kern -4.95pt\alpha}

\newcommand\nn{\nonumber}
\renewcommand{\log}{\ln}

\newcommand{\mt}{{\mathfrak t}}
\newcommand{\TbT}{\text{T}\bar{\text{T}}}
\newtheorem*{theorem}{Theorem}
\usepackage{authblk}

\usepackage{fbb}
\emergencystretch=\maxdimen
\begin{document}
\title{Geometric aspects of the ODE/IM correspondence}
\author{Patrick Dorey$^1$, Clare Dunning$^2$, Stefano Negro$^3$ and Roberto Tateo$^4$\\
\vspace{1cm}
\small{$^1$ Department of Mathematical Sciences, Durham University, Durham DH1 3LE, United Kingdom\\\vspace{0.1cm}$^2$ School of Mathematics, Statistics and Actuarial Science, University of Kent, Canterbury CT2 7FS, United Kingdom\\\vspace{0.1cm}
$^3$ C.N. Yang Institute for Theoretical Physics, New York Stony Brook, NY 11794-3840, USA.\\\vspace{0.1cm}
$^4$ Dipartimento di Fisica and Arnold-Regge Center, Universit\`a di Torino, and INFN, Sezione di Torino, Via P. Giuria 1, I-10125 Torino, Italy\\ 
\vspace{1cm}
\texttt{\href{mailto:p.e.dorey@durham.ac.uk}{p.e.dorey@durham.ac.uk},}\\\texttt{\href{mailto:t.c.dunning@kent.ac.uk}{t.c.dunning@kent.ac.uk},}\\ \texttt{\href{mailto:stefano.negro@stonybrook.edu}{stefano.negro@stonybrook.edu},}\\\texttt{\href{mailto:tateo@to.infn.it}{tateo@to.infn.it}.}}}
\date{\vspace{-5ex}}
\maketitle

\begin{abstract}
This review describes a link between Lax operators, embedded surfaces and  Thermodynamic Bethe Ansatz equations for integrable quantum field theories.   This surprising connection between classical and quantum models is undoubtedly one of the most striking discoveries that emerged from the off-critical generalisation of the ODE/IM  correspondence, which initially involved only conformal invariant quantum field theories.  We will mainly focus of the KdV and sinh-Gordon models. However,  various aspects of other interesting systems, such as affine Toda field theories and non-linear sigma models, will be mentioned.  We also discuss the implications of these ideas in the AdS/CFT context, involving minimal surfaces and Wilson loops.
This work  is  a  follow-up of the ODE/IM review published  more than ten years ago by JPA, before the discovery of its off-critical
generalisation and the corresponding geometrical interpretation.\\
(Partially based on  lectures given at the ``Young Researchers Integrability School 2017'', in Dublin.)
\end{abstract}

\newpage

\tableofcontents

\newpage
\section{Introduction}
\label{sec:intro}
There is a deep connection between integrable equations in two dimensions and the embedding of surfaces in higher-dimensional manifolds. The simplest instance of this relation appeared in the works of 19th-century geometers \cite{Bour_862, Liou_853} on the description of pseudo-spherical and minimal surfaces sitting in $3$-dimensional Euclidean space $\mathbb R^3$. The structural equations describing their embedding, the Gauss-Mainardi-Codazzi (GMC) system, are today known as the sine-Gordon and Liouville equations, respectively. More recently, in the works of Lund, Regge, Pohlmeyer and Getmanov \cite{Pohl_76,Lund_Regg_76,Getm_77}, a general correspondence has been suggested and subsequently formalised by Sym \cite{Sym_82,Sym_83,Sym_84_1,Sym_84_2,Sym_84_3}. These results showed that any integrable field theory, with associated linear problem based on a semi-simple Lie algebra $\mathfrak g$, could be put in the form of a GMC system for a surface embedded in a $\textrm{dim}(\mathfrak g)$-dimensional space.

The connection between embedded surfaces and integrable models has proven especially fruitful in the context of the AdS/CFT correspondence. In this framework, the semiclassical limit of a string worldsheet theory in an AdS$_{n+1}$ space can be exploited to compute certain observables of conformal field theory (CFT) living on the boundary of that space.  
The canonical example of this correspondence deals with AdS$_5\times$S$_5 $. In this case, semiclassical worldsheet solutions are used to describe, in the dual CFT, states with large quantum numbers \cite{Gubser:2002tv}, expectation values of Wilson loop operators \cite{Maldacena:1998im,Rey:1998ik} and universal properties of  Maximally Helicity Violating (MHV) gluon scattering amplitudes \cite{Alday:2007hr,Alday:2008yw}. The connection with integrable models allows these quantities to be related to certain known universal structures of integrability, such as the Y-system or the corresponding set of Thermodynamic Bethe Ansatz (TBA) equations \cite{Alday:2010vh, Hatsuda:2010cc}.\\
Generally speaking, the ODE/IM correspondence, discovered in \cite{Dorey:1998pt}, is instead a link between quantum  Integrable Models, studied within the formalism of \cite{Bazhanov:1994ft,Bazhanov:1996dr} where analytic properties and functional relations are the main ingredients, and the theory of Ordinary Differential Equations in the complex domain \cite{Sibuya:1975,Voros1983}. The relationship is far more general than initially thought,  with concrete ramifications in string theory,  AdS/CFT, and aspects of the recently-discovered correspondences between supersymmetric gauge theories and integrable models \cite{Gaiotto:2009hg, Alday:2009dv, Gaiotto:2014bza, Ito:2017ypt,Ito:2018eon, Grassi:2019coc, Fioravanti:2019vxi, Fioravanti:2019awr,Komatsu:2019xzz}. The ODE/IM correspondence relies on an exact equivalence between spectral determinants associated with certain generalised Sturm-Liouville problems, and the Baxter T and Q  functions emerging within the Bethe Ansatz framework. Currently, the link mainly involves the finite volume/temperature  Bethe Ansatz equations associated with 2D integrable quantum field theories. However, there are mild hopes that it can be generalised to accommodate also integrable lattice models \cite{Dorey:2003sv}.

The primary purpose of this review is to describe the deep connection existing between the ODE/IM correspondence and the theory of embedded surfaces in higher-dimensional manifolds.

The rest of the article is organised as follows.
A brief review on the  KdV theory and associated integrals of motion, at both the classical and quantum level, is contained in sections \ref{sec:KdVbn} and \ref{sec:quantum}.
Section \ref{sec:intro1} contains a  preliminary discussion of the  ODE/IM correspondence for the quantum KdV (mKdV/sinh-Gordon) hierarchy, the relevant Schr\"odinger equation is introduced, and some general facts about the correspondence are described. 
Section \ref{sec:excited} is devoted to a schematic  derivation of the Baxter TQ relation from the Schr\"odinger equation 
(more details can be found in the original works \cite{Dorey:1998pt}, \cite{Bazhanov:2003ni}, \cite{Dorey:1999uk} and in the review \cite{Dorey:2007zx}). Section \ref{sec:WKBbn} describes how the local integrals of motion emerge from the  semiclassical quantisation. A short discussion 
of generalisations to excited states and to models related to higher-rank algebras is contained in  section \ref{sec:gen}.

The problem associated with the off-critical variant of the ODE/IM correspondence, the connection with the sinh-Gordon model (shG) and surfaces embedded in AdS spaces is discussed in section \ref{sec:class}. In particular, section \ref{sec:surf_in_AdS} contains a general introduction to embedded surfaces in AdS$_{n+1}$, while in section \ref{subsec:CMC_min_surf_ADS3}  the specific case of minimal surfaces in AdS$_3$ is discussed in more detail, together with their relation with Lax equations and the modified sinh-Gordon model  (mshG).  In section \ref{subsec:boundary_p},  the generalised potential appearing in the modified sinh-Gordon model  
is interpreted  within a  Wilson loop type setup while in sections
\ref{subsec:ass_lin_prob}--\ref{eq:Tsystem} the associated linear problem is linked, also with the help of a WKB analysis, to the T- and  Y-systems.  Starting from the Y-system and the WKB asymptotics, the corresponding Thermodynamic Bethe Ansatz equations are derived in section \ref{sec:TBA} and the interpretation of the surface area in terms of the free energy is given in section \ref{sec:Area}. Finally, section \ref{sec:conclusions} contains our conclusions.
\section{Classical and quantum KdV, the light-cone shG model, and local integrals of motion}
\label{sec:KdVbn0}
The starting point of the  work \cite{Bazhanov:1994ft} by  Bazhanov, Lukyanov and Zamolodchikov (BLZ) is the Korteweg-de Vries equation\footnote{In the following, we will denote partial derivatives with subscripts after a comma:
\beq
F_{,x_1 x_2, \dots}\left(x_1, x_2, \dots \right) = \frac{\partial}{\partial x_1} \frac{\partial}{\partial x_2} \dots F\left(x_1, x_2, \dots \right) = 
\partial_{x_1} \partial_{x_2} \dots F\left(x_1, x_2, \dots \right)\,.
\eeq} 
\beq
u_{, \mt}(x, \mt) + 12 \,u_{,x}(x,\mt)  u(x,\mt) +2 \, u_{,xxx}(x,\mt)=0\,,\quad 
\label{eq:KdV2}
\eeq
on a segment of length $L=2 \pi$ with  periodic boundary conditions 
$u(x+ 2\pi,\mt)=u(x,\mt)$. In the following we will often omit the time dependence of $u$, since we will mainly work within the Hamiltonian formalism.  
It is well-known  (see, for example,  \cite{sasaki1988}) that from the point of view of integrability, the KdV equation is also deeply connected with the light-cone classical sinh-Gordon model
\beq
\phi_{,x t}(x,t) + \sinh\left(\phi(x,t) \right)=0\,, 
\label{eq:LsG}
\eeq
since they formally share the same set of local integrals of motion.
Note that we have used different font styles for the KdV time parameter $\mt$ in equation (\ref{eq:KdV2}) and the sinh-Gordon time $t$ in equation (\ref{eq:LsG}). As will become apparent from later considerations, this is to underline the fact that the corresponding Hamiltonians, when considered as part of the same hierarchy of conserved charges for one of the two models,  evolve field configurations along  different  `generalised time directions'.
\subsection{Lax pair and classical conserved charges}
\label{sec:KdVbn}
The purpose of this section is to derive the expression of the classical integrals of motion for the KdV model  through the introduction of  a pair of Lax operators which depend on a spectral parameter. We will essentially sketch the derivation presented in the book \cite{babelon2003introduction}, to which the interested reader is addressed for further details.

First of all, notice that the KdV equation (\ref{eq:KdV2}) can be written as a Zero Curvature Condition (ZCC)
\beq
 A_{\mt,x} - A_{x,\mt} - [A_x,A_\mt]=0\,,
\label{eq:zeroc}
\eeq
%with connections $A_x$ and $A_\mt$ given by
for the $\mathfrak{sl}\left(2\right)$ connection\footnote{That is, an $\mathfrak{sl}\left(2\right)$-valued one-form.} $A = A_x \, dx + A_\mt \, d\mt$, with components
\begin{equation}
     A_x =\left(\begin{array}{c c}
        0 & 1 \\
        \lambda^2-u & 0
    \end{array}\right)\;, \qquad A_\mt = -2 \left(\begin{array}{c c}
        - u_{,x} & 4 \, \lambda^2 +2\, u \\
        4 \lambda^4 -2 \lambda^2 \, u -  u_{,x x} - 2\, u^2 & u_{,x}
    \end{array}\right)\;,
\label{Ax}
\end{equation}
where $\lambda$ is the spectral parameter. In turn, equation  (\ref{eq:zeroc})  coincides with  the compatibility condition of the following pair of linear systems of (first-order) differential equations:
\begin{equation}
    (\mathbbm{1}\partial_x - A_x)\left(\begin{array}{c}
        \Psi  \\
        \chi 
    \end{array}\right)=0\;,\qquad (\mathbbm{1}\partial_\mt - A_\mt) \left(\begin{array}{c}
        \Psi  \\
        \chi 
    \end{array}\right)=0\;.
\label{eq:flat0}
\end{equation}
The first equation in (\ref{eq:flat0}) gives
$
\chi= \Psi_{,x}\,,
$
together with the  Schr\"odinger-type  equation
\beq
(\mathbf{L}  -\lambda^2) \Psi=0\;,\quad \mathbf{L} =\partial_x^2 + u\,.  
\label{eq:L1}
\eeq
The second relation in (\ref{eq:flat0}) leads instead to the time-evolution equation
\beq
(\partial_\mt - \mathbf{M} ) \Psi=0\;,\quad  \mathbf{M}= - 2 ( \partial_x^3 + 3 \,u \,\partial_x + 3\, u_{,x})\,.   
\label{eq:M1}
\eeq
The compatibility between equations (\ref{eq:L1}) and (\ref{eq:M1}) gives
\beq
\mathbf{L}_{,\mt} - [\mathbf{M},\mathbf{L} ]=0\,, 
\eeq
a constraint which is also equivalent to the original KdV equation (\ref{eq:KdV2}).

A direct consequence of the zero-curvature condition (\ref{eq:zeroc}), which 
involves the arbitrary parameter $\lambda$, is the existence of an infinite tower of independent conserved charges. The generator of  these quantities is the trace
\beq
\mathcal{T}(\lambda)= \textrm{tr}(\mathcal{M}(\lambda))\,,
\label{eq:TraceM}
\eeq
of the so-called monodromy matrix 
\beq
\mathcal{M}(\lambda) = \overleftarrow{\exp} \left (  \int_0^{2 \pi} dx \, A_x(x,\mt,\lambda) \right) = \lim_{\delta x \rightarrow 0}  (1+\delta x \, A_x(x_n,\mt,\lambda))\dots  (1+\delta x \, A_x(x_1,\mt,\lambda))\,.
\label{eq:mono0}
\eeq
In (\ref{eq:mono0}), the symbol $\overleftarrow{\exp}$ denotes the path-ordered exponential and $x_1=0 < x_2 < \dots < x_n=2 \pi$.

Since $A_x$ and $A_\mt$ belong to the $sl(2)$ algebra we can introduce the matrices
\begin{equation}
    \mathbf{H} =\left(\begin{array}{cc}
        1 & 0 \\
        0 & -1
    \end{array}\right)\,, \quad \mathbf{E}_+ =\left(\begin{array}{cc}
        0 & 1 \\
        0 & 0
    \end{array}\right)\,,\quad
    \mathbf{E}_- =\left(\begin{array}{cc}
        0 & 0 \\
        1 & 0
    \end{array}\right)\,,
\label{eq:flat}
\end{equation}
with $[\mathbf{H},\mathbf{E}_{\pm}]=\pm 2 \mathbf{E}_{\pm}$, $[\mathbf{E}_+,\mathbf{E}_-]=\mathbf{H}$ and,
expand the connection $A_x$ over the basis $\{\mathbf{H},\mathbf{E}_-,\mathbf{E}_+\}$ as
\beq
A_x = A_h \,\mathbf{H}+ A_- \, \mathbf{E}_- +A_+ \, \mathbf{E}_+\,.
\label{eq:exBasis}
\eeq
Notice  that $\mathcal{T}(\lambda)$, defined in (\ref{eq:TraceM}),  is  invariant under (periodic) gauge transformations of  $A_x$
\beq
A_x \rightarrow {}^gA_x = g^{-1} \, A_x \,g - g^{-1} \,g_{,x}\,.
\eeq
Therefore,  we can gauge transform (\ref{eq:exBasis})  such that ${}^gA_-={}^gA_+ =0$. We first perform the gauge transformation $g_1= \exp(f_- \,\mathbf{E}_-)$, which leads to
\beq
{}{}^{g_1} A_x = (A_h + A_+ f_-) \, \mathbf{H} -
( f_{-,x} +2 \, A_h \, f_- +A_+ \, f^2_- -A_-)\, \mathbf{E}_- +A_+ \, \mathbf{E}_+\,.
\label{eq:g1}
\eeq
Setting
\beq
f_- = \frac{1}{A_+}\left(\nu  -\mathcal{A}\right)\,,\quad \mathcal{A}= A_h -\frac{1}{2} \partial_x \log A_+\,, 
\eeq
the vanishing of the coefficient $A_-$ of $\mathbf{E}_-$ in (\ref{eq:g1}) becomes equivalent to the solution of the following Riccati equation:
\beq
\nu_{,x} + \nu^2 = V\,,\quad V=\mathcal{A}_{,x} +\mathcal{A}^2 +  \, \mathcal{A}_- \, \mathcal{A}_+ \,,
\eeq
that, with the standard  replacement $\nu(x) = \partial_x \log y(x)$, can be recast into the Schr\"odinger-type form
\beq
\left( \partial_x^2   - V(x,\lambda) \right) y(x) =0\,. 
\label{eq:schKdV}
\eeq

Since the potential in (\ref{eq:schKdV}) is periodic,  $V(x+2\pi,\lambda) = V(x,\lambda)$, we can introduce a pair of independent Bloch solutions $\{y_+, y_-\}$ such that the corresponding Wronskian $W[y_+,y_-]=1$ and
\beq
y_{\pm} (x +2 \pi,\lambda) = \exp(\pm \mathcal{P}(\lambda)) y_{\pm} (x,\lambda)\,,
\eeq
where $\mathcal{P}$ is the  quasi-momentum:
\beq
\mathcal{P}(\lambda) = \log \left( \frac{y_+(2 \pi,\lambda)}{y_+(0,\lambda)}\right) = \int_0^{2 \pi} dx \, \nu(x,\lambda)\,.
\eeq
However, in  (\ref{eq:g1}),  the coefficient $A_+$ is still  unfixed and $A_h$ may still depend on the coordinate $x$. Following \cite{babelon2003introduction}, we can perform  two further independent gauge transformations, $g_2$ and $g_3$, without spoiling the $A_-=0$ constraint. In fact, the combined transformation  $g=g_1\,g_2\,g_3$ with 
\beq
g_2= \exp (f_+ \,\mathbf{E}_+)\,,\quad  g_3= \exp ( h \, \mathbf{H})\,,
\eeq
and
\beq
f_+ = A_+ \, y_+ \,  y_- \,,\quad  h= \frac{1}{2} \log \left(A_+ \, y_+^2 \exp\left(-2 \, \mathcal{P}(\lambda)\frac{x}{2 \pi} \right)   \right) \,,
\eeq
leads to 
\beq
{}{}^{g}A_x = \frac{1}{2 \pi} \mathcal{P}(\lambda) \, \mathbf{H}\,,\
\eeq
giving
\beq
\mathcal{T}(\lambda)= \textrm{tr}(\mathcal{M}(\lambda)) = 2 \cosh \left( \mathcal{P}(\lambda) \right)\,.
\eeq

For the KdV model under consideration, we have (cf. (\ref{Ax}), (\ref{eq:flat}) and (\ref{eq:exBasis}) )
\beq
A_h=0\,,\quad A_-= \lambda^2 - u\,, \quad A_+=1 \,,
\eeq
while the Riccati  and the Schr\"odinger equations are
\beq
\nu_{,x} + \nu^2 =\lambda^2 - u\,,\quad (\mathbf{L} -\lambda^2) y=0\,. 
\label{eq:ricSh}
\eeq
To find the local conserved charges, we expand $\nu$ as series in the spectral parameter around $\lambda^2=\infty$:
\beq
\nu= \lambda + \sum_{n=0}^\infty (-1)^n \frac{\nu_n}{\lambda^{n}}\,,
\eeq
and therefore
\beq
\mathcal{P}(\lambda) = \int_0^{2 \pi} dx\,\nu(x) = 2 \pi \lambda  + \sum_{n=0}^{\infty} \frac{(-1)^n}{\lambda^n}  \int_0^{2 \pi} dx \, \nu_n(x)\,.
\label{eq:AsyLambda}
\eeq
Finally, plugging (\ref{eq:AsyLambda}) into (\ref{eq:ricSh}) we find the recursion relation
\beq
\nu_{n+1} =  \frac{1}{2} \left(\nu_{,x} + \sum_{p=0}^{n} \nu_p \nu_{n-p}\right) \,,\quad \nu_0=0\,, \quad \nu_1= \frac{1}{2} u\,. 
\label{eq:recursion0}
\eeq
The first few coefficients are 
\begin{eqnarray}
\nu_1 &=& \frac{1}{2} u\,,\quad  \nu_2= \frac{1}{4} u_{,x}\,,\quad
 \nu_3= \frac{1}{8} (u^2 + u_{,x x})\,,\quad \nonumber \\
 \nu_4 &=& \frac{1}{2} \nu_{3,x} + \frac{1}{8} u \, u_{,x}\,,\quad \nu_5= \frac{1}{2} \nu_{4,x} + \frac{1}{32} (u_{,x})^2 +\frac{1}{16} u\, u_{,x x}+\frac{1}{16} u^3\,,\quad
\end{eqnarray}
which correspond, when normalised as in \cite{Bazhanov:1994ft} and up to total derivatives, to the following integrals of motion:
\begin{eqnarray}
I_1^{(cl)}=I_1^{[{\bf KdV}]}&=&\int_0^{2 \pi}  \frac{dx}{2 \pi} \, u(x)  \,,\quad I_3^{(cl)}= I_3^{[{\bf KdV}]} =\int_0^{2 \pi}  \frac{dx}{2 \pi} \, u^2(x)\,,\nonumber \\
\, I_5^{(cl)}=I_5^{[{\bf KdV}]}&=& \int_0^{2 \pi} \,  \frac{dx}{2 \pi} \left(u^3(x) - \frac{1}{2} u_{,x}^2(x) \right)\,.
\end{eqnarray}

The relation between the KdV and the modified KdV (mKdV) equations  emerges through  the  Miura transformation
\beq
u(x,\mt)= -v^2(x,\mt) -v_{,x}(x,\mt)\,,
\label{eq:miura}
\eeq
which implies 
\beq
u_{,\mt} + 2 \, u_{,x x x} + 12 \, u u_{,x}= -(2 \, v +  \partial_x) \left( v_{,\mt} + 2 \, v_{,x x x} - 12 \, v^2 \, v_{,x}\right)=0\,.   
\eeq
Hence a solution $v(x,\mt)$ of the mKdV equation
\beq
v_{,\mt}(x,\mt) + 2 \, v_{,x x x}(x,\mt) - 12 \, v^2(x,\mt) \, v_{,x}(x,\mt) =0\,,
\label{eq:mkdv0}
\eeq
can be mapped into a KdV  solution through the Miura transformation (\ref{eq:miura}). A straightforward consequence of this fact is that the quantities  $I_n^{(cl)}$ coincide with the  integrals of motion $I_n^{[{\bf mKdV}]}$  of the mKdV theory
\beq
I_n^{[{\bf mKdV}]}[v] =-I_n^{[{\bf KdV}]}[u= -v^2 -v_{,x}]\,,
\eeq
that is 
\beq
I_1^{[{\bf mKdV}]}= \int_0^{2 \pi}  \frac{dx}{2 \pi} \, v^2(x)  \,,\quad I_3^{[{\bf mKdV}]} =-\int_0^{2 \pi}  \frac{dx}{2 \pi} \, \left( v^4(x) + (v_{,x}(x))^2 \right) \,,\,\dots
\eeq
Furthermore,  the  sinh-Gordon model (\ref{eq:LsG}) also possesses the same set of local charges, provided the formal identification 
$v(x,\mt)= \phi_{,x}(x,t)/2$  is made at fixed times $\mt$ and $t$:
\beq
I_n^{[{\bf shG}]}[\phi] =I_n^{[{\bf mKdV}]}[v=\fract{1}{2} \phi_{,x}]\,.
\eeq

In fact, the  sinh-Gordon Lagrangian in light-cone coordinates is
\beq
\mathcal{L}^{[\bf shG]}= \frac{1}{2 \pi} \left(\phi_{,t}(x,t) \, \phi_{,x}(x,t) - \cosh(\phi(x,t))+1 \right)\,,
\label{eq:lsg1}
\eeq
and the conjugated momentum and Hamiltonian are
\beq
{\it \pi}(x,t)=\frac{1}{2 \pi} \phi_{,x}(x,t)\,,\quad   H^{[\bf shG]}= \int_{0}^{2 \pi} \frac{dx}{2 \pi} \, \left(\cosh \phi(x,t)-1\right)\;.
\label{eq:cmom}
\eeq
Then $\{ \phi(x,t) ,\pi(x',t) \} = \delta(x-x')$, and the sinh-Gordon equations of motion can be written as
\beq
\phi_{,x t}(x,t) = 2\, v_{,t}(x,\mt,t) = 2 \, \{ v(x,\mt,t), H^{[\bf shG]} \}\,.  
\label{eq:tt}
\eeq
Notice that in (\ref{eq:tt}),  $t$ denotes the sinh-Gordon time, which differs from the KdV (mKdV) time $\mt$ appearing in  (\ref{eq:KdV2})  and (\ref{eq:mkdv0}).\footnote{ 
At least formally, relation (\ref{eq:tt}) can be regarded as  a particular instance of  the  KdV/mKdV  hierarchy of equations \cite{Lax:1968fm}:
\beq
v_{,\mt_{2k-1}}( \{\mt_{i} \})=\{ I_{2k-1}^{[{\bf mKdV}]},v(\{\mt_{i} \}) \}\, ,\quad \nonumber
\eeq
where $\{\mt_{i} \}$, with $i \in 2 \ZZ +1$, is the set of generalised time directions with the identifications $\mt_1 = x$, $\mt_3 = \mt$ and also  $\mt_{-1}=t$, i.e. $I_{-1}^{[{\bf mKdV}]}=H^{[\bf shG]}$  (see, for example \cite{Fioravanti:1998ha, Fioravanti:2000np} ).
}

In addition,
imposing  periodic boundary conditions $\phi(x+2 \pi,t) = \phi(x ,t)$
and using the equation of motion, it is not difficult to prove that 
\beq
\{I_{2n+1}^{(cl)}[v= \frac{1}{2} \phi_{,x}  ], H^{[\bf shG]} \}=0\,,\quad (\forall n \in \mathbb Z_{\geq})\,.
\eeq
For example:
\beq
\{I_1^{(cl)}[v= \frac{1}{2} \phi_{,x}  ], H^{[\bf shG]} \}= \int_{0}^{2 \pi} \frac{dx}{4 \pi} \phi_{,x} \, \phi_{,x t}= - \int_{0}^{2 \pi} \frac{dx}{4 \pi} \partial_x  \cosh(\phi(x,t))=0\,. 
\eeq
Therefore, and as mentioned in the previous section,  the KdV conserved charges $\{ I^{(cl)}_n \}$  are also integrals of motion for the sinh-Gordon model (\ref{eq:lsg1}). We will see later that the off-critical field theory  generalisation 
of the  ODE/IM correspondence described in this review is naturally based  on the sinh-Gordon perspective of this connection. \\ 
\subsection{Quantisation of the local conserved charges}
\label{sec:quantum}
It is well known (cf. \cite{babelon2003introduction}) that the  KdV model admits two equivalent Hamiltonian structures. The first Hamiltonian  is 
\beq
H= I_3^{(cl)} =\int_0^{2 \pi}  \frac{dx}{2 \pi} \, u^2(x)\,,
\eeq
with  Poisson bracket
\beq
\frac{1}{2 \pi}  \{ u(x), u(y) \} = 2 (u(x)+u(y))  \delta_{,x}(x-y) + \delta_{,x x x}(x-y)\,.
\label{eq:clVir}
\eeq
The second possibility is instead
\beq
{H'}= I_5^{(cl)} =\int_0^{2 \pi} \,  \frac{dx}{2 \pi} \left(u^3(x) - \frac{1}{2} (u_{,x}(x))^2 \right)\,,
\eeq
with  Poisson bracket 
\beq
\frac{1}{2 \pi}  \{ u(x), u(y) \}' = 2 \, \delta_{,x}(x-y)\,.
\eeq
Both options lead to the KdV  equation:
\begin{eqnarray}
\partial_\mt u =\{ H , u \} =\{ H', u \}'= -12 \, u \, u_{,x} - 2 \, u_{,x x x} \,.
\end{eqnarray}
Furthermore, through the change of variables $u(x)= -(\phi_{,x}(x))^2 - \phi_{,x x}(x)$, the first Poisson bracket (\ref{eq:clVir}) reduces to
\beq
\frac{1}{2 \pi}  \{ \phi(x), \phi(y) \} = \frac{1}{2} \epsilon(x-y)\,,
\eeq
with $\epsilon(x)=n$ for $2 \pi n<x<2 \pi (n+1)$ and $n \in \ZZ$. This is the standard Poisson bracket involving a single bosonic field $\phi(x,t)$ with periodic boundary conditions and conjugated momenta $\pi(x,t)$ as in (\ref{eq:cmom}).

The quantisation of (\ref{eq:clVir}) is then  achieved by performing  the following replacements  \cite{Gervais:1982nw}:
\beq
\frac{1}{2 \pi}  \{\,,\,\} \rightarrow \frac{\mi c}{6 \pi} [\,,\,]\,,\quad u(x) \rightarrow - \frac{6}{c} {\bf T}(x)\, .
\eeq
Expanding
\beq
{\bf T}(x) = \frac{c}{24} + \sum_{n=-\infty}^{\infty} {\bf L}_{-n} e^{\mi n x}\,,
\eeq
we see, from (\ref{eq:clVir}),  that  the operators ${\bf L}_n$ satisfy the Virasoro algebra
\beq
[{\bf L}_n ,{\bf L}_m ] = (n-m) {\bf L}_{n+m} + \frac{c}{12} (n^3-n) \delta_{n+m,0}\,. 
\label{eq:virQ}
\eeq
Alternatively, performing  first a quantum Miura transformation
\beq
- \beta^2 {\bf T}(x)=\, : \hat{\phi}_{,x}(x)^2: + (1-  \beta^2) \hat{\phi}_{,x x}(x) + \frac{\beta^2}{24} \,,
\label{eq:Vir}
\eeq
and expanding the fundamental quantum field $\hat{\phi}(x)$ in plane-wave modes  as 
\beq
\hat \phi(x) =  \mi{\bf Q} +  \mi {\bf P}\, x + \sum_{n \ne 0} \frac{{\bf a}_{-n}}{n} e^{\mi n x}\,,
\eeq
we obtain  the Heisenberg algebra
\beq
[{\bf Q},{\bf P}] = \frac{\mi}{2} \beta^2,\,\quad [{\bf a}_n,{\bf a}_m] = \frac{n}{2} \beta^2 \delta_{n+m,0}\,.
\label{eq:Hei}
\eeq
The relation between the central charge $c$ appearing in the Virasoro algebra (\ref{eq:virQ}) and the parameter $\beta$ in equation (\ref{eq:Hei}) is  
\beq
\beta = \sqrt{\frac{1-c}{24}} - \sqrt{\frac{25-c}{24}}\,. 
\label{eq:betac}
\eeq
The  highest weight (vacuum) vector $| \pp \ket$ over the Heisenberg algebra is defined by   
\beq
{\bf P} {| \pp \ket} =\pp {| \pp \ket}\,,\quad a_n{| \pp \ket}=0\,,\quad (\forall n>0).
\eeq
In terms of the Virasoro representation, the states ${| \pp \ket} $ are highest weights with conformal dimensions
\beq
\Delta = \left( \frac{\pp}{\beta} \right)^2 + \frac{c-1}{24}\,,
\label{eq:Delta}
\eeq
\beq
{\bf L}_{0} {| \pp \ket} = \Delta   {| \pp \ket}\,,\quad {\bf L}_n  {| \pp \ket} =0,\quad (\forall n>0)\,.
\eeq

The quantum charges were first determined in \cite{sasaki1988} under the replacement of classical fields with the corresponding operators ($\phi \rightarrow \hat \phi$), and by following the scheme 
\begin{enumerate}
\item
${\bf I}_n = \, : I_n^{(cl)}:$\,,\quad  $(n=1,3) $;
\item
${\bf I}_n = \, :I_n^{(cl)}: + \sum_{k=1}^{n}  (\beta)^{2k} :I_n^{(k) } : \,,\quad (n=5,7, \dots) $;
\item
The quantum corrections $:I_n^{(k)}:$ do not contain any of the $: I_m^{(cl)}:$ as a part (see \cite{sasaki1988} for more details.);
\item
$[{\bf I}_n,{\bf I}_m] = 0$\,,\quad $(\forall \, n,m \in 2 \, \mathbb Z_{\geq} + 1)$\,.
\end{enumerate}
The first three non-vanishing local integrals of motion, written in terms of the generators of the Virasoro algebra (\ref{eq:Vir}), are: 
\begin{eqnarray}
{\bf I}_1 &=& {\bf L}_0 - \frac{c}{24}\,,\quad {\bf I}_3= 2 \sum_{n=1}^{\infty} {\bf L}_{-n} {\bf L}_n + {\bf L}_0^2 - \frac{c+2}{12} {\bf L}_0 +  \frac{c\,(5\,c +22)}{2880}\,, \nonumber \\
{\bf I}_5 &=& \sum_{n_1+n_2+n_3=0} : {\bf L}_{n_1}{\bf L}_{n_2} {\bf L}_{n_3} :  + \sum_{n=0}^{\infty} \left( \frac{c +11}{6} n^2 -1 - \frac{c}{4}  \right){\bf L}_{-n} {\bf L}_{n} + \frac{3}{2} \sum_{n=0}^{\infty} {\bf L}_{1-2 n} {\bf L}_{2 n -1} \nonumber \\
&-& \frac{c+4}{8} {\bf L}_0^2 + \frac{(c+2)(3\, c +20)}{576} {\bf L}_0 - \frac{c\,(3\,c +14)(7\,c+68)}{290304}\,.
\label{eq:quantumIM}
\end{eqnarray}
In equation (\ref{eq:quantumIM}), the normal ordering $:\,\,:$ means that  the operators ${\bf L}_{n_i}$ with larger $n_i$ are placed to the right. The corresponding expectation values $I^{vac}_n= {\bra \pp |} {\bf I}_n {| \pp \ket}$  on the vacuum states are
\begin{eqnarray}
I^{vac}_1 &=& \Delta - \frac{c}{24}\,,\quad I^{vac}_3= \Delta^2 - \frac{c+2}{12} \Delta + \frac{ c\,(5\, c +22)}{2880}\,, \nonumber \\
I^{vac}_5 &=& \Delta^3 -  \frac{c+4}{8} \Delta^2 - \frac{(c+2)(3 \, c +20)}{576} \Delta - \frac{c\, (3\,c +14)(7\,c+68)}{290304}\,,
\label{eq:IM}
\end{eqnarray}
where $c$ and $\Delta$  are related to $\pp$ and $\beta$ through
equations (\ref{eq:betac}) and (\ref{eq:Delta}). An alternative, but more sophisticated, method leading to the same result (\ref{eq:quantumIM}) is described in \cite{Bazhanov:1994ft}.  
\section{The ODE/IM correspondence for the quantum KdV-shG hierarchy}
\label{sec:intro1}
The simplest instance of the ODE/IM correspondence involves, on the ODE side, the second order differential equation \cite{Dorey:1998pt,Bazhanov:1998wj}
\beq
\Bigl(-\partial^2_x+ P(x) \Bigr)\chi(x)=0
\label{eq:sh0}
\eeq
with
\beq
P(x)=P_0^{\text{ \bf [KdV]}}(x,E,l,M)= \left( x^{2M}+
\frac{l(l+1)}{x^2} - E \right).
\label{eq:ShKdV}
\eeq
The  generalised potential $P$ and wavefunction  $\chi$  depend, therefore, on three extra parameters: the energy or spectral parameter $E$, the `angular-momentum' $l$, and the exponent $M$.
For simplicity,  throughout this review, $M$ and $l$ will be kept real with $M \geq 0$. However, there are no serious limitations forbidding the extension of both $M$ and $l$ to the complex domain. The range  $-1 \leq M \leq 0$ is essentially equivalent, by a simple change of variables, to the  $M>0$  regime \cite{Bazhanov:1998wj,Dorey:1999uk}.\footnote{In fact, with the identification  $\beta^{-2}=M+1$,  the equivalence   $(-1 \leq M \leq 0) \leftrightarrow (M \geq 0)$ coincides with the
$\beta^2 \rightarrow \beta^{-2}$,
duality in the integrals of motion in the  quantum KdV model (see, for example,  \cite{Bazhanov:1996dr}).}  We will see that for $M \geq -1$ equation (\ref{eq:sh0}) is  related, through the ODE/IM correspondence, to the conformal field theory with central charge $c \le 1$ associated  to the quantisation of the  KdV-shG theory.\footnote{The regime $M<-1$ is also interesting, since it is related 
to the Liouville field theory \cite{ZamLiouville}.}

The ODE/IM correspondence is based on the observation that the CFT version of Baxter's TQ  equation \cite{Baxter:1971} for the six-vertex model, and the quantum Wronskians  introduced  in the works by  BLZ \cite{Bazhanov:1994ft}, exactly match the Stokes relations and Wronskians between independent  solutions of  (\ref{eq:sh0}). 
BLZ  introduced a continuum analogue of the
lattice transfer matrix $\TT$ for the quantum KdV equation, an
operator-valued  function $\T(\lambda,\pp)$, together with the  Baxter  $\Q_{\pm}(\lambda,\pp)$ operators with $\Q(\lambda,\pp) \equiv  \Q_{+}(\lambda,\pp) = \Q_{-}(\lambda,-\pp)$, where $\pp$ is the quasi-momentum \cite{Bazhanov:1994ft}.
Both the $\Q$ and $\T$ operators are entire  in the spectral parameter $\lambda$ with 
\beq
[\T(\lambda,\pp),\Q_{\pm}(\lambda,\pp)]=0\,.
\eeq
All the descendent CFT states in the Verma module associated to the  highest-weight vector $| \pp \rangle$ are characterised by the real parameter $\pp$. 
Since $\T$ and $\Q_{\pm}$ commute, we can work directly  with their eigenvalues
\beq
T(\lambda,\pp) = \langle \pp|  \T(\lambda,\pp) | \pp \rangle\,,\quad
Q_{\pm}(\lambda,\pp)= \langle \pp | \lambda^{\mp \pp/\beta^2 } \Q_{\pm}(\lambda,\pp) | \pp \rangle
\eeq
which satisfy the TQ relation \cite{Bazhanov:1996dr}
\beq
T(\lambda,\pp) Q_{\pm}(\lambda,\pp)=
e^{\mp \mi 2 \pi \pp}Q_{\pm}(q^{-1} \lambda,\pp)+
e^{\pm \mi 2 \pi \pp}Q_{\pm}(q \lambda,\pp)
\label{eq:tq1}
\eeq
with $q =\exp(  i \pi \beta^2)$.

It turns out that equation (\ref{eq:tq1})
exactly matches a \emph{Stokes relation}, i.e.\ a connection formula, for  particular solutions of the  ODE (\ref{eq:sh0}). 
The precise correspondence between the parameters in (\ref{eq:tq1}) and those in (\ref{eq:sh0}) is: 
\beq
\beta^{-2}= M+1\,,\quad \pp=\frac{2l+1}{4 M +4}\,,\quad
\lambda=  (2M{+}2)^{-2M/(M+1)} \,
\Gamma\left(\frac{M}{M+1}\right)^{-2}\,E\,.
\label{eq:ident1}
\eeq
Supplemented with the analytic requirement that both   T and Q are entire in $\lambda$,  (\ref{eq:tq1}) leads to the Bethe Ansatz equations. At a zero $\lambda=\lambda_i$ of $Q(\lambda,\pp)=Q_+(\lambda,\pp)$, the RHS of (\ref{eq:tq1}) vanishes since $T(\lambda_i,\pp)$ is finite, and hence
\beq
\frac{Q(q^{-1}\lambda_i,\pp)}{Q(q \;\lambda_i,\pp)}= -e^{\mi 4 \pi \pp}\,.
\label{eq:BAE0}
\eeq
As a result, the link between (\ref{eq:sh0}) and the Baxter relation (\ref{eq:tq1}) for the quantum KdV model is more than formal: the resulting  T and Q functions emerging from these two -- apparently disconnected -- setups are exactly the same. 
\subsection{Derivation of Baxter's TQ relation from the ODE}
\label{sec:excited}
Consider the ODE (\ref{eq:sh0}), where
we will henceforth allow $x$ to be complex, living on a suitable cover $C$ of the punctured complex plane $\CC^*=\CC\setminus \{0\}$ so as to render the equation and its solutions single-valued. 
A straightforward WKB analysis shows  that for large $x$  close to the positive real axis a generic solution has a growing leading asymptotic of the form
\beq
\chi(x) \sim c_+ \,  P(x)^{-1/4}
\exp \left(\int^x dx'\;\sqrt{P(x')}\right)\,,\quad (\textrm{Re}[x] \rightarrow + \infty)\,.
\label{eq:yas1}
\eeq
Even at fixed normalisation $c_+$, this asymptotic does not uniquely characterise the solution, since an exponentially decreasing  contribution  can always be added to $\chi(x)$  without spoiling the large-$x$ behaviour (\ref{eq:yas1}). The exponentially small term can explicitly emerge from the asymptotics  only if the nontrivial solution to (\ref{eq:sh0}) is carefully chosen such that the coefficient of the exponentially  growing term vanishes. In this special situation
\beq
\chi(x) \sim c_- \,  P(x)^{-1/4}
\exp \left(-\int^x dx'\; \sqrt{P(x')} \right)\,,\quad (\textrm{Re}[x] \rightarrow + \infty)\,.
\label{eq:yas2}
\eeq
Apart for the arbitrariness of the overall normalisation factor $c_-$, the asymptotic (\ref{eq:yas2}) now uniquely specifies the solution of (\ref{eq:sh0}). This was formalised  by  Sibuya and collaborators in the following statement, which holds not only on the real axis but also in an $M$-dependent wedge of the complex plane:
the ODE (\ref{eq:sh0}) has a basic solution $y(x,E,l)$ with the following properties, which fix it uniquely:
\begin{enumerate}
\item
$y(x,E,l)$ is an entire function of $E$, 
and a holomorphic function of $x\in C$, where $C$ is
a suitable cover of the punctured complex plane $\CC^*=\CC \setminus\{0\}$\,;
\item
the asympotic behaviour of $y(x,E,l)$ for  $|x|\to\infty$ with $|\arg(x)\,|<3\pi/(2M{+}2)$ is
\beq
y\sim \frac{1}{\sqrt{2\mi}}\,
x^{-\fract{M}{2}}\exp\left(-\frac{x^{M{+}1}}{M{+}1}\right),\,\,
y_{,x}\sim {-\frac{1}{\sqrt{2 \mi}}}\,
x^{\fract{M}{2}}\exp\left(-\frac{x^{M{+}1}}{M{+}1}\right),
\label{yas}
\eeq
though there are small modifications in the asymptotics (\ref{yas}) for $M\leq 1$ (see, for example, \cite{Dorey:1999uk}).\\
\end{enumerate}

To proceed with our analysis, it is necessary to continue $x$ even further into the complex plane, beyond the wedge where Sibuya's initial result applies.
We define general rays in the complex plane by setting $x= \varrho e^{\mi \vartheta}$ with $\varrho$ and $\vartheta$ real. Substituting into the WKB formulas (\ref{eq:yas1}) and (\ref{eq:yas2}), we detect two possible asymptotic behaviours
\beq
\chi_{\pm} \sim P^{-1/4}\exp\left(\pm\frac{1}{M{+}1}e^{\mi \vartheta(1{+}M)}
\varrho^{1{+}M}\right).
\label{eq:twosol}
\eeq
For most values of $\vartheta$, one of these solutions
will be exponentially growing, or \emph{dominant}, and the other exponentially decaying, or \emph{subdominant}. However, for
\beq
\textrm{Re}\left[e^{\mi \vartheta(1{+}M)}\right]=0
\label{eq:antiS}
\eeq
both solutions oscillate, and neither dominates the other. The values 
\beq
\vartheta=
\pm\frac{\pi}{2M{+}2}~,~
\pm\frac{3\pi}{2M{+}2}~,~
\pm\frac{5\pi}{2M{+}2}~,~\dots~,
\eeq
where this happens, and the two solutions  (\ref{eq:twosol}) exchange r\^oles, are called \emph{anti-Stokes lines}.\footnote{We are following here the convention used, for example, in \cite{BerryStokes}. Unfortunately, the lines characterised by the condition (\ref{eq:antiS}) are sometimes called instead \emph{Stokes
lines}.}
The  \emph{Stokes lines} are instead the lines along which $\chi$ either grows or shrinks the fastest, and in the current case they lie  right in the middle, between adjacent anti-Stokes lines, and are characterised by 
\beq
\textrm{Im}\left[e^{\mi\vartheta(1{+}M)}\right]=0\,.
\label{eq:StoL}
\eeq
The wedges between  adjacent  anti-Stokes lines are called Stokes sectors, and we will label them as 
\beq
\CS_k =\left\{ x\in\mathbb{C} : \left|\arg(x)-
\frac{2\pi k}{2M{+}2}\right|<\frac{\pi}{2M{+}2}\right\}\,.
\eeq
In this notation the region of validity of the asymptotic (\ref{yas}) is the union of wedges
\beq
\CS_{\text{WKB}}=\CS_{-1}\cup\overline{\CS_0}\cup\CS_1
\label{eq:region}
\eeq
where $\overline{\CS_0}$ is the closure of $\CS_0$. 
\begin{figure}[t!]
 \centering
  \includegraphics[scale=0.48]{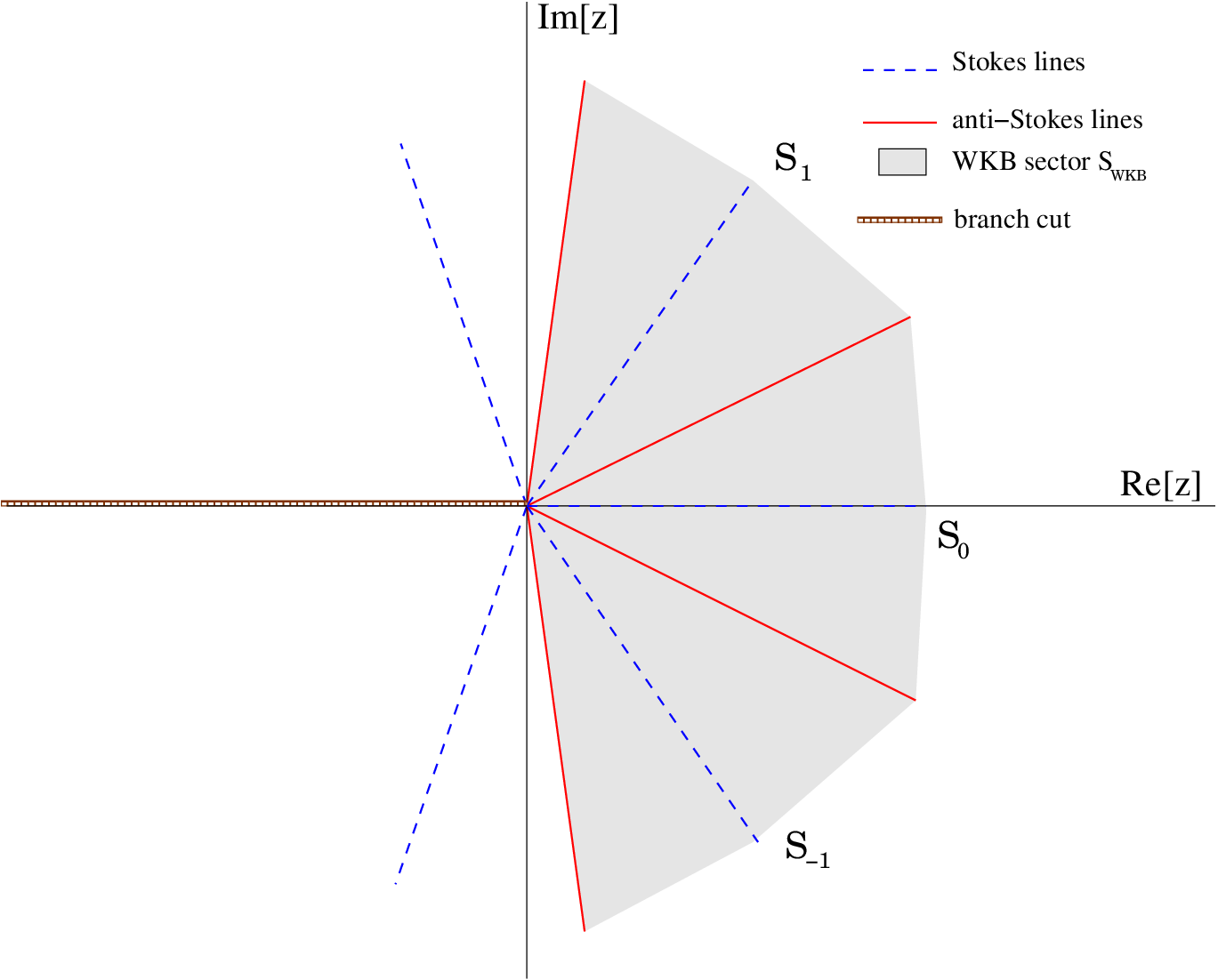}
\caption{\small Stokes, WKB  sectors and convention for the branch cut when $2M \notin \mathbb Z_{\geq}$.}
\label{fig:StokesS}
\end{figure}

Finding  the large $|x|$ behaviour of the particular solution  $y(x,E,l)$ outside the region (\ref{eq:region}) is a non-trivial task: the continuation
of a limit is not in general the same as the limit of a
continuation, and so (\ref{yas}) no longer holds once $\CS_{\text{WKB}}$ is left. This issue is related to the so-called
\emph{Stokes phenomenon}, wherein the  quantities of principal interest are the
\emph{Stokes multipliers}, encoding the switching-on of small (subdominant) exponential terms as Stokes lines are crossed~\cite{BerryStokes}. 

Thus far we have discussed the behaviour of solutions to (\ref{eq:sh0}) when $|x|$ is large. 
Consider now the region $x \simeq  0$. For $M> -1$, 
the origin  corresponds to a regular  singularity, and the associated indicial equation shows that a generic solution to (\ref{eq:sh0}) behaves as a linear combination of $x^{l+1}$ and $x^{-l}$ as  $x \rightarrow 0$. This allows a special solution $\psi(x,E,l)$ to be  specified  by the requirement
\beq
\psi(x,E,l) \sim x^{l+1} + \mathcal O(x^{l+3})\,.
\label{psidefn}
\eeq
This boundary condition defines $\psi(x,E,l)$ uniquely provided $\textrm{Re}[l] > -3/2$. A second solution can be obtained from $\psi(x,E,l)$
by noting that, since the  differential equation (\ref{eq:sh0}) is invariant under the analytic continuation $l\to -1{-}l$, $\psi(x,E,-1{-}l)$ is also a solution. Near the origin, $\psi(x,E,-1{-}l) \sim   x^{-l} +O(x^{-l+2})$,  therefore for generic values of the  angular momentum $l$ the two solutions
\beq
\psi_+(x,E)=\psi(x,E,l)\,,\quad \psi_-(x,E) =\psi(x,E,-1{-}l)\,,
\label{psipmdef0}
\eeq
are linearly independent, i.e. the Wronskian  $W[\psi_+,\psi_-]$ is non-vanishing.  Some subtleties arise at the isolated points
\beq
l+ \frac{1}{2}= \pm \left( m_1 + (M+1) m_2 \right)\,,\quad (m_1,m_2 \in \mathbb Z_{\geq})\,,
\eeq
where $\{\psi_+,\psi_-\}$ fails to be a  basis of solutions \cite{Dorey:1999uk}. For $2M \in \mathbb Z_{\geq}$, this is just the standard resonant phenomenon in the Frobenius method, which predicts  that one of the two independent solutions may acquire a logarithmic component, when  the two roots of the indicial equation differ by an integer.
For the remainder of this review we will steer clear of such points, but see \cite{Dorey:1999uk} for some further discussion of the issue.

A natural eigenproblem for a Schr\"odinger  equation, the so-called {\it radial} or {\it central} problem, is to look
for values of $E$ at which there exists a solution that vanishes as $x\to +\infty$, and behaves as $x^{l+1}$ at origin.
For $\textrm{Re}[l]>  -1/2$, this boundary condition  is equivalent to demanding the square integrability of the solution on the half line, and for $\textrm{Re}[l]> 0$ to the requirement that the divergent  
$x^{-l-1}$ term  is absent. For $\textrm{Re}[l]\le -1/2$, the problem can be defined by analytic continuation. 

Addressing the reader to \cite{Dorey:1999uk} and \cite{Dorey:2007zx}  for more details, we proceed by adopting a trick due to Sibuya \cite{Sibuya:1975}. Starting from  the uniquely-defined solution $y(x,E,l)$, subdominant in the Stokes sector $\CS_0$, we generate a set of functions
\beq
y_k(x,E,l)= \omega^{k/2}
y(\omega^{-k}x,\;\omega^{2k}E,\;l)
{}~~,\quad \omega = e^{\frac{2\pi \mi}{2M+2}}
{}~~,\quad (k\in\ZZ)\,,
\label{yks}
\eeq
all of which solve~(\ref{eq:sh0}). Notice that the asymptotic expansion
\beq
y_{\pm 1}(x,E,l) \sim \pm \sqrt{\mathbbm i}
\frac{x^{-M/2}}{\sqrt{2}}
\exp \lf( \frac{x^{M+1}}{M+1} \ri),
\label{yas3}
\eeq
is valid in the Stokes sector $\mathcal S_0$ containing the real line. Hence, we can compute the Wronskians $W\left[y,y_{\pm1}\right]$ using the expansions (\ref{yas}) and (\ref{yas3}), finding that they are non-zero: $W\left[y,y_{\pm1}\right] = \pm 1$. As a consequence $\{y,y_{\pm1}\}$ are bases of the
two-dimensional space of solutions to the ODE (\ref{eq:sh0}). More generally, a similar consideration shows that $W\left[y_k,y_{k+1}\right]=1$
and hence any pair $\{y_k,y_{k+1}\}$ constitutes a basis.
In particular, $y_{-1}$ can be written as a linear combination of the basis elements $y=y_0$ and $y_{1}$ as $y_{-1}=Cy+\tilde{C}y_1$, or equivalently
\beq
C(E,l)\,y(x,E,l)= y_{-1}(x,E,l)-
\tilde{C}(E,l)\, y_{\;1}(x,E,l)\,,
\label{eq:Ty1}
\eeq
where the connection coefficients  $\tilde{C}$ and $C$ are the Stokes multipliers.  For the right-hand side of of (\ref{eq:Ty1}) to match the exponentially decreasing behaviour on the left, we must set $\tilde{C}=-1$ (cf. equation (\ref{yas3})) and so
\beq
C(E,l)\,y_0(x,E,l)=y_{-1}(x,E,l)+
y_{1}(x,E,l)\,,
\label{Ty}
\eeq
where the sole non-trivial Stokes multiplier $C(E,l)$ takes, in the chosen normalisations (\ref{yas}) and (\ref{yks}) for $y(x)$ and $y_k(x)$, the simple form:
\beq
C(E,l)=W[\,y_{-1}\,, y_1]/W[\,y_0,y_1]=
W[\,y_{-1}\,, y_1]\,.
\label{tform}
\eeq

We now project  $y(x,E,l)$  onto another solution,
defined by its asymptotics as $x \to 0$. 
Taking the Wronskian of both sides of (\ref{Ty})
with $\psi(x,E,l)$ results in the $x$-independent equation
\beq
C(E,l) W[y_0,\psi](E,l)=W[y_{-1},\psi](E,l)+W[y_{1},\psi](E,l)\,.
\label{tqi}
\eeq
To relate the objects on the right-hand side of this equation back to
$W[y_0,\psi]$, we first define another set of `rotated' solutions, by analogy
with (\ref{yks}):
\beq
\psi_k(x,E,l)= \omega^{k/2}
\psi(\omega^{-k}x,\;\omega^{2k}E,\;l)\,,\quad (k\in\ZZ)\,.
\label{psiks}
\eeq
The functions (\ref{psiks}) also solve (\ref{eq:sh0}) and a consideration of their behaviour as
$x\to 0$ shows that
\beq
\psi_k(x,E,l)= \omega^{-(l+1/2)k}\psi(x,E,l)\,.
\eeq
In addition,
\beq
W[y_k,\psi_k](E,l)=
\omega^k\,W[y(\omega^{-k}x,\omega^{2k}E,l),
\psi(\omega^{-k}x,\omega^{2k}E,l)]
=
W[y,\psi](\omega^{2k}E,l)\,.
\eeq
Combining these results,
\beq
W[y_k,\psi](E,l)=\omega^{(l{+}1/2)k}W[y,\psi](\omega^{2k}E,l)\,,
\eeq
and setting
\beq
D(E,l)=W[y,\psi](E,l)\,,
\label{dy}
\eeq
the projected Stokes relation (\ref{tqi}) becomes
\beq
C(E,l) D(E,l)=
\omega^{-(l+1/2)}D(\omega^{-2}E,l)+
\omega^{(l+1/2)}D(\omega^{2}E,l)\,.
\label{eq:tq}
\eeq
Therefore, as anticipated at the end of  section \ref{sec:intro1}, with the identifications $T = C$ and $Q = D$ and (\ref{eq:ident1}), the Stokes equation (\ref{eq:tq}) exactly matches the Baxter TQ relation  (\ref{eq:tq1}) for the quantum KdV theory described in \cite{Bazhanov:1998wj}. Finally, the constraint $W[\,y_{k}\,, y_{k+1}]=1$, becomes 
\beq
    \det\left( \begin{array}{c c}
        \omega^{-\frac{2 l+1}{4}} D_-(\omega^{-1} E)&\omega^{\frac{2 l+1}{4}}  D_-(\omega E) \\
        \omega^{\frac{2 l+1}{4}}  D_+(\omega^{-1} E)& \omega^{-\frac{2 l+1}{4}} D_+(\omega E)
    \end{array}\right) = (2l+1)\;,
\label{eq:QQ-system0}
\eeq
with $D_-(E)=D(E,l)$ and $D_+(E)=D(E,-l-1)$.
Equation (\ref{eq:QQ-system0}) is known in the literature as \emph{quantum Wronskian} \cite{Bazhanov:1994ft}, and is a special case of the  QQ-systems of \cite{Kazakov:2015efa}. In turn, the QQ-systems  are $x$-independent versions of the  $\psi$-systems  of \cite{Dorey:2006an}.
\subsection{All orders semiclassical expansion and the quantum integrals of motion}
\label{sec:WKBbn}
We first note that with a simple change of variables \cite{Dorey:2004fk}, the Schr\"odinger  equation (\ref{eq:sh0}) can be recast into the form
\beq
\left( - \varepsilon^2 \partial^2_w  + Z(w) \right) y(w) =0\,,
\label{simp1}
\eeq
where
\beq
Z(w)= \frac{1}{4\clambda^2}\, w^{1/\clambda-2}
(w^{M/\clambda}-1)\,, \quad \clambda= l +\frac{1}{2}\,,\quad \varepsilon=
{E^{-(M{+}1)/2M}} \,.
\label{newpot}
\eeq
A key feature of equations (\ref{simp1}) and (\ref{newpot}) is that
the $E$-dependence, contained in $\varepsilon$, has been
factored out of the transformed potential $Z(w)$.
Suppose now that (\ref{simp1}) has a  solution of the form
\beq
y(w)=\exp \left(\frac{1}{\varepsilon}\sum_{n=0}^{\infty}
\varepsilon^nS_n(w) \right).
\label{wkbsol}
\eeq
For equation (\ref{simp1}) to be fulfilled order-by-order in $\varepsilon$,
the derivatives 
$S_{n,w}(w)$ must obey the following recursion relation:
\beq
S_{0,w}(w)= - \sqrt{Z(w)}~,~~~2 \, S_{0,w} \, S_{n,w} +\sum_{j=1}^{n-1} S_{j,w} \, S_{n-j,w}
+S_{n-1, w w}=0\,,\quad (n \ge 1)\,.
\label{rec}
\eeq
The first few terms of the solution are
\beqa
S_{1,w} &=& - \frac{\,Z_{,w}}{4Z}~,\quad S_{2,w} = - \frac{1}{48} \left( \frac{Z_{,w w}}{Z^{3/2}} +5\,
\partial_w \left(\frac{Z_{,w}}{Z^{3/2}} \right)\right)\,,  \nn \\ 
S_{3,w} &=& - \frac{~Z_{,w w}}{16Z^2}+\frac{5(Z_{,w})^2}{64Z^3}\,=\,
\partial_w \left(\frac{5(Z_{,w})^2}{64Z^3}-\frac{Z_{,w w}}{16Z^2}\right),
\eeqa
and further terms are very easily obtained using, for example, Mathematica.
Keeping only the first two contributions, $S_0$ and $S_1$, corresponds to the standard physical optics or WKB approximation.
Near the turning points $Z=0$  the approximation breaks down, and further work is needed to find the 
connection formulae for the continuation of WKB-like solutions of
given order from one region of non-vanishing $Z$ to another (see, for example,  section 10.7 of \cite{bender2013advanced}).

In cases where $Z(w)$ is an entire function of the coordinate $w$, with
just a pair of well-separated simple zeros on the real axis,
Dunham~\cite{Dunham:PhysRev.41.713} 
found a remarkably simple formulation of the final quantisation condition, 
valid to all orders in $\varepsilon$:
\beq
\frac{1}{\mi}\ointop_{\gamma} dw\; \left(\sum_{n=0}^{\infty} \varepsilon^{n-1}
S_{n,w}(w)\right) =2\pi\,k\ \, ,\quad (k \in \mathbb Z_{\geq})\,.
\label{quanta}
\eeq
In (\ref{quanta}), the contour $\gamma$ encloses the two turning points; it closes
because for such a $Z$
all of the functions $S_{n,w}$ derived from (\ref{rec}) 
are either entire or else have a pair of square root branch
points which can be connected by a branch cut along the real axis.
Notice that the contour $\gamma$ can be taken to lie far from the two turning
points where the WKB series breaks down and so there is no need to
worry about connection formulae.
All of the terms $S_{2n+1,w}$, $n\ge 1$, turn out to be total
derivatives and can, therefore, be discarded,
while the contribution of $\frac{1}{2\mi}S_{1,w}=-\frac{1}{8\mi}Z_{,w}/Z$ is 
a simple factor $\pi/2$, when integrated round the two zeros of $Z$.
Dunham's condition then becomes 
\beq
\frac{1}{\mi}\ointop_{\gamma} dw\; \left(\sum_{n=0}^{\infty}\varepsilon^{2n-1}
 S_{2n,w}(w) \right)=(2k{+}1) \pi\,,\quad (k \in\mathbb Z_{\geq})\,.
\label{quant}
\eeq

\begin{figure}[t!]
 \centering
  \includegraphics[scale=0.5]{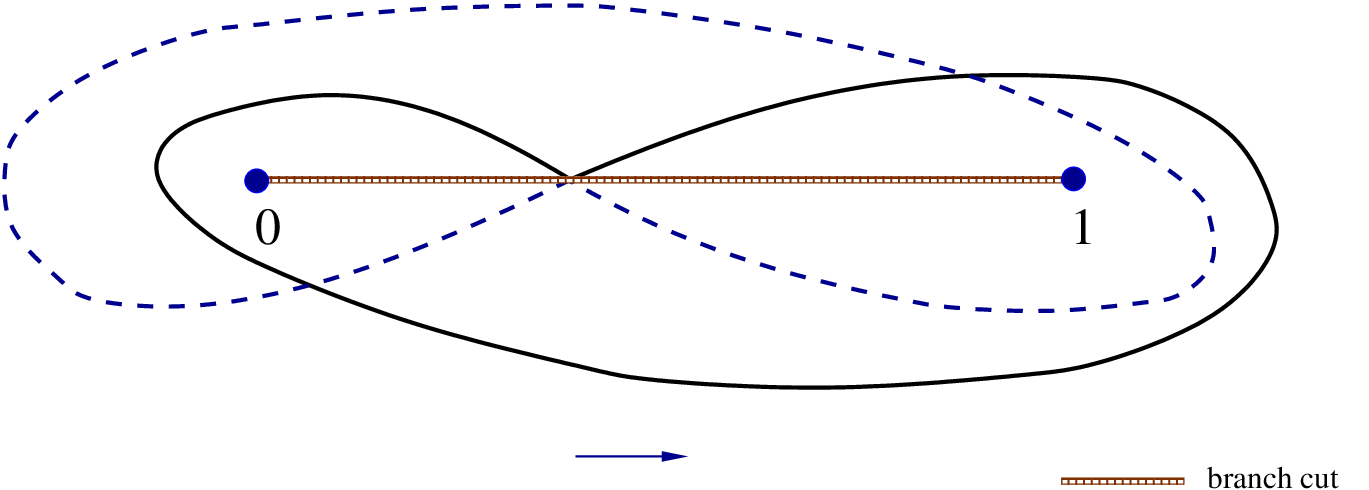}
\caption{\small The  Pochhammer contour $\gamma_P$.}
\label{fig:Pochhammer}
\end{figure}

In the current situation, we are interested in the radial connection problem, where the integration contour runs initially on the segment $w \in (0,1)$:
\beq
\ointop_{\gamma} dw \; S_{2n,w}(w) \rightarrow 2\int_{0}^{1} dw\; 
 S_{2n,w}(w)\,.
\label{eq:quant0}
\eeq
However, for generic values of $\clambda$,  $M$ and $n$ the integrand in (\ref{eq:quant0}) is  divergent at $w=0$ and/or at $w=1$. We need, therefore, a consistent regularisation prescription. To this end we replace the integration on the segment $w \in (0,1)$ with an integral over  the Pochhammer contour $\gamma_P$, represented in figure \ref{fig:Pochhammer}, around   the branch points at $w=0$ and $w=1$. To proceed, we first perform a change of variable $z = w^{M/\clambda}$,
\beq
\check{I}_{2n-1}(M,\clambda)=\frac{2}{\mi}\int_{0}^{1} dw\;
 S_{2n,w}(w)=\frac{2}{\mi} \frac{\clambda}{M}\int_{0}^{1} dz\;  S_{2n,w}\left(z^{\clambda/M}\right)\, z^{\clambda /M -1}\,.
\label{eq:quant}
\eeq
Setting
\beq
\tilde S_{2n}(z) = \frac{2}{\mi} \frac{\clambda}{M} S_{2n,w}\left(z^{\clambda/M}\right)\, z^{\clambda /M -1}\,,
\eeq
the monodromies around $z=0$ and $z=1$ are:
\beq
 \tilde S_{2n}(z e^{\mi 2 \pi} ) \rightarrow e^{\mi \frac{\pi}{M} (1-2 n)} \tilde S_{2n}(z)\,,\quad  \tilde S_{2n}( (z-1)e^{\mi 2 \pi} +1  ) \rightarrow  - \tilde S_{2n}(z)\,.
\eeq
Therefore, we can replace the integral over the interval $(0,1)$ with an integral over $\gamma_P$, provided  the extra contribution introduced by integrating over the Pochhammer contour is properly balanced by a normalisation factor. The result is
\beq
 \check{I}_{2n-1}(M,\clambda)= \frac{1}{2\left(1-  e^{\mi \frac{\pi (1-2 n)}{M}}\right)} 
 \ointop_{\gamma_P} dz\; 
 \tilde S_{2n}(z) \;,  
\label{eq:quant1}
\eeq
which is now well defined for generic values of $M$ and $\lambda$ and can always be written as a finite sum of Euler Beta functions. The explicit  outcome is:
\beq
\check{I}_{2n-1}(M,\clambda)
= (-1)^{n}\frac{\sqrt{\pi}\,\Gamma\Big(1{-}\frac{(2n-1)}{2M}\Big)}
{\Gamma\Big(\frac{3}{2}{-}n{-}\frac{(2n-1)}{2M}\Big)}
\frac{(4M{+}4)^n}{(2n{-}1)\,n!}\,I_{2n-1}(M,\clambda)\,,
\label{localexpcoeff}
\eeq
where $I_{-1}=1$, while the coefficients $I_{2n-1}(M,\clambda)$, with $n>0$, coincide with the local KdV  conserved  charges for the vacuum states (\ref{eq:IM}), provided  the following identifications are made:
\beq
c= 1- \frac{6 M^2}{M+1}\,,\quad
\Delta=\frac{(2l
+ 1)^2 -4 M^2}{16 (M + 1)}~.
\label{cdelta}
\eeq
The exact link between the all-order WKB coefficients and the integrals of motion (\ref{eq:IM}) is another striking result of the ODE/IM correspondence. 
\subsection{Simple generalisations}
\label{sec:gen}
First of all, the  link between the ODE (\ref{eq:sh0}) and the vacuum states of the quantum  KdV model in finite volume $L=2\pi$ can be generalized to accommodate the whole tower of excited states \cite{Bazhanov:2003ni} (see also \cite{Fioravanti:2004cz}). The basic replacement is to send
$
P_0^{\text{ \bf [KdV]}} \rightarrow P_{\text{exc}}^{\text{ \bf [KdV]}}
$
in (\ref{eq:ShKdV}) with
\beq
P_{\text{exc}}^{\text{ \bf [KdV]}}(x,E,l,M,\{z_k\})= \left(x^{2M}+ \frac{l(l+1)}{x^2} -2 \partial_x^2\left(  \sum_{k=1}^K \ln(x^{2M+2}-z_k)\right) -E \right),
\label{ge2}
\eeq
where the constants $\{z_k\}$ satisfy the auxiliary Bethe Ansatz type equations:  
\beq
\sum_{{j=1\atop j \ne k}}^{{K}}
\frac{z_k(z_k^2+(M{+}3)(2M{+}1) z_k z_j +M(2M{+}1)
z_j^2)}{(z_k -z_j)^3} - \frac{M z_k}{4(M{+}1)} + \Delta=0\,.
\eeq
Generalisations of the ODE/IM correspondence for both the vacuum and the excited states involving families of higher-order differential operators  were studied in \cite{Dorey:1999pv,Suzuki:2000fc, JS2000,Bazhanov:2001xm, Dorey:2006an,Masoero:2015lga,Masoero:2015rcz,Masoero:2018rel}.

In the following, instead of describing the setup of \cite{Bazhanov:2003ni} or
\cite{Dorey:1999pv,Suzuki:2000fc, JS2000,Bazhanov:2001xm, Dorey:2006an,Masoero:2015lga,Masoero:2015rcz,Masoero:2018rel,Lacroix:2018fhf} we shall focus on an off-critical variant,  which is related to the classical problem of embedded surfaces in AdS$_3$ and also to polygonal Wilson loops \cite{Alday:2009dv,Alday:2007hr}. As a preliminary remark, we notice that a natural generalisation of the Sturm-Liouville  problem associated with (\ref{eq:sh0})-(\ref{eq:ShKdV}) corresponds to  polynomial potentials of the form
\beq
P_0^{\text{ \bf [HsG]}}(x, \{ x_k\}) = \prod_{k=1}^{2N} (x -x_k)\,,\quad (2 N \in \mathbb Z_>)\,,
\label{eq:polPot}
\eeq
where $x_1$ can be set to zero by shifting $x$, while the remaining constants $x_k$ ($i=2,\dots, 2N$) are free parameters. It was argued in \cite{Dorey:2007ti} that the choice (\ref{eq:polPot}), is connected 
to  the  Homogeneous sine-Gordon model (hsG) in its CFT limit or equivalently  to the  $SU(2N)_2/U(1)^{2N-1}$ parafermions \cite{Kuniba:1993cn,CastroAlvaredo:1999em, Dorey:2004qc}. The specific choices of the set  $x_k$ which lead to
\beq
P_0^{\text{ \bf [Vir]}}(x,m,m') = x^{m-2} (x^{m'-m} - \tilde{E})\,,
\label{eq:potM}
\eeq
correspond to the Virasoro minimal models M$_{m,m'}$. As described in \cite{Dorey:2007ti}, the generalised potential  (\ref{eq:potM}) is related to the original instance of the ODE/IM correspondence,  discussed in the previous sections, by a simple change of variables.

We shall see in the remaining part of this review that the polynomial potentials (\ref{eq:polPot}) appear naturally in the description of Wilson loops in AdS$_3$ with polygonal boundaries. 
\section{Classical integrable equations and embedded surfaces     }
\label{sec:class}
In this section we wish to recall the general properties of minimal and constant mean curvature (CMC) surfaces embedded in AdS$_{n+1}$ and explain how a linear differential system arises as a structural constraint on the functions describing the embedding of these surfaces. We will then focus on the simplest non-trivial case of minimal surfaces embedded in AdS$_3$. Here a single field $\tilde \varphi$ is present, parametrizing the conformal factor of the metric. This field satisfies the modified sinh-Gordon equation \cite{Lukyanov:2010rn,Alday:2009dv, Bazhanov:2013cua,Bazhanov:2013oya}, with (anti)-holomorphic potentials $A$ and $\bar{A}$,\footnote{As shown in section \ref{sec:surf_in_AdS}, these functions intuitively measure how `curved' the surface is, and enter in the definition of the Gauss curvature.}  whose singularity structure has profound effects on the shape of the embedded surface. In particular, the presence of an irregular singularity (e.g. when $A$ is a polynomial) corresponds to the presence of a Stokes phenomenon in the linear differential system which then translates into the existence of light-like edges of the surface at the conformal boundary of AdS$_3$. For $A$ and $\bar{A}$ polynomials of order $2N \in \mathbb Z_>$, the embedded surface will sit on a light-like $4(N+1)$-gon on the conformal boundary. Finally, we will explain how to encode the full information of this embedding into a set of finite difference equations, the T-system and the Baxter TQ equation, which can then be converted into non-linear integral equation form.  
\subsection{Surfaces embedded in AdS$_{n+1}$}
\label{sec:surf_in_AdS}

The $(n+1)$-dimensional anti de-Sitter space AdS$_{n+1}$ can be described by a pseudo-spherical restriction of the pseudo-Riemannian flat space $\mathbb R^{2,n}$. More precisely, consider $\vec{Y} = \left(Y^{-1}\;,Y^0\;,\cdots \;,Y^n\right)^{\textrm{T}}\in\mathbb R^{2,n}$, where the superscript T denotes the operation of matrix transposition; then the condition
\beq
    \vec{Y}\cdot\vec{Y} \equiv -\left(Y^{-1}\right)^2-\left(Y^0\right)^2+\sum_{k=1}^n\left(Y^k\right)^2  =-\alpha^2\;,\qquad (\alpha\in\mathbb R)\;,
\label{eq:AdS_vector_constraint}
\eeq
represents an immersion of AdS$_{n+1}$ with radius $\alpha$ inside $\mathbb R^{2,n}$. Here and below we use the dot to denote the scalar product of vectors in $\mathbb R^{2,n}$:
\beq
    \vec{Y}\cdot \vec{Y}' = \eta_{AB}Y^A Y'^B\;,\qquad \eta_{AB} = \textrm{diag}\left(-1\;,-1\;,\underbrace{1\;,\ldots\; ,1}_{n}\right)\;.
\eeq
Concerning the indices we will adopt the convention
\begin{subequations}
    \begin{align}
    &A,B,C,\ldots =-1,0,1,\ldots,n\;,&  \mu,\nu,\ldots = 0,1\;, \\
    &j,k,l,\ldots=1,2,\ldots ,n\;,& a,b,\ldots = 1,2\;.
    \end{align}
\end{subequations}
The AdS$_{n+1}$ space can be parametrised by global coordinates $\left(\rho,\tau,\theta_1,\ldots,\theta_{n-1}\right)$ as
\beqa
    &&Y^{-1} = \alpha \cosh(\rho)\,\cos(\tau)\;, \notag \\
    &&Y^0 = \alpha \cosh(\rho)\,\sin(\tau)\;, \label{eq:ADS_globalcoordinates}
\\
    &&Y^j = \alpha \sinh(\rho)\, \cos(\theta_{n-j+1}) \prod_{k=1}^{n-j}\sin(\theta_{k})\,\;,\qquad \theta_n=0\;. \notag
\eeqa
From the last equations  we can read the standard AdS metric
\beq
    ds^2 = \alpha^2\left(-\cosh^2(\rho) \,d\tau^2+ d\rho^2 + \sinh^2(\rho) \,d\Omega^2_{n-1}\right)\;,
\eeq
where $d\Omega^2_{n-1}$ is the metric of the unit $(n-1)$-dimensional sphere. The conformal boundary of AdS$_{n+1}$ can be reached by taking the limit $\rho \rightarrow \infty$ jointly with a rescaling of the arc-length $ds\rightarrow ds/\sinh(\rho)$. The resulting metric is that of a cylinder in $\mathbb R^{1,n}$\,:
\beq
    ds_{\partial}^2 = \alpha^2\left(-d\tau^2+d\Omega^2_{n-1}\right)\;.
\eeq

Let us mention another useful parametrization of the space AdS$_{n+1}$: the \emph{Poincar\'{e} coordinates} $\left\{\mathsf{r},\mathsf{t},\vec{\mathsf{x}}\right\}$
\beqa
    &&Y^{-1} = \frac{\alpha^2}{2\mathsf{r}}+\mathsf r \frac{\alpha^2 + \vert\vec{\mathsf{x}}\vert^2 - \mathsf{t}^2}{2\alpha^2}\;, \notag \\
    &&Y^{n} = -\frac{\alpha^2}{2\mathsf{r}}+\mathsf r \frac{\alpha^2 - \vert\vec{\mathsf{x}}\vert^2 + \mathsf{t}^2}{2\alpha^2}\;, \label{eq:ADS_poincarecoordinates}
\\
    &&Y^0 = \frac{\mathsf{r}}{\alpha}\mathsf{t}\;,\qquad Y^{j} = \frac{\mathsf{r}}{\alpha}\mathsf{x}^j\;,\quad 1\leq j<n\;. \notag
\eeqa
In these coordinates the metric reads
\beq
    ds^2 = \frac{\alpha^2}{\mathsf{r}^2} d\mathsf{r}^2 - \frac{\mathsf{r}^2}{\alpha^2} d\mathsf{t}^2 + \frac{\mathsf{r}^2}{\alpha^2} \vert d\vec{\mathsf{x}}\vert^2\;,
\eeq
from which we see that $\mathsf{r} \rightarrow \infty$ approaches the boundary $\partial$AdS$_{n+1}$. The singularity $\mathsf{r} = 0$ is an apparent one, called \emph{Poincar\'{e}-Killing horizon} and shows that the Poincar\'{e} coordinates are not global. 

Now that we have defined our embedding space, AdS$_{n+1}$, we move on to the construction of the embedded surface $\Sigma$. Here we have a choice to make: we need to decide whether the time-like direction of AdS$_{n+1}$ lies in the tangent space $T\Sigma$, in which case we will have what is known as a time-like surface, or is orthogonal to it which will yield a space-like surface. This choice will dictate the type of reality conditions we need to impose on the parametrisation of $\Sigma$. For time-like surfaces we will need to describe the surface with Minkowski coordinates $\xi^\mu$ or, equivalently, with light-cone coordinates $\left(\xi^+=\xi^0+\xi^1,\xi^-=\xi^0-\xi^1\right)\in\mathbb R^2$. On the contrary, space-like surfaces will be parametrised by Euclidean coordinates $x^a$ or, which is the same, complex coordinates $\left(z=x^1+\mathbbm{i}\, x^2,\bar z = x^1 - \mathbbm{i}\, x^2\right)\in\mathbb C$. In the following we will concentrate on the latter type of surfaces. The same type of analysis can be carried over with some modifications for time-like surfaces. As is usual when dealing with the Euclidean plane, we will let the coordinates $\left(z,\bar z\right)$ take values in the full two dimensional complex space $\mathbb C^2$ while keeping the real slice condition $z^\ast = \bar z$ in the back of our minds, imposing it only when we see fit. Furthermore, we will continue to denote partial derivatives with subscripts after a comma, i.e.:
\beq
f_{,z}\left(z,\bz\right) = \frac{\partial}{\partial z}f\left(z,\bz\right) = \partial f\left(z,\bz\right)\;, \quad  f_{,\bz}\left(z,\bz\right) = \frac{\partial}{\partial \bz}f\left(z,\bz\right) = \bar\partial f\left(z,\bz\right)\;.
\eeq
Finally, whenever it is not necessary, we will drop the explicit dependence on the coordinates.

The description of the embedding of $\Sigma$ in AdS$_{n+1}$ is carried by the embedding function $\vec Y\,:\,\mathbb C^2\,\longrightarrow \,\mathbb R^{2,n}$, such that $\vec Y\left(z,\bz\right)\cdot \vec Y\left(z,\bz\right)=-\alpha^2$. From it we can immediately construct the tangent space $T_p\Sigma$ at any point $p\in\Sigma$ as the span of the two vectors $\vec Y_{,z}$ and $\vec Y_{,\bz}$, and compute the metric tensor, also known as \emph{first fundamental form}:
\beq
    \textrm{I} = ds^2 = g_{zz}\left(dz\right)^2 + 2\,g_{z\bz}\, dz\,d\bz + g_{\bz\bz}\left(d\bz\right)^2\;,\qquad g=\left(\begin{array}{c c}
        \vec Y_{,z} \cdot \vec Y_{,z} & \vec Y_{,z} \cdot \vec Y_{,\bz} \\ 
        \vec Y_{,z} \cdot \vec Y_{,\bz} & \vec Y_{,\bz} \cdot \vec Y_{,\bz}
    \end{array}\right)\;.
\eeq
It is an established fact \cite{Korn1914,Lich_916,Chern_1955,douady_buff_2000} that, at least locally, one can choose isothermal coordinates $\left(z',\bz'\right)$ such that
\beq
ds^2 = 2\, g_{z'\bz'}' \,dz' \, d\bz'\;.
\eeq
In the following we will fix these coordinates and drop the primes. The requirements $\vec Y_{,z} \cdot \vec Y_{,z} =\vec Y_{,\bz} \cdot \vec Y_{,\bz} = 0$ are known as Virasoro constraints and we see that these immediately imply that the (real) vectors $\vec Y_{,1} = \vec Y_{,z} + \vec Y_{,\bz}$ and $\vec Y_{,2} = -\mathbbm i \vec Y_{,z} +\mathbbm i \vec Y_{,\bz}$ satisfy the following identities
\beq
    \vec Y_{,1} \cdot \vec Y_{,1} = \vec Y_{,2} \cdot \vec Y_{,2}\;,\qquad \vec Y_{,1} \cdot \vec Y_{,2} = 0\;.
\eeq
As a consequence, since we already have one independent time-like vector $\vec Y$ and in $\mathbb R^{2,n}$ there can be at most $2$, we conclude that
\beq
    \vec Y_{,1} \cdot \vec Y_{,1} > 0 \;,\qquad \vec Y_{,2} \cdot \vec Y_{,2} > 0 \quad \Longrightarrow \quad \vec Y_{,z} \cdot \vec Y_{,\bz} > 0\;.
\eeq

Due to the AdS constraint $\vec Y \cdot \vec Y = -\alpha^2$, we see that the triple $\left(\vec Y, \vec Y_{,z}, \vec Y_{,\bz} \right)$ spans, at any point of $\Sigma$, a three-dimensional subspace of AdS$_{n+1}$. In order to understand the structure of the embedding, we now need to augment the above triple to a full basis of $\mathbb R^{2,n}$ and we can do this by introducing the following set of orthonormal real vectors,\footnote{To have a basis of $\mathbb R^{2,n}$ we need $2$ time-like vectors. One, $\vec Y$, we already have, the other has to be one of these normals. We choose it to be $\vec N_1$.}
\beq
\left\lbrace\vec N_j\right\rbrace_{j=1}^{n-1}\;,\qquad \vec N_i \cdot \vec N_j = \eta_{ij}\;,\qquad \eta_{ij} = \textrm{diag}\left(-1\;,1\;,\ldots \;,1\right)\;,
\eeq
spanning, together with $\vec Y$, the normal space $\left(T_p\Sigma\right)^{\bot}$ at any point $p\in\Sigma$\,:
\beq
    \vec N_i \cdot \vec Y = \vec N_i \cdot \vec Y_{,z} = \vec N_i \cdot \vec Y_{,\bz} = 0\;.
\eeq
For each of these vectors there exists a \emph{second fundamental form} $\textrm{II}_j$, defined as
\beqa
    &&\textrm{II}_j = \left(d_j\right)_{zz}\left(dz\right)^2 + 2\left(d_j\right)_{z\bz}dz\,d\bz + \left(d_j\right)_{\bz\bz}\left(d\bz\right)^2\;,\\
    &&d_j = \left(\begin{array}{c c}
        \vec Y_{,zz} \cdot \vec N_j & \vec Y_{,z\bz} \cdot \vec N_j \\
        \vec Y_{,z\bz} \cdot \vec N_j & \vec Y_{,\bz\bz} \cdot \vec N_j
    \end{array}\right)\;. \notag
\eeqa
Note that while in principle we should also have a fundamental form associated to the normal direction $\vec Y$,\footnote{We will identify this direction with the index $0$.}  this turns out to be trivial:
\beq
    d_0 = \left(\begin{array}{c c}
        \vec Y_{,zz} \cdot \vec Y & \vec Y_{,z\bz} \cdot \vec Y \\
        \vec Y_{,z\bz} \cdot \vec Y & \vec Y_{,\bz\bz} \cdot \vec Y
    \end{array}\right) = \left(\begin{array}{c c}
        -\vec Y_{,z} \cdot \vec Y_{,z} & -\vec Y_{,z} \cdot \vec Y_{,\bz} \\
        -\vec Y_{,z} \cdot \vec Y_{,\bz} & -\vec Y_{,\bz} \cdot \vec Y_{,\bz}
    \end{array}\right) = - g\;.
\eeq
It is now a good point to simplify the notation by introducing the following functions
\begin{subequations}
\begin{align}
    e^{\tilde{\varphi}} &= \vec Y_{,z} \cdot \vec Y_{\bz}\;,\qquad H_j = e^{-\tilde{\varphi}} \vec Y_{,z\bz} \cdot \vec N_j\;, \\
    A_j &= \vec Y_{,zz} \cdot \vec N_j\;,\quad\;\; \bar A_j = \vec Y_{,\bz\bz} \cdot \vec N_j\;.
\end{align}
\end{subequations}
The field $\tilde{\varphi}\in\mathbb R$ is sometimes called the Pohlmeyer field. From the first and the second fundamental forms one can construct the shape operators
\beq
    w_j = d_j g^{-1} = \left(\begin{array}{c c}
         H_j & e^{-\tilde{\varphi}} A_j \\
         e^{-\tilde{\varphi}}\bar A_j & H_j
    \end{array}\right)\;,
\eeq
whose invariants compute the total Gauss curvature $K$ and the components $H_j$ of the mean curvature vector $\vec H$
\begin{subequations}
\begin{align} \label{subeq:Mean_curvature}
    H_j &= \frac{1}{2} \textrm{tr} \left(w_j\right) = \frac{\vec Y_{,z\bz} \cdot \vec N_j}{\vec Y_{,z} \cdot \vec Y_{,\bz}} = H_j\;,  \\ \label{subeq:Gauss_curvature}
    K &= \sum_{j=1}^{n-1} \det\left(w_j\right) = \sum_{j=1}^{n-1} \left(H_j H_j - e^{-2\tilde{\varphi}} A_j\bar A_j \right)\;.
\end{align}
\end{subequations}

Now we have, at any point $p\in\Sigma$, a complete set of orthogonal vectors in $\mathbb R^{2,n}$ which we collect as the rows of a matrix $\boldsymbol{\sigma}$
\beq
    \boldsymbol{\sigma} = \left(\begin{array}{c c c c c c}
        \vec Y\;, & \vec Y_{,z}\;, & \vec Y_{,\bz}\;, & \vec N_1\;, & \cdots\;, & \vec N_{n-1}
    \end{array}\right)^{\textrm{T}}\;.
\eeq
This object is known as the \emph{frame field} or \emph{moving frame} and is anchored on the surface $\Sigma$. Consequently, its motion along the surface has to satisfy certain constraints and, since $\boldsymbol{\sigma}$ provides a basis everywhere on $\Sigma$, these take the form of a set of linear equations, called the Gauss-Weingarten (GW) system:
\beq
    \boldsymbol{\sigma}_{,z} = \mathcal U \boldsymbol{\sigma}\;,\qquad \boldsymbol{\sigma}_{,\bz} = \bar{\mathcal U} \boldsymbol{\sigma} \;.
\label{eq:GW_system}
\eeq
Finally, this system immediately implies a consistency condition which, in the geometry literature, is known as the Gauss-Codazzi-Mainardi (GMC) equation
\beq
    \mathcal U_{,\bz} - \bar{\mathcal U}_{,z} +\left[\mathcal U, \bar{\mathcal U}\right] = 0\;.
\eeq

The above equation represents a set of structural conditions for the surface, imposing non-linear constraints on the functions defining the shape and properties of $\Sigma$. Its functional form is completely general and appears as a condition for every surface embedded in any space, the details of the particular problem at hand being contained in the form of the matrices $\mathcal U$ and $\bar{\mathcal U}$. In a more geometrical language, $\mathcal U$ and $\bar{\mathcal U}$ are the components of a connection one-form $\mathcal U dz + \bar{\mathcal U} d\bar{z}$ and the GMC equation above is a vanishing condition on the curvature two-form associated to said connection, completely analogous to the ZCC (\ref{eq:zeroc}) which appeared in the case of the KdV equation.
In our case, for a generic surface embedded in AdS$_{n+1}$, $\mathcal U$ and $\bar{\mathcal U}$  are $\left(n+2\right) \times \left(n+2\right)$ matrices, which depend on
\begin{itemize}
    \item the real Pohlmeyer field $\tilde{\varphi}$,
    \item the $n-1$ real mean curvatures $H_j$,
    \item the $n-1$ complex functions $A_j$,
    \item the $\frac{1}{2}n\left(n-1\right)$ complex functions $B_{ij} = -B_{ji}$, describing the rotation of the normal space $\left(T\Sigma\right)^{\bot}$ under motion along the surface:
    \beq
        B_{ij} = \vec N_{i,z} \cdot \vec N_j = -\vec N_i \cdot \vec N_{j,z}\;.
    \eeq
\end{itemize}
The curvatures $H_j$ and the functions $A_j$ are usually treated as inputs, identifying the type of surface one is dealing with. An interpretation of the functions $A_j$ for the case $n=2$ is presented in section \ref{subsec:boundary_p}. On the other hand, the Pohlmeyer field $\tilde{\varphi}$ and the functions $B_{ij}$ are to be treated as proper dynamical variables.

We will not give the explicit expressions, in the general case, for the matrices $\mathcal U$ and $\bar{\mathcal U}$ nor for the GMC equation, as the case of interest of this review, presented below, is $n=2$. The reader can easily extract them by derivation from the various constraints amongst the vectors in $\boldsymbol{\sigma}$. We wish however to note that for general $n$ the matrices $\mathcal U$ and $\bar{\mathcal U}$ entering the GW system (\ref{eq:GW_system}) can be seen to belong to the affine untwisted Ka\v{c}-Moody algebra of type $B$ or $C$. By appropriately redefining the quantities listed above, one can connect this system with the corresponding Toda field theory. 
Off-critical generalisations of the ODE/IM correspondence associated to higher-rank algebras 
have been discussed in \cite{Dorey:2012bx, Ito:2013aea, Adamopoulou:2014fca, Negro:2017xwc, Ito:2015nla, Ito:2016qzt, Ito:2018wgj,Vicedo:2017cge}, although without specific analysis of the connection with surface embedding.
The case we focus on here, that is $n=2$, is particularly simple as the associated algebra turns out to be $B_1^{(1)} = \mathfrak{so}^{(1)}_3 \equiv A_1^{(1)} = \mathfrak{su}^{(1)}_2 $.

\subsection{Minimal surfaces in AdS$_3$}
\label{subsec:CMC_min_surf_ADS3}

While in section \ref{sec:surf_in_AdS} the description of embedded surfaces in AdS$_{n+1}$ was reviewed, here we concentrate on the simple case of minimal surfaces embedded in AdS$_3$.\footnote{In three dimensions, a minimal surface is defined by the vanishing of the mean curvature $H\equiv H_1 =0$.}  The number of functions we have to deal with collapses now to two: the real Pohlmeyer field $\tilde \varphi$ and the complex function $A_1 = A$. The former will be our unknown function, while we will consider $A$ as a given. 

As mentioned in section \ref{sec:surf_in_AdS}, the structural data of an embedded surface $\Sigma\subset\textrm{AdS}_{3}$ is contained in a pair of $4\times 4$ matrices $\mathcal U$ and $\bar{\mathcal U}$ satisfying the Gauss-Codazzi-Mainardi equation
\beq
    \mathcal U_{,\bar{z}} - \bar{\mathcal U}_{,z}+\left[\mathcal U,\bar{\mathcal U}\right] = 0\;.
\eeq
These matrices depend on the complex variables $\left(z,\bar{z}\right)$ through the Pohlmeyer field $\tilde{\varphi}$, its derivatives and the function $A$. In the case of a minimal surface in AdS$_{3}$ they take the following explicit form
\beq
    \mathcal U = \left(\begin{array}{c c c c}
            0 & 1 & 0 & 0 \\
            0 & {\tilde\varphi}_{,z} & 0 & -A \\
            \frac{1}{\alpha^2}e^{\tilde{\varphi}} & 0 & 0 & 0 \\
            0 & 0 & -e^{-\tilde{\varphi}}A & 0
        \end{array}\right)\;, \qquad
        \bar{\mathcal U} = \left(\begin{array}{c c c c}
            0 & 0 & 1 & 0 \\
            \frac{1}{\alpha^2}e^{\tilde{\varphi}} & 0 & 0 & 0 \\
            0 & 0 & \tilde{\varphi}_{,\bz} & -\bar{A} \\
            0 & -e^{-\tilde{\varphi}}\bar{A} & 0 & 0
        \end{array}\right)\;,
\eeq
and the GMC equation reduces to the non-linear partial differential equation
\beq
    \label{eq:ADS3_GMC}
    \tilde{\varphi}_{,z\bz} = \frac{1}{\alpha^2}e^{\tilde{\varphi}} - A\bar A e^{-\tilde{\varphi}}\;,\qquad A_{,\bz} = \bar A_{,z} = 0\;.
\eeq
This can be further simplified by introducing the quantities
\beq
    \varphi = \tilde{\varphi} - \log( 2\alpha^2)\;,\qquad P\left(z\right) =  \frac{1}{2\mathbbm i\alpha} A(z)\;,\qquad \bar P\left(\bz\right) =-\frac{1}{2\mathbbm i\alpha} \bar A(\bz)\;,
\label{eq:shG_quantities_rescaling}
\eeq
in terms of which the matrices $\mathcal U$ and $\bar{\mathcal U}$ read
\beq
    \mathcal U = \left(\begin{array}{c c c c}
        0 & 1 & 0 & 0 \\
        0 & \varphi_{,z} & 0 & -2\mathbbm i \alpha P \\
        2e^{\varphi} & 0 & 0 & 0\\
        0 & 0 &  -\mathbbm i\frac{e^{-\varphi}}{\alpha} P & 0
    \end{array}\right)\;,\quad \bar{\mathcal U} = \left(\begin{array}{c c c c}
        0 & 0 & 1 & 0 \\
        2e^{\varphi} & 0 & 0 & 0 \\
        0 & 0 & \varphi_{,\bz} & 2\mathbbm i \alpha \bar P \\
        0 & \mathbbm i \frac{e^{-\varphi}}{\alpha}  \bar P & 0 & 0
    \end{array}\right)\;,
\label{eq:ADS3_Lax_Matricex_simple}
\eeq
and the GMC equation takes the form of the so-called \emph{modified sinh-Gordon equation}
\beq
    \label{eq:ADS3_GMC_simple}
    \frac{1}{2}\varphi_{,z\bz} = e^{\varphi}-P\bar P e^{-\varphi}\;.
\eeq
This equation can be written in the form (\ref{eq:LsG}) by a shift of the field $\varphi$ together with a redefinition of the variables $z,\bar{z}$
\begin{subequations}
    \begin{align}
    &\varphi\left(z,\bar{z}\right) \;\longrightarrow \; \varphi\left(w\left(z\right),\bar{w}\left(\bar{z}\right)\right) +\frac{1}{2} \log\left(P\left(z\right)\bar{P}\left(\bar{z}\right)\right)\;, \\
    &w\left(z\right) = 2\intop^z \sqrt{P\left(z'\right)} dz'\;,\qquad \bar{w}\left(\bar{z}\right) = 2\intop^{\bar{z}} \sqrt{\bar{P}\left(\bar{z}'\right)} d\bar{z}'\;.
    \end{align}
\end{subequations}
We wish to remark that the above transformation, making (\ref{eq:ADS3_GMC_simple}) into (\ref{eq:LsG}), does alter the geometry on which the equation is considered. Moreover, equation (\ref{eq:ADS3_GMC_simple}) is defined on the space $\mathbb C^2$, on which we impose the real slice condition $\bar{z} = z^{\ast}$; on the other hand, equation (\ref{eq:LsG}) is defined on $\mathbb R^2$. Hence the two equations are not to be considered equivalent.

Although it is not immediately evident, the above pair (\ref{eq:ADS3_Lax_Matricex_simple}) can be gauge rotated to a tensor product form:\footnote{It is an easy exercise to verify that the GMC equations (and thus the structural data of $\Sigma$) is invariant under the gauge rotation $$ \left(\mathcal U,\bar{\mathcal U}\right) \longrightarrow \left(\Gamma^{-1}\mathcal U\Gamma - \Gamma^{-1}\Gamma_{,z},\Gamma^{-1}\bar{\mathcal U}\Gamma - \Gamma^{-1}\Gamma_{,\bz}\right)\;, $$ where $\Gamma$ is some $4\times 4$ matrix depending on $\left(z,\bz\right)$.}
\beq
    \mathcal U' = U_{\textrm{L}}\otimes \mathbbm 1_2 + \mathbbm 1_2 \otimes U_{\textrm{R}}\;,\qquad \bar{\mathcal U}' = \bar U_{\textrm{L}} \otimes \mathbbm 1_2 + \mathbbm 1_2 \otimes \bar U_{\textrm{R}}\;,
\eeq
where
\beq
    \mathcal U' = \Gamma^{-1}\mathcal U \Gamma -\Gamma^{-1}\Gamma_{,z}\;,\qquad \bar{\mathcal U}' = \Gamma^{-1}\mathcal U \Gamma -\Gamma^{-1}\Gamma_{,\bz}\;.
\eeq
The explicit expressions for the $2\times 2$ $U_{\textrm{R}}$, $U_{\textrm{L}}$, $\bar U_{\textrm{R}}$ and $\bar U_{\textrm{L}}$ matrices are as follows:
\begin{subequations}
    \begin{align}
    U_{\textrm{L}} &= \left(\begin{array}{c c}
        -\frac{1}{2} \varphi_{,z} & 1 \\ 
        P & \frac{1}{2} \varphi_{,z}
    \end{array}\right)\;,\qquad \bar U_L = \left(\begin{array}{c c}
        0 & \bar P e^{-\varphi}\\ 
        e^{\varphi} & 0
    \end{array}\right)\;, \\
    U_{\textrm{R}} &= \left(\begin{array}{c c}
        -\frac{1}{2} \varphi_{,z} & \mathbbm i \\ 
         \mathbbm i P & \frac{1}{2} \varphi_{,z}
    \end{array}\right)\;,\qquad \bar U_R = \left(\begin{array}{c c}
        0 & - \mathbbm i \bar P  e^{-\varphi}\\ 
        - \mathbbm i e^{\varphi} & 0
    \end{array}\right)\;,
    \end{align}
\end{subequations}
while the rotation matrix is
\beq
    \Gamma = \left(\begin{array}{c c c c}
        0 &  \mathbbm i \alpha & \alpha & 0 \\
        0 & 0 & 0 & 2 \mathbbm i \alpha \\
        2\alpha e^{\varphi} & 0 & 0 & 0 \\
        0 & -1 & -\mathbbm i & 0
    \end{array}\right)\;.
\eeq
One can further rotate both left and right pairs as
\beq
    L_{\textrm{L}} = e^{\frac{1}{4}\varphi \sigma^3} U_{\textrm{L}} e^{-\frac{1}{4}\varphi \sigma^3} - e^{\frac{1}{4}\varphi \sigma^3} \partial e^{-\frac{1}{4}\varphi \sigma^3}\;,\qquad \sigma^3 = \left(\begin{array}{c c}
        1 & 0 \\
        0 & -1
    \end{array}\right)\;,
\eeq
and similarly for the other three matrices, obtaining the more symmetric form
\begin{subequations}
\label{eq:rotated_pair}
    \begin{align}
    L_{\textrm{L}} &= \left(\begin{array}{c c}
        -\frac{1}{4} \varphi_{,z} & e^{\frac{\varphi}{2}} \\ 
        P e^{-\frac{\varphi}{2}} & \frac{1}{4} \varphi_{,z}
    \end{array}\right)\;,\qquad \bar L_{\textrm{L}} = \left(\begin{array}{c c}
        \frac{1}{4} \varphi_{,\bz} & \bar P e^{-\frac{\varphi}{2}}\\ 
        e^{\frac{\varphi}{2}} & -\frac{1}{4} \varphi_{,\bz} 
    \end{array}\right)\;, \\
    L_{\textrm{R}} &= \left(\begin{array}{c c}
        -\frac{1}{4} \varphi_{,z} & \mathbbm i e^{\frac{\varphi}{2}} \\ 
        \mathbbm i P e^{-\frac{\varphi}{2}} & \frac{1}{4} \varphi_{,z}
    \end{array}\right)\;,\qquad \bar L_{\textrm{R}} = \left(\begin{array}{c c}
        \frac{1}{4} \varphi_{,\bz} & -\mathbbm i \bar P e^{-\frac{\varphi}{2}}\\ 
        -\mathbbm i e^{\frac{\varphi}{2}} & -\frac{1}{4} \varphi_{,\bz} 
    \end{array}\right)\;.
    \end{align}
\end{subequations}
As a consequence of the above decomposition, the rotated frame $ \boldsymbol{\sigma}' = \left(e^{\frac{1}{4}\varphi \sigma^3}\otimes e^{\frac{1}{4}\varphi \sigma^3}\right)\Gamma^{-1}\boldsymbol{\sigma}$ is also decomposed as
\beq
    \boldsymbol{\sigma}' = \boldsymbol{\Psi} \mathbf{M}_0\;,\qquad \boldsymbol{\Psi} = \boldsymbol{\Psi}_{\textrm{L}} \otimes \boldsymbol{\Psi}_{\textrm{R}}\;,
\eeq
where $\mathbf{M}_0$ is a constant $4\times 4$ matrix, while $\boldsymbol{\Psi}_{\textrm{L}}$ and $\boldsymbol{\Psi}_{\textrm{R}}$ are solutions to their respective linear problems
\begin{subequations}
    \label{eq:L_R_ADS3_linear_problem}
    \begin{align}
        \boldsymbol{\Psi}_{\textrm{L},z} &= L_{\textrm{L}} \boldsymbol{\Psi}_{\textrm{L}}\;,\qquad \boldsymbol{\Psi}_{\textrm{L},\bz} = \bar L_{\textrm{L}} \boldsymbol{\Psi}_{\textrm{L}}\;, \label{subeq:L_ADS3_linear_problem}\\
        \boldsymbol{\Psi}_{\textrm{R},z} &= L_{\textrm{R}} \boldsymbol{\Psi}_{\textrm{R}}\;,\qquad \boldsymbol{\Psi}_{\textrm{R},\bz} = \bar L_{\textrm{R}} \boldsymbol{\Psi}_{\textrm{R}}\;. \label{subeq:R_ADS3_linear_problem}
    \end{align}
\end{subequations}

Recapitulating, given two solutions of the above systems (\ref{subeq:L_ADS3_linear_problem},\ref{subeq:R_ADS3_linear_problem}), one can reconstruct the corresponding embedding function $\vec Y$ for the minimal surface in AdS$_3$ as
\beqa
\label{eq:embedding_vector}
    \vec Y \equiv \vec e_1^{\;\textrm{T}} \boldsymbol{\sigma} = \vec e_1^{\;\textrm{T}} \Gamma \left(e^{-\frac{1}{4}\varphi \sigma^3}\otimes e^{-\frac{1}{4}\varphi \sigma^3}\right) \left(\boldsymbol{\Psi}_{\textrm{L}}\otimes \boldsymbol{\Psi}_{\textrm{R}}\right) \mathbf{M}_0 \;, \\
    \vec e_1^{\;\textrm{T}} = \left(\begin{array}{c c c c} 1\;, & 0\;, & 0\;, & 0\end{array}\right)\;. \notag
\eeqa
Let us also mention that the matrix $\mathbf{M}_0$ is not completely general. In fact its form can be almost entirely fixed by considering the orthogonality and normalisation conditions on the scalar products of the basis vectors, which in terms of $\boldsymbol{\sigma}$ can be written as
\beqa
    \boldsymbol{\sigma}\left(\begin{array}{cccc}
         -1 & 0 & 0 & 0 \\
         0 & -1 & 0 & 0 \\
         0 & 0 & 1 & 0 \\
         0 & 0 & 0 & 1
    \end{array}\right) \boldsymbol{\sigma}^{\textrm{T}} &=& \left(\begin{array}{cccc}
         \vec Y \cdot \vec Y & \vec Y \cdot \vec Y_{,z} & \vec Y \cdot \vec Y_{,\bz} & \vec Y \cdot \vec N \\
         \vec Y_{,z} \cdot \vec Y_{,z} & \vec Y \cdot \vec Y_{,z} & \vec Y_{,z} \cdot \vec Y_{,\bz} & \vec Y_{,z} \cdot \vec N \\
         \vec Y_{,\bz} \cdot \vec Y & \vec Y_{,\bz} \cdot \vec Y_{,z} & \vec Y_{,\bz} \cdot \vec Y_{,\bz} & \vec Y_{,\bz} \cdot \vec N \\
         \vec N \cdot \vec Y & \vec N \cdot \vec Y_{,z} & \vec N \cdot \vec Y_{,\bz} & \vec N \cdot \vec N
    \end{array}\right)  \notag \\
    &=&\left(\begin{array}{cccc}
         -\alpha^2 & 0 & 0 & 0 \\
         0 & 0 & e^{\tilde{\varphi}} & 0 \\
         0 & e^{\tilde{\varphi}} & 0 & 0 \\
         0 & 0 & 0 & -1
    \end{array}\right)\;.
\eeqa
One then has
\beq
    \left(\boldsymbol{\Psi}_{\textrm{L}}\otimes \boldsymbol{\Psi}_{\textrm{R}}\right) \mathbf{M}_0 \left(\sigma^3\otimes \mathbbm 1_2\right) \mathbf{M}_0^{\textrm{T}} \left(\boldsymbol{\Psi}_{\textrm{L}}\otimes \boldsymbol{\Psi}_{\textrm{R}}\right)^{\textrm{T}}  = \frac{\mathbbm i}{2} \left(\begin{array}{cccc}
         0 & 0 & 0 & 1 \\
         0 & 0 & -1 & 0 \\
         0 & -1 & 0 & 0 \\
         1 & 0 & 0 & 0
    \end{array}\right)\;,
\eeq
or, equivalently,
\beq
    \mathbf{M}_0\left(\sigma^3\otimes \mathbbm 1_2\right)\mathbf{M}_0^{\textrm{T}} = \frac{\mathbbm i/2}{\det(\boldsymbol{\Psi}_{\textrm{L}})\,\det(\boldsymbol{\Psi}_{\textrm{R}})}\left(\begin{array}{cccc}
         0 & 0 & 0 & 1 \\
         0 & 0 & -1 & 0 \\
         0 & -1 & 0 & 0 \\
         1 & 0 & 0 & 0
    \end{array}\right)\;.
    \label{eq:M_normalisation}
\eeq

It is a matter of straightforward computation to verify that the following matrix
\beq
    \mathbf{M}_{\textrm{spec}} = \frac{1}{2\sqrt{\det(\boldsymbol{\Psi}_{\textrm{L}})\,\det(\boldsymbol{\Psi}_{\textrm{R}})}} \left(\begin{array}{cccc}
         0 & \mathbbm i b & \mathbbm i b & 0 \\
         -\frac{1}{c} & 0 & 0 & \frac{1}{c} \\
         \mathbbm i c & 0 & 0 & \mathbbm i c \\
         0 & \frac{1}{b} & -\frac{1}{b} & 0
    \end{array}\right)\;,
\eeq
represents a particular solution to the equation \eqref{eq:M_normalisation}. In order to derive the general solution, we can reason as follows. Let $\mathbf{M}$ be a solution to \eqref{eq:M_normalisation} and $\mathbf{R} \in GL(4)$ a generic non-singular matrix. Then we can write $\mathbf{M} = \mathbf{R}\mathbf{M}_{\textrm{spec}}$. Due to both matrices  solving the same equation, the matrix $\mathbf{R}$ has to satisfy the following relation
\beq
    \mathbf{R} \left(\boldsymbol{\varsigma}\otimes \boldsymbol{\varsigma}\right)\mathbf{R}^t = \left(\boldsymbol{\varsigma}\otimes \boldsymbol{\varsigma}\right)\;,\qquad \boldsymbol{\varsigma} = \left(\begin{array}{c c}
        0 & 1 \\
        -1 & 0
    \end{array}\right)\;.
\eeq
Expanding this relation in $2\times 2$ blocks, we obtain the following three equations
\beqa
    \mathbf{R}_{11} \boldsymbol{\varsigma} \mathbf{R}_{12}^{t} & = & - \left(\mathbf{R}_{11} \boldsymbol{\varsigma} \mathbf{R}_{12}^{t}\right)^{t}\;, \nonumber \\
    \mathbf{R}_{21} \boldsymbol{\varsigma} \mathbf{R}_{22}^{t} & = & - \left(\mathbf{R}_{21} \boldsymbol{\varsigma} \mathbf{R}_{22}^{t}\right)^{t}\;, \label{eq:gen_const_mat_middle_relations}\\
    \mathbf{R}_{11} \boldsymbol{\varsigma} \mathbf{R}_{22}^{t} & + & \left(\mathbf{R}_{21} \boldsymbol{\varsigma} \mathbf{R}_{12}^{t}\right)^{t} = \boldsymbol{\varsigma} \;, \nonumber
\eeqa
where, evidently, $\mathbf{R}_{ij}$ are the $2\times2$ blocks of the matrix $\mathbf{R}$. The first two relations are solved by
\beq
    \mathbf{R}_{11} = a \boldsymbol{\varsigma}\left(\mathbf{R}_{12}^{t}\right)^{-1} \boldsymbol{\varsigma}^{-1} = \frac{a}{\det\left(\mathbf{R}_{12}\right)} \mathbf{R}_{12} \;,\qquad \mathbf{R}_{21} = a' \boldsymbol{\varsigma}\left(\mathbf{R}_{22}^{t}\right)^{-1} \boldsymbol{\varsigma}^{-1}  = \frac{a'}{\det\left(\mathbf{R}_{22}\right)} \mathbf{R}_{22}\;,
\eeq
where $a$ and $a'$ are some undetermined constants. Plugging the above solutions into the third equation of \eqref{eq:gen_const_mat_middle_relations}, we have
\beq
    a \boldsymbol{\varsigma}\left[\left(\mathbf{R}_{12}\mathbf{R}_{22}^{-1}\right)^t\right]^{-1} - a' \mathbf{R}_{12}\mathbf{R}_{22}^{-1} \boldsymbol{\varsigma} = \boldsymbol{\varsigma}\;,
\eeq
or, equivalently,
\beq
    \left(a\frac{\det\left(\mathbf{R}_{22}\right)}{\det\left(\mathbf{R}_{12}\right)} - a'\right) \mathbf{R}_{12}\mathbf{R}_{22}^{-1} = \mathbbm{1}_{2}\;,
\eeq
from which we deduce
\beq
    \mathbf{R}_{22} = a'' \mathbf{R}_{12} \quad,\quad 
a a'' - \frac{a'}{a''}=1\;.
\eeq
From these manipulations we conclude that
\beq
    \mathbf{R} = \left(\begin{array}{c c}
        \frac{a}{\det\left(\mathbf{R}_{12}\right)} & 1 \\
        \frac{b}{c \det\left(\mathbf{R}_{12}\right)} & c
    \end{array}\right)\otimes \mathbf{R}_{12}\;.
\eeq

We have found that we can write the general solution to \eqref{eq:M_normalisation} as follows
\beq
    \mathbf{M}_0 = \left(\mathbf{M}_L \otimes \mathbf{M}_R\right) \mathbf{M}_{\textrm{mix}}\;,
\label{eq:constant_matrix_decomposition}
\eeq
where $\mathbf{M}_L$ and $\mathbf{M}_R$ are $SL(2)$ matrices that rotate, respectively, the solutions $\Psi_L$ and $\Psi_R$, while $\mathbf{M}_{\textrm{mix}}$ takes the following form
\beq
    \mathbf{M}_{\textrm{mix}} = \frac{1}{2\sqrt{\det\left(\boldsymbol{\Psi}_L^M\right)\det\left(\boldsymbol{\Psi}_R^M\right)}}\left(\begin{array}{cccc}
         0 & \mathbbm i b & \mathbbm i b & 0 \\
         -\frac{1}{c} & 0 & 0 & \frac{1}{c} \\
         \mathbbm i c & 0 & 0 & \mathbbm i c \\
         0 & \frac{1}{b} & -\frac{1}{b} & 0
    \end{array}\right)\;,
\label{eq:mix_matrix_expression}
\eeq
with $\boldsymbol{\Psi}_L^M = \boldsymbol{\Psi}_L \mathbf{M}_L$ and similarly for the right one. We thus see that a generic constant matrix $\mathbf{M}_0$ in (\ref{eq:embedding_vector}) is determined by $10$ complex parameters, $4$ for each $SL(2)$ rotation $\mathbf{M}_{L/R}$ and an additional pair for the matrix $\mathbf{M}_{\textrm{mix}}$. Note that $10$ is the real dimension of the isometry group of the space $\mathbb R^{2,2}$, in which AdS$_3$ is immersed. 
A further condition on the constant matrix $\mathbf{M}_0$ comes from the reality properties of the basis vectors
\beq
    \boldsymbol{\sigma}^\ast = \left(\begin{array}{cccc}
         1 & 0 & 0 & 0 \\
         0 & 0 & 1 & 0 \\
         0 & 1 & 0 & 0 \\
         0 & 0 & 0 & 1
    \end{array}\right)\boldsymbol{\sigma}\;,
\eeq
which implies
\beq
    \left(\boldsymbol{\Psi}_{\textrm{L}}\otimes \boldsymbol{\Psi}_{\textrm{R}}\right)^{\ast} \mathbf{M}_0^{\ast} = \mathbbm i \left(\begin{array}{cccc}
         0 & 0 & 0 & 1 \\
         0 & 0 & 1 & 0 \\
         0 & 1 & 0 & 0 \\
         1 & 0 & 0 & 0
    \end{array}\right) \left(\boldsymbol{\Psi}_{\textrm{L}}\otimes \boldsymbol{\Psi}_{\textrm{R}}\right) \mathbf{M}_0\;.
\label{eq:reality_condition_on_solutions}
\eeq
and reduces the $10$ complex parameter determining $\mathbf{M}_0$ to $10$ real ones. Hence our embedded surface determined by (\ref{eq:embedding_vector}) is uniquely determined up to isometries of $\mathbb R^{2,2}$.

Finally, let us also mention that minimal surfaces are naturally related to string theory. The very fact of being minimal implies the possibility of obtaining their defining relations by means of the minimisation of some quantity which, as it turns out, is nothing but the action of a non-linear sigma model
\beq
    \mathscr A_{\textrm{NLSM}} = \intop_{\Sigma} dz\,d\bz \left( \vec Y_{,z} \cdot \vec Y_{,\bz} + \Lambda \left(\vec Y \cdot \vec Y + \alpha^2\right)\right)\;,
\eeq
where the Lagrange multiplier $\Lambda$ imposes the constraint (\ref{eq:AdS_vector_constraint}), forcing the target space to be AdS$_3$. The equations of motion
\beq
    \vec Y_{,z\bz} = \frac{1}{\alpha^2}\left(\vec Y_{,z} \cdot \vec Y_{,\bz}\right)\vec Y\;,\qquad \vec Y_{,z}\cdot \vec Y_{,z} = \vec Y_{,\bz} \cdot \vec Y_{,\bz} = 0\;,
\eeq
are rather easily connected with (\ref{eq:ADS3_GMC}) \cite{Jevicki:2007aa,Dorn:2009gq,Miramontes:2008wt}. The area $\mathcal A$ of the worldsheet is then computed thanks to the metric $g$ as follows
\beq
\label{eq:Area_formula}
    \mathcal A = \intop_{\Sigma} dz\,d\bz \sqrt{-\det(g)} = \intop_{\Sigma} dz\,d\bz \left(\vec Y_{,z} \cdot \vec Y_{,\bz}\right) = \intop_{\Sigma} dz\,d\bz \, e^{\tilde{\varphi}}\;.
\eeq
Note that, due to the modified sinh-Gordon equation (\ref{eq:ADS3_GMC_simple}), one has
\beq
\label{eq:Area_formula_2}
    \mathcal A  = 2\alpha^2\intop_{\Sigma} dz\,d\bz \left(\varphi_{,z\bz}+P\bar P e^{-\varphi}\right) = 2\alpha^2\intop_{\Sigma} dz\,d\bz\; P\bar P e^{-\varphi} + \textrm{total derivatives}\;,
\eeq
where the total derivative term is a constant independent of the kinematics. This area is divergent and needs to be regularized. As will be explained below, the asymptotic behaviour as $\vert z\vert\rightarrow\infty$ of the modified sinh-Gordon field is $\varphi \sim \log\vert P\vert$ and one can define a regularized area
\beq
\label{eq:Area_formula_regularized}
    \mathcal A_{\textrm{reg}}  = 2\alpha^2\intop_{\Sigma} dz\,d\bz \left(P\bar P e^{-\varphi} - \left(P \bar{P}\right)^{\frac{1}{2}} \right)\;.
\eeq

\subsection{A boundary interpretation of the function $P$ and the Wilson loop}
\label{subsec:boundary_p}

Let us recall that the function $P$ -- equivalently $A$ (\ref{eq:shG_quantities_rescaling}) -- is related to the Gauss curvature through equation (\ref{subeq:Gauss_curvature}). In the current case we have
\beq
    K = -e^{-2\tilde{\varphi}}A\bar{A} = -\frac{1}{\alpha^2}e^{-2\varphi}P\bar P\;.
\label{eq:Gauss_in_term_of_potential}
\eeq
Thus, since we wish the surface $\Sigma$ to be everywhere regular, we must demand for solutions to (\ref{eq:ADS3_GMC}) to compensate for divergences of $P$. More concretely, we impose that
\beq
    \lim_{\left(z,\bz\right)\rightarrow \left(z_{\textrm{c}},\bar{z}_{\textrm{c}}\right)} \frac{1}{\left\vert P\right\vert} = 0\quad \Longrightarrow \quad \varphi \underset{\left(z,\bz\right)\rightarrow \left(z_{\textrm{c}},\bar{z}_{\textrm{c}}\right)}{\sim} \log\left\vert P \right\vert\;.
\label{eq:varphi_asymptotics}
\eeq
Note that this asymptotic behaviour at the singularities of $P$ is consistent with equation (\ref{eq:ADS3_GMC}). 
From now on we will assume that the function $P$ is a polynomial of order $2N$, then the only singular point is $\left\vert z\right\vert\rightarrow\infty$. The Gaussian curvature is, therefore, asymptotically a constant
\beq
    K_{\infty} =\lim_{\left\vert z\right\vert\rightarrow\infty} K = -\frac{1}{\alpha^2}\;,
\eeq
and in this limit the matrices of the linear system (\ref{eq:L_R_ADS3_linear_problem}) become
\begin{subequations}
    \begin{align}
    L_{\textrm{L}} \sim \left(\begin{array}{c c}
        0 & z^{N/2}\bz^{N/2} \\ 
        z^{3N/2}\bz^{-N/2} & 0
    \end{array}\right)\;,\qquad &\bar L_{\textrm{L}} \sim \left(\begin{array}{c c}
        0 & z^{-N/2}\bz^{3N/2}\\ 
        z^{N/2}\bz^{N/2} & 0 
    \end{array}\right)\;,& \\
    L_{\textrm{R}} \sim \left(\begin{array}{c c}
        0 & \mathbbm i z^{N/2}\bz^{N/2} \\ 
        \mathbbm i z^{3N/2}\bz^{-N/2} & 0
    \end{array}\right)\;,\quad\; &\bar L_{\textrm{R}} \sim \left(\begin{array}{c c}
        0 & -{\mathbbm i} z^{-N/2}\bz^{3N/2}\\ 
        -{\mathbbm i} z^{N/2}\bz^{N/2} & 0 
    \end{array}\right)\;.&
    \end{align}
\end{subequations}

In order to study what happens to the boundary of AdS$_3$ we need to jump ahead of ourselves and consider the first order in the WKB expansion of the solutions $\boldsymbol{\Psi}_{\textrm{L}}$ and $\boldsymbol{\Psi}_{\textrm{R}}$. A more detailed analysis of the WKB solutions and the Stokes phenomenon will be given in section \ref{subsec:Stokes_phenomenon}; here we will just present some facts which will be useful in deriving the boundary of the minimal surface. A simple WKB analysis (cf. section \ref{sec:excited}) yields 
\begin{subequations}
    \begin{align}
        \boldsymbol{\Psi}_{\textrm{L}} &\propto \left(\begin{array}{cc}
             e^{\frac{2 \,\varrho^{N+1}}{N+1}\cos\left(\left(N+1\right)\vartheta\right)} & -e^{-\mathbbm i N\vartheta} e^{-\frac{2 \,\varrho^{N+1}}{N+1}\cos\left(\left(N+1\right)\vartheta\right)}  \\
             e^{\mathbbm i N \vartheta} e^{\frac{2 \, \varrho^{N+1}}{N+1}\cos\left(\left(N+1\right)\vartheta\right)} & e^{-\frac{2 \, \varrho^{N+1}}{N+1}\cos\left(\left(N+1\right)\vartheta\right)} 
        \end{array}\right)\;, \\
        \boldsymbol{\Psi}_{\textrm{R}} &\propto \left(\begin{array}{cc}
             e^{\frac{2 \,\varrho^{N+1}}{N+1}\sin\left(\left(N+1\right)\vartheta\right)} & e^{-\mathbbm i N\vartheta} e^{-\frac{2 \,\varrho^{N+1}}{N+1}\sin\left(\left(N+1\right)\vartheta\right)} \\
             -e^{\mathbbm i N \vartheta} e^{\frac{2 \,\varrho^{N+1}}{N+1}\sin\left(\left(N+1\right)\vartheta\right)} & e^{-\frac{2 \, \varrho^{N+1}}{N+1}\sin\left(\left(N+1\right)\vartheta\right)} 
        \end{array}\right)\;,
    \end{align}
\end{subequations}
with $z=\varrho e^{\mathbbm i\vartheta}$ and $\bz=\varrho e^{ -\mi\vartheta}$.
\begin{figure}[t!]
  \centering
  \includegraphics[scale=0.4]{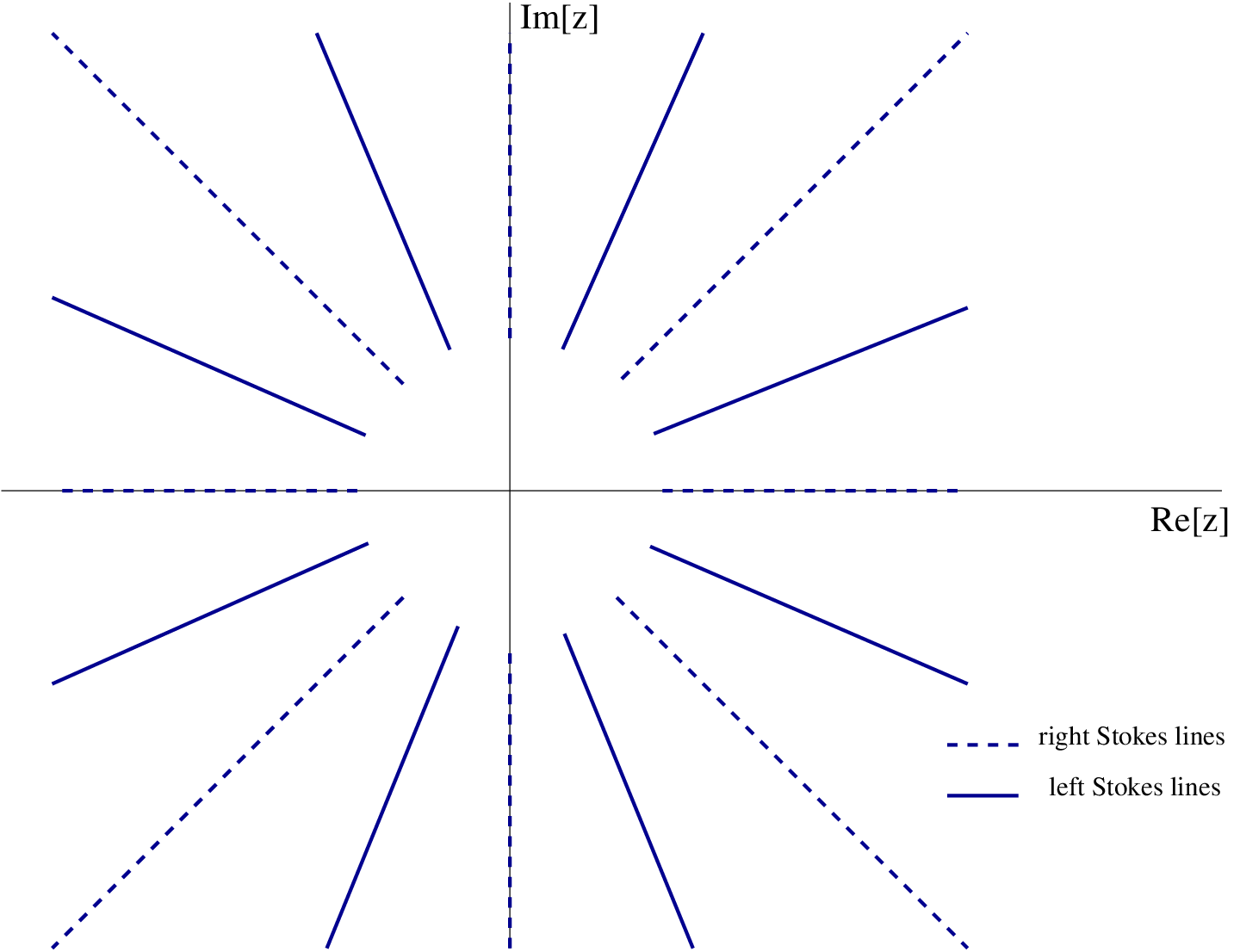}
\caption{\small Representation of the Stokes sectors and lines in the complex $(z,\bz)$ plane, for the linear system (\ref{eq:L_R_ADS3_linear_problem}), with $P\sim z^{2N}$ and $N=3$.}
\label{fig:Stokes_n_3}
\end{figure}
We see that the linear problem displays a Stokes phenomenon at $\varrho \rightarrow \infty$, meaning that we can pin down the asymptotic of a specific solution only in certain sectors of the complex plane (see figure \ref{fig:Stokes_n_3}). These sectors, which we denote by $\mathcal S_{\textrm{L}}^{(i)}$ and $\mathcal S_{\textrm{R}}^{(i)}$, are bounded by the anti-Stokes lines which are given by $\cos\left(\left(N+1\right)\vartheta\right) =\textrm{Re}\left[z^{N+1}\right]= 0$ for the left solution and by $\sin\left(\left(N+1\right)\vartheta\right) = \textrm{Im}\left[z^{N+1}\right] = 0$ for the right one.

Now, we choose a solution $\boldsymbol{\Psi}_{\textrm{L}}^{(i)} \otimes \boldsymbol{\Psi}_{\textrm{R}}^{(i)}$ having the above asymptotic behaviour in a definite sector of the complex plane, which happens to be the overlap of $\mathcal S_{\textrm{L}}^{(i)}$ with $\mathcal S_{\textrm{R}}^{(i)}$. Suppose that we rotate our solution in the complex plane and, at some point, we cross a left anti-Stokes line. Then the asymptotic of our solution will change, since the diverging solution might obscure the presence of a smaller decaying solution. In mathematical terms,
\beq
    \boldsymbol{\Psi}_{\textrm{L}}^{(i)} \otimes \boldsymbol{\Psi}_{\textrm{R}}^{(i)} = \left( \boldsymbol{\Psi}_{\textrm{L}}^{(i+1)} \mathbf{S}\left(\gamma_{\textrm{L}}^{(i)}\right)\right) \otimes \boldsymbol{\Psi}_{\textrm{R}}^{(i)}\;,\qquad \mathbf{S}\left(\gamma\right) = \left(\begin{array}{cc}
         0 & -1 \\
         1 & \gamma
    \end{array}\right)\;.
    \label{eq:Stokes_rotation}
\eeq
A similar jump will happen for the right solution at the right anti-Stokes lines, meaning we have $4(N+1)$ parameters $\left\lbrace \gamma_{\textrm{L}}^{(i)},\gamma_{\textrm{R}}^{(i)}\right\rbrace_{i=1}^{2(N+1)}$, one for each anti-Stokes line.

Now let us consider what happens to the surface embedding function $\vec Y$ for $\left\vert z\right\vert\rightarrow\infty$.
We will see things more clearly by working in Poincar\'{e} coordinates (\ref{eq:ADS_poincarecoordinates}):
\beq
    \mathsf{r} = Y_{-1} + Y_{2}\;,\qquad \mathsf{x}^{\pm} = \mathsf{x}\pm\mathsf{t} = \frac{Y_1\pm Y_0}{Y_{-1}+Y_2}\;,
\eeq
where we have introduced the light-cone Poincar\'{e} coordinates $\mathsf{x}^{\pm}$. Some simple but tedious computation shows that these coordinates have the following expression\footnote{Note that we have not implemented the reality condition (\ref{eq:reality_condition_on_solutions}) in the above expression. When doing so, these embedding functions will be, clearly, real.} for our embedding (\ref{eq:embedding_vector})
\beqa
    \mathsf{r} &=& \mathbbm i\alpha c \frac{\boldsymbol{\Psi}_{L,22}^M \boldsymbol{\Psi}_{R,11}^M + \mathbbm i \boldsymbol{\Psi}_{L,12}^M \boldsymbol{\Psi}_{R,21}^M}{\sqrt{\det\left(\boldsymbol{\Psi}_{L}^M\right)\det\left(\boldsymbol{\Psi}_{R}^M\right)}}\;, \notag \\
    \mathsf{x}^+ &=& \frac{b}{c} \frac{\boldsymbol{\Psi}_{L,21}^M \boldsymbol{\Psi}_{R,11}^M + \mathbbm i \boldsymbol{\Psi}_{L,11}^M \boldsymbol{\Psi}_{R,21}^M}{\boldsymbol{\Psi}_{L,22}^M \boldsymbol{\Psi}_{R,11}^M + \mathbbm i \boldsymbol{\Psi}_{L,12}^M \boldsymbol{\Psi}_{R,21}^M}\;, \label{eq:Poincare_embedding}\\
    \mathsf{x}^- &=& \frac{1}{\mathbbm ib\,c} \frac{\boldsymbol{\Psi}_{L,22}^M \boldsymbol{\Psi}_{R,12}^M + \mathbbm i \boldsymbol{\Psi}_{L,12}^M \boldsymbol{\Psi}_{R,22}^M}{\boldsymbol{\Psi}_{L,22}^M \boldsymbol{\Psi}_{R,11}^M + \mathbbm i \boldsymbol{\Psi}_{L,12}^M \boldsymbol{\Psi}_{R,21}^M}\;, \notag
\eeqa
where we used (\ref{eq:constant_matrix_decomposition}) and (\ref{eq:mix_matrix_expression}) while $\boldsymbol{\Psi}_{L,ij}^M$ and $\boldsymbol{\Psi}_{R,ij}^M$ are the components of the rotated solutions $\boldsymbol{\Psi}_{L} M_L$ and $\boldsymbol{\Psi}_{R} M_R$, respectively.

Let us suppose we are in a Stokes sector, away from Stokes lines; in the next few expressions, in order to lighten the notation, we will omit the superscript $(i)$ specifying the Stokes sector. Then, as $\vert z\vert \rightarrow \infty$, the components $\boldsymbol{\Psi}_{L,ij}^M$ and $\boldsymbol{\Psi}_{R,ij}^M$ will be naturally expressed by a superposition of a growing and a decaying solution:
\beq
    \boldsymbol{\Psi}_{L,ij}^M = c_{L,j}^{\textrm{large}} \psi_{L,i}^{\textrm{large}} + c_{L,j}^{\textrm{small}} \psi_{L,i}^{\textrm{small}}\;,
\label{eq:rotated_sol_as_large_small_superposition}
\eeq
where the functions $\psi_{L/R,i}^{\textrm{large}}$ and $\psi_{L/R,i}^{\textrm{small}}$ are the components of two arbitrary vector solutions to the linear system (\ref{eq:L_R_ADS3_linear_problem}) respectively diverging and decaying\footnote{In sec. \ref{subsec:Stokes_phenomenon} we will define more precisely solutions to the linear problem according to their asymptotic behaviour. There we will refer to them as \emph{dominant} and \emph{subdominant}. For the moment, however, we content ourselves with this intuitive definition as it will be sufficient to gain a qualitative understanding of the asymptotic behaviour of the embedded surface. For this same reason we follow the example of \cite{Alday:2009yn} and denote them as \emph{large} and \emph{small}.} as $\vert z\vert\rightarrow \infty$ in our chosen Stokes sector. We easily verify that
\beq
    c_{L,j}^{\textrm{large}} = \frac{\det \left(\begin{array}{c c} \boldsymbol{\Psi}_{L,1j}^{M} & \psi_{L,1}^{\textrm{small}} \\ \boldsymbol{\Psi}_{L,2j}^{M} & \psi_{L,2}^{\textrm{small}} \end{array}\right)}{\det \left(\begin{array}{c c} \psi_{L,1}^{\textrm{large}} & \psi_{L,1}^{\textrm{small}} \\ \psi_{L,2}^{\textrm{large}} & \psi_{L,2}^{\textrm{small}} \end{array}\right)}\;,\qquad c_{L,j}^{\textrm{small}} = -\frac{\det \left(\begin{array}{c c} \boldsymbol{\Psi}_{L,1j}^{M} & \psi_{L,1}^{\textrm{large}} \\ \boldsymbol{\Psi}_{L,2j}^{M} & \psi_{L,2}^{\textrm{large}} \end{array}\right)}{\det \left(\begin{array}{c c} \psi_{L,1}^{\textrm{large}} & \psi_{L,1}^{\textrm{small}} \\ \psi_{L,2}^{\textrm{large}} & \psi_{L,2}^{\textrm{small}} \end{array}\right)}\,.
\label{eq:aux_constants_for_boundary}
\eeq
Equivalent expressions hold for the constants $c_{R,j}^{(\textrm{large})/(\textrm{small})}$. Finally plugging (\ref{eq:rotated_sol_as_large_small_superposition}) into (\ref{eq:Poincare_embedding}), we see that in the limit $\vert z\vert \rightarrow \infty$, the Poincar\'{e} radius diverges\footnote{Indeed, the numerator of $\mathsf{r}$ in (\ref{eq:Poincare_embedding}) is dominated by $\psi_{L,i}^{\textrm{large}}$ and $\psi_{L,i}^{\textrm{small}}$, while the denominator is a constant.} $\mathsf{r}\rightarrow \infty$ -- signalling that we are indeed approaching the boundary $\partial$AdS$_3$ -- while the light cone coordinates take the following simple form
\beq
    \mathsf{x}^+ = \frac{b}{c}\frac{\det \left(\begin{array}{c c} \boldsymbol{\Psi}_{L,11}^{M} & \psi_{L,1}^{\textrm{small}} \\ \boldsymbol{\Psi}_{L,21}^{M} & \psi_{L,2}^{\textrm{small}} \end{array}\right)}{\det \left(\begin{array}{c c} \boldsymbol{\Psi}_{L,12}^{M} & \psi_{L,1}^{\textrm{small}} \\ \boldsymbol{\Psi}_{L,22}^{M} & \psi_{L,2}^{\textrm{small}} \end{array}\right)}\;,\quad \mathsf{x}^- = \frac{1}{\mathbbm i b c}\frac{\det \left(\begin{array}{c c} \boldsymbol{\Psi}_{R,12}^{M} & \psi_{R,1}^{\textrm{small}} \\ \boldsymbol{\Psi}_{R,22}^{M} & \psi_{R,2}^{\textrm{small}} \end{array}\right)}{\det \left(\begin{array}{c c} \boldsymbol{\Psi}_{R,11}^{M} & \psi_{R,1}^{\textrm{small}} \\ \boldsymbol{\Psi}_{R,21}^{M} & \psi_{R,2}^{\textrm{small}} \end{array}\right)}\;.
\eeq
Note that, while the expressions (\ref{eq:aux_constants_for_boundary}) depend on the choice of normalization for the functions $\psi_{L/R,i}^{\textrm{large}}$ and $\psi_{L/R,i}^{\textrm{small}}$, the boundary light-cone coordinates above are independent of it.

Given these results, we can easily see what happens when a Stokes line, say a left one, is crossed. Let us  reinstate the explicit index for the sector: $\mathsf{x}^{+}_{(i)}$ and $\mathsf{x}^{-}_{(i)}$ are given by the above expressions, where each of the components of the solutions $\boldsymbol{\Psi}_{L}^M$, $\boldsymbol{\Psi}_{R}^M$, $\psi_{L}^{\textrm{small}}$, $\psi_{L}^{\textrm{large}}$ are defined in the overlap of the $i$-th Stokes sectors $\mathcal S_L^{(i)} \cap \mathcal S_R^{(i)}$. Looking back at (\ref{eq:Stokes_rotation}), we notice that crossing a left Stokes line, only the light-cone coordinate $\mathsf{x}^+_{(i)}$ is influenced, while $\mathsf{x}^-_{(i)}$ is the same on both sides of the left Stokes line. In other words, in $\mathcal S_L^{(i)} \cap \mathcal S_R^{(i)}$ we have light-cone boundary coordinates $\left(\mathsf{x}^+_{(i)},\mathsf{x}^-_{(i)}\right)$, while in $\mathcal S_L^{(i+1)} \cap \mathcal S_R^{(i)}$ they are $\left(\mathsf{x}^+_{(i+1)},\mathsf{x}^-_{(i)}\right)$. The same exact reasoning repeats for the crossing of a right Stokes line. Hence we conclude that points on the boundary determined by solutions lying in neighboring Stokes sectors are light-like separated. 
\begin{figure}[t!]
 \centering
  \subfloat[]{\includegraphics[scale=0.5]{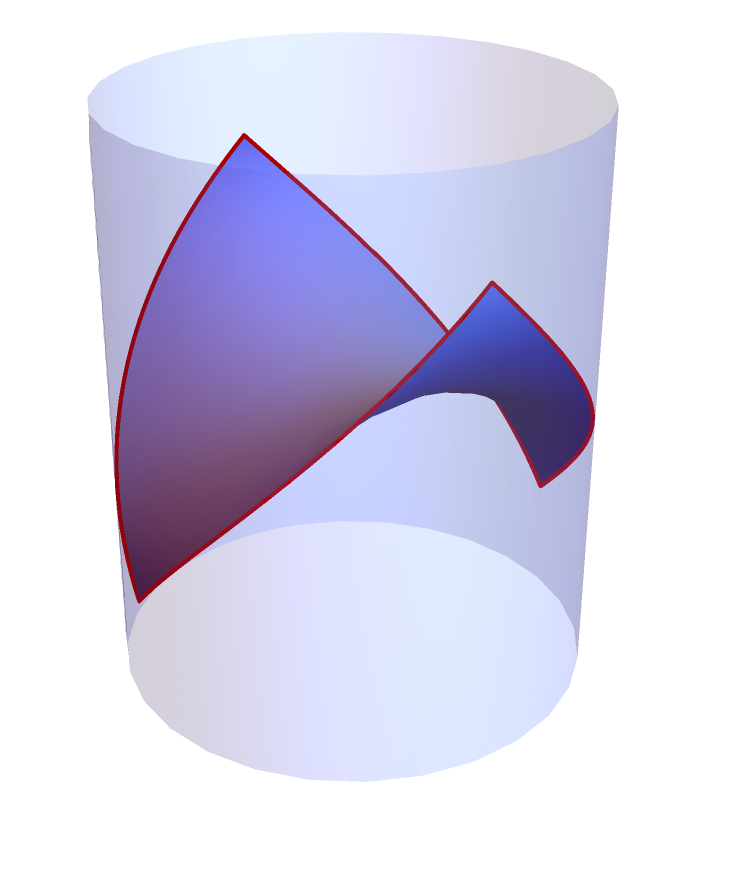}
  \label{fig:ADS3_quadrangle}}
  \hspace{1cm}
  \subfloat[]{\includegraphics[scale=0.6]{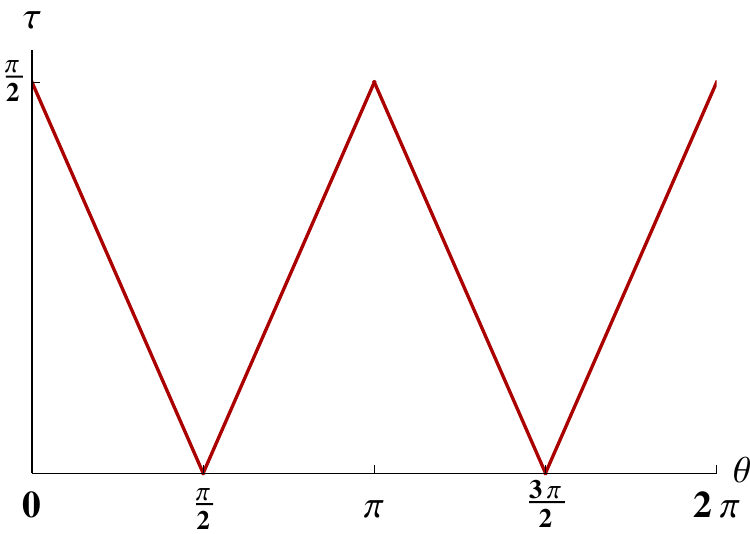}
  \label{fig:ADS3_quadrangle_WL}}
\caption{\small Minimal surface for the case $P = \bar P = 1$, $\alpha = 1$ and $\varphi = 0$ in AdS$_3$ and its Wilson loop. Figure \ref{fig:ADS3_quadrangle} is a representation with $\tanh(\rho)$ as a radius, $\tau$ as a vertical direction and $\theta$ as an angle where $\left(\rho,\tau,\theta\right)$ are AdS$_3$ global coordinates (\ref{eq:ADS_globalcoordinates}). The shaded cylinder is the conformal boundary and the red line is the Wilson loop. Figure \ref{fig:ADS3_quadrangle_WL} is a plot of the Wilson loop on the plane $(\theta,\tau)$ corresponding to the boundary $\tanh\rho =1$. 
}
\label{fig:ADS3_quadrilateral_min_surf}
\end{figure}

Recapitulating, we have seen that the order $2N$ polynomial $P$ defines $4(N+1)$ distinct Stokes sectors on the $\left(z,\bz\right)$ plane and, consequently, $4(N+1)$ points on the boundary of AdS$_3$. These are connected by $4(N+1)$ light-like lines, forming a light-like $4(N+1)$-gon on the boundary of AdS$_3$. In figure \ref{fig:ADS3_quadrilateral_min_surf} we plotted the minimal surface, along with its Wilson loop, for the simplest possible case $P = \bar P = 1$, $\alpha = 1$ and $\varphi = 0$. The polygon on the boundary has the interpretation, in the CFT living on $\partial$AdS$_3$, as a light-like Wilson loop and, according to the proposal of \cite{Rey:1998ik,Maldacena:1998im}, we can measure its expectation value by computing the area of the minimal surface $\Sigma$ in AdS$_3$ having the Wilson loop as its boundary. Moreover, as explained in \cite{Alday:2007hr,Alday:2010vh}, this same area can be used to compute the gluon scattering amplitude, at leading order in strong coupling, in the boundary theory.

We will now turn to a more in-depth analysis of the solutions to the linear problem (\ref{eq:L_R_ADS3_linear_problem}). As we will see, the presence of the Stokes phenomenon, instead of being a hindrance, will allow us to derive a closed set of functional equations for a collection of functions $Y_k$\footnote{These functional equations form a closed set only if $P(z)$ lives on a finite cover of $\mathbb C$. This can be understood intuitively from the fact that there exists a function $Y_k$ for each generator of the first homology group $\textrm{H}_1\left(\mathcal{R}_{\textrm{WKB}},\mathbb Z\right)$ of the Riemann surface $\mathcal{R}_{\textrm{WKB}}$ associated to $\sqrt{P}$. If we allow non-rational powers in $P$, then the 
first homology group of this Riemann surface will not be finitely
generated and we will have to deal with an infinite set of functions $Y_k$. From a physical point of view, in this case on the boundary of AdS$_3$ there will be an infinity of light-like lines, never closing themselves into a polygon.}. These can then be exploited to reconstruct the solutions $\boldsymbol{\Psi}_{\textrm{L}}$ and $\boldsymbol{\Psi}_{\textrm{R}}$ and compute the area (\ref{eq:Area_formula_regularized}) of the minimal surface.

\subsection{The associated linear problem, the spectral parameter and the WKB solutions}
\label{subsec:ass_lin_prob}

The left and right pair of matrices (\ref{eq:rotated_pair}) are, essentially, the Lax operators for the modified sinh-Gordon model appearing in \cite{Lukyanov:2010rn}:
\beq
\label{eq:shG_Lax}
    \mathcal L\left(\lambda\right) = \left(\begin{array}{c c}
        -\frac{1}{4} \varphi_{,z} & \lambda e^{\frac{\varphi}{2}} \\ 
        \lambda P e^{-\frac{\varphi}{2}} & \frac{1}{4} \varphi_{,z}
    \end{array}\right)\;,\qquad \bar{\mathcal L}\left(\lambda\right) = \left(\begin{array}{c c}
        \frac{1}{4} \varphi_{,\bz} & \frac{1}{\lambda} \bar P e^{-\frac{\varphi}{2}}\\ 
        \frac{1}{\lambda} e^{\frac{\varphi}{2}} & -\frac{1}{4} \varphi_{,\bz} 
    \end{array}\right)\;.
\eeq
 The only missing element in the pairs (\ref{eq:rotated_pair}) is the spectral parameter $\lambda$. However we immediately notice that by specialising the value of $\lambda$ one has
\begin{subequations}
\begin{align}
\label{eq:specialization}
	L_{\textrm{L}} &= \mathcal L\left(\lambda = 1\right)\;,\qquad \bar L_{\textrm{L}} = \bar{\mathcal L}\left(\lambda = 1\right)\;, \\
	L_{\textrm{R}} &= \mathcal L\left(\lambda = \mathbbm i\right)\;,\qquad \bar L_{\textrm{R}} = \bar{\mathcal L}\left(\lambda = \mathbbm i\right)\;.
\end{align}
\end{subequations}
The analysis of the Lax pair (\ref{eq:shG_Lax}) has been carried out in \cite{Lukyanov:2010rn} for the particular case of the function $P\left(z\right) = z^{2M} - s^{2M}$. There it was shown that the generalised monodromy data for the linear problem
\beq
\label{eq:shG_linear_problem}
    \boldsymbol{\Phi}_{,z} = \mathcal L \, \boldsymbol{\Phi} \;,\qquad \boldsymbol{\Phi}_{,\bz} = \bar{\mathcal L} \,\boldsymbol{\Phi} \;,
\eeq
is connected with the integrable 
structures of the quantum sine-Gordon (for $M>0$) or sinh-Gordon (for $M<-1$) models. As mentioned above, in what follows we will think of $P\left(z\right)$ as a polynomial function of order $2N$.\footnote{We might think of considering more general multi-valued potentials, e.g. $P\left(z\right) = z^{2N} - s^{2N}$ where $N\notin \frac{1}{2}\mathbb Z$ but we still ask that $N\in\mathbb Q$. The presence of non-integer powers in the function $P\left(z\right)$ would  force us to consider the linear problem on an appropriate finite covering of the complex plane. Since the substance of our analysis would not change, we will avoid this complication.} For further simplicity, we will concentrate on polynomials having only real roots; hence, from now on we will set
\beq
    P\left(z\right) = z^{2N} + \sum_{k=0}^{2N-1} P_k \, z^k = \prod_{k=1}^{2N}\left(z-z_k\right)\;,\qquad (z_k,P_k\in\mathbb R)\;.
\eeq

The first thing we notice about the linear problem (\ref{eq:shG_linear_problem}) is that it possesses a $\mathbb Z_2$ symmetry
\beq
	\left(\mathcal L\left(z,\bar{z}\vert\lambda\right),\bar{\mathcal L}\left(z,\bar{z}\vert\lambda\right)\right) = \left(\sigma^3\mathcal L\left(z,\bar{z}\vert-\lambda\right)\sigma^3,\sigma^3\bar{\mathcal L}\left(z,\bar{z}\vert-\lambda\right)\sigma^3\right)\;,
\label{eq:Z2_symmetry}
\eeq
which implies that, given a solution $\boldsymbol{\Phi}\left(z,\bar{z}\vert\lambda\right)$, then $\sigma^3 \boldsymbol{\Phi}\left(z,\bar{z}\vert e^{\mathbbm i\pi}\lambda\right)$ is also a solution. This fact will be useful momentarily, when we discuss the Stokes phenomenon associated with our linear problem. A simple way to study the linear problem (\ref{eq:shG_linear_problem}) is to gauge rotate it by the matrix $\exp\left( \frac{1}{4}\varphi \sigma^3 \right)$, so that one obtains
\beq
    \tilde{\boldsymbol{\Phi}}_{,z} = \tilde{\mathcal L} \, \tilde{\boldsymbol{\Phi}}\;,\qquad \tilde{\boldsymbol{\Phi}}_{,\bar{z}} = \tilde{\bar{\mathcal L}} \, \tilde{\boldsymbol{\Phi}}\;,\qquad \tilde{\boldsymbol{\Phi}} = e^{-\frac{1}{4}\varphi \sigma^3} \, \boldsymbol{\Phi}\;,
\label{eq:shG_linear_problem_again}
\eeq
where
\beq
    \tilde{\mathcal L} = e^{-\frac{1}{4}\varphi \sigma^3}\, \mathcal L \, e^{\frac{1}{4}\varphi \sigma^3} - e^{-\frac{1}{4}\varphi \sigma^3}\partial \left(e^{\frac{1}{4}\varphi \sigma^3}\right) = \left(
    \begin{array}{c c}
        -\frac{1}{2}\varphi_{,z} & \lambda \\
        
        \lambda P & \frac{1}{2}\varphi_{,z}
    \end{array}
    \right)\;,
\eeq
and
\beq
    \tilde{\bar{\mathcal L}} = e^{-\frac{1}{4}\varphi \sigma^3}\bar{\mathcal L} e^{\frac{1}{4}\varphi \sigma^3} - e^{-\frac{1}{4}\varphi \sigma^3}\bar{\partial}e^{\frac{1}{4}\varphi \sigma^3} = \left(
    \begin{array}{c c}
        0& \frac{1}{\lambda} \bar{P} e^{-\varphi} \\
        
        \frac{1}{\lambda} e^{\varphi} & 0
    \end{array}
    \right)\;.
\eeq
With this form of the linear problem, it is easier to obtain the WKB expansion.

We start from the following ansatz
\beq
    \tilde{\boldsymbol{\Phi}} = \frac{1}{\sqrt{S_{,z}}}\left(
    \begin{array}{c c}
        1 & 1 \\
        \left( S + \frac{\varphi - \log(\partial S)}{2\lambda}\right)_{,z} & \left(- S + \frac{\varphi - \log(\partial S)}{2\lambda}\right)_{,z}
    \end{array}
    \right)
    \cdot e^{-\lambda S \sigma^3}\;,
\label{eq:WKB_solution_ansatz}
\eeq
where $S$ is a function of the variables $\left(z,\bar{z}\right)$ and of the square of the spectral parameter $\lambda$, with asymptotic expansion as $\lambda^2 \rightarrow \infty$\, 
\beq
    S = S\left(z,\bar{z}\vert\lambda^2\right) = \sum_{k=0}^{\infty} \lambda^{-2k} S_k\left(z,\bar{z}\right)\;.
\label{eq:WKB_S_expansion_asymptotic}
\eeq
The solution $\tilde{\boldsymbol{\Phi}}$ is normalized in such a way that
\beq
	\det(\tilde{\boldsymbol{\Phi}}) = -2\quad \Longrightarrow \quad \det(\boldsymbol{\Phi})= -2\;.
\label{eq:WKB_normalization}
\eeq
The linear system (\ref{eq:shG_linear_problem_again}) then reduces to a pair of equations for the function $S$,
\begin{subequations}
\begin{align}
    &S_{,z}^2 - \frac{1}{2\lambda^2}\left\lbrace S,z\right\rbrace = \frac{\varphi_{,z}^2 - 2\varphi_{,zz}}{4\lambda^2} + P\;,& \left\lbrace S,z\right\rbrace = \frac{S_{,zzz}}{S_{,z}} - \frac{3}{2}\left(\frac{S_{,zz}}{S_{,z}}\right)^2\;, \\
    &S_{,\bar{z}} - \frac{\bar{P}}{\lambda^2} e^{-\varphi} S_{,z} = 0 \;,
    \end{align}
\end{subequations}
which, as one can easily check, are mutually compatible. Exploiting the series representation (\ref{eq:WKB_S_expansion_asymptotic}) we  turn this pair of equations into an infinite triangular system for the coefficients $S_k$, which we then solve by iteration, the first few equations being
\begin{subequations}
\begin{align}
    &S_{0,z}^2 = P\;,\qquad\qquad\qquad\qquad\qquad\qquad\qquad\qquad\qquad\;\; S_{0,\bar{z}} = 0\;, \\
    &S_{1,z} = \frac{1}{8\sqrt{P}}\left( \frac{P_{,zz}}{P} - \frac{5}{4}\left(\frac{P_{,z}}{P}\right)^2 + \varphi_{,z}^2 - 2\varphi_{,zz} \right)\; ,\qquad S_{1,\bar{z}} = e^{-\varphi}\sqrt{P}\bar{P}\;, \\
    &~~~\cdots \; , \qquad\qquad\qquad\qquad\qquad\qquad\qquad\qquad\qquad\qquad\quad \cdots \;.\nonumber
\end{align}
\label{eq:WKB_triangular_system}
\end{subequations}

We thus have expressed the solution to the linear problem (\ref{eq:shG_linear_problem}) as an expansion around $\lambda\rightarrow \infty$ as follows:
\beq
    \boldsymbol{\Phi} = e^{\frac{1}{4}\varphi\sigma^3}
    \left(
    \begin{array}{c c}
        e^{-\lambda S_0 - \frac{1}{4}\log P + \frac{1}{\lambda} S_1 +\mathcal O\left(\lambda^{-2}\right)} & e^{\lambda S_0 - \frac{1}{4}\log P - \frac{1}{\lambda} S_1 +\mathcal O\left(\lambda^{-2}\right)} \\
        e^{-\lambda S_0 + \frac{1}{4}\log P + \frac{1}{\lambda} \left(S_1 + \frac{\varphi_{,z}}{2\sqrt{P}}- \frac{P_{,z}}{4P^{3/2}}\right) +\mathcal O\left(\lambda^{-2}\right)} & -e^{\lambda S_0 + \frac{1}{4}\log P - \frac{1}{\lambda} \left(S_1 + \frac{\varphi_{,z}}{2\sqrt{P}}- \frac{P_{,z}}{4P^{3/2}}\right) +\mathcal O\left(\lambda^{-2}\right)}
    \end{array}
    \right)\;,
\label{eq:WKB_solution_large_lambda}
\eeq
with
\beq
    S_0 = \intop_{z_\ast} dz\, \sqrt{P}\;,\qquad S_1 = \intop_{z_\ast} \frac{dz}{8\sqrt{P}}  \left( \frac{P_{,zz}}{P} - \frac{5}{4}\left(\frac{P_{,z}}{P}\right)^2 + \varphi_{,z}^2 - 2\varphi_{,zz} \right)\;,
\eeq
and $z_{\ast}$ some arbitrarily-chosen base point.

A similar analysis for the linear system (\ref{eq:shG_linear_problem}), gauge rotated with the matrix $\exp\left(-\frac{1}{4}\varphi\sigma^3\right)$, yields the small-$\lambda$ behaviour
\beq
    \boldsymbol{\Phi} = e^{-\frac{1}{4}\varphi\sigma^3}
    \left(
    \begin{array}{c c}
        e^{-\frac{1}{\lambda} \bar{S}_0 + \frac{1}{4}\log \bar{P} + \lambda\left( \bar{S}_1 + \frac{\varphi_{,\bar{z}}}{2\sqrt{\bar{P}}}- \frac{\bar{P}_{,\bar{z}}}{4\bar{P}^{3/2}}\right) +\mathcal O\left(\lambda^{2}\right)} & -e^{\frac{1}{\lambda} \bar{S}_0 + \frac{1}{4}\log \bar{P} - \lambda\left( \bar{S}_1 + \frac{\varphi_{,\bar{z}}}{2\sqrt{\bar{P}}}- \frac{\bar{P}_{,\bar{z}}}{4\bar{P}^{3/2}}\right) +\mathcal O\left(\lambda^{2}\right)} \\
        e^{-\frac{1}{\lambda} \bar{S}_0 - \frac{1}{4}\log \bar{P} + \lambda \bar{S}_1 +\mathcal O\left(\lambda^{2}\right)} & e^{\frac{1}{\lambda} \bar{S}_0 - \frac{1}{4}\log \bar{P} - \lambda \bar{S}_1 +\mathcal O\left(\lambda^{2}\right)}
    \end{array}
    \right)\;,
\label{eq:WKB_solution_small_lambda}
\eeq
with
\beq
    \bar{S}_0 = \intop_{z_\ast} d\bar{z} \sqrt{\bar{P}}\;,\qquad \bar{S}_1 = \intop_{z_\ast} \frac{d\bar{z}}{8\sqrt{\bar{P}}}  \left( \frac{\bar{P}_{,\bar{z}\bar{z}}}{\bar{P}} - \frac{5}{4}\left(\frac{\bar{P}_{,\bar{z}}}{\bar{P}}\right)^2 + \varphi_{,\bar{z}}^2 - 2\varphi_{,\bar{z}\bar{z}} \right)\;.
\eeq

\subsection{WKB geometry, Stokes sectors and subdominant solutions}
\label{subsec:Stokes_phenomenon}

\begin{figure}[t!]
 \centering
  \includegraphics[scale=0.4]{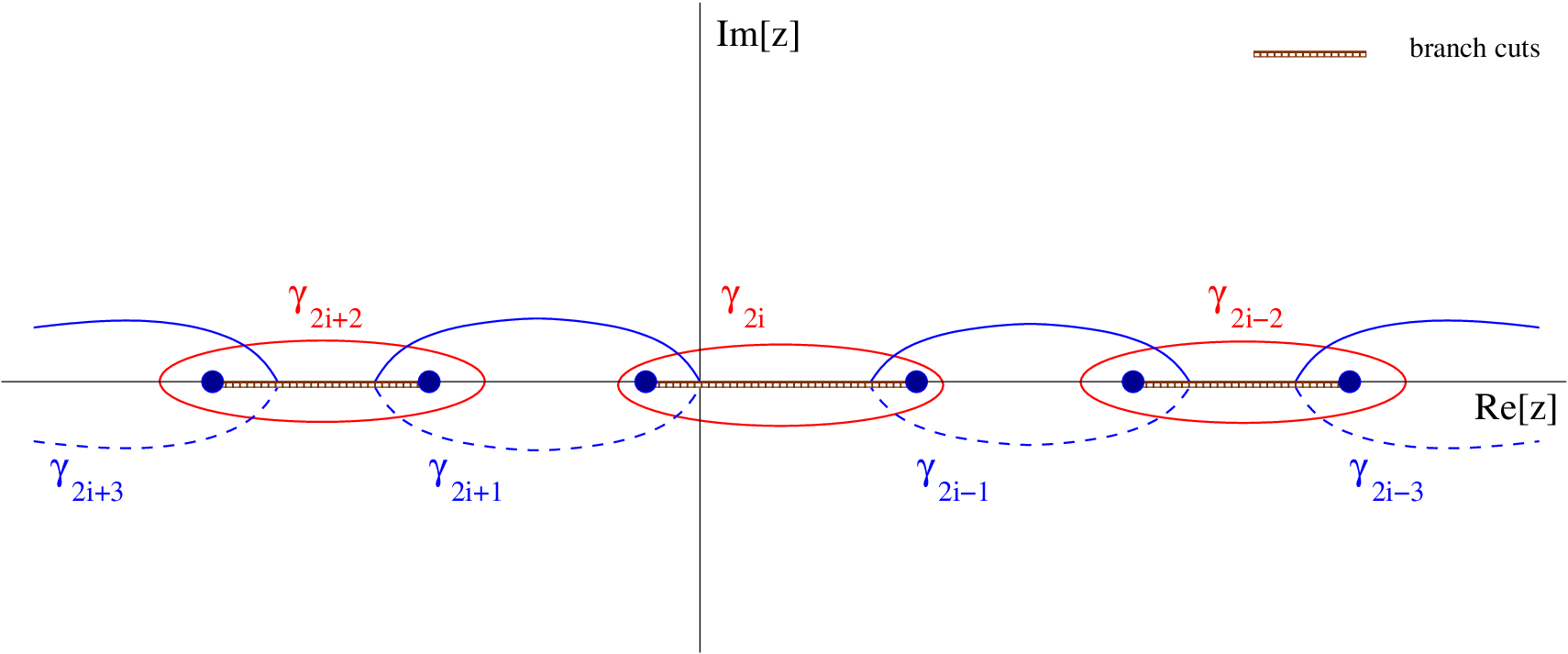}
\caption{\small An example of the (first sheet of the) Riemann surface $\mathcal R_{\textrm{WKB}}$ for a polynomial $P\left(z\right)$ having real roots, and a basis $\{\gamma_i\}$ of cycles on that surface.}

%Now we wish to show that the regularized area is really the \emph{Free Energy} associated to the TBA equation (\ref{eq:TBA_equation}) -- or, more generally, (\ref{eq:complex_mass_TBA}). In order to do so we will take a route which might appear to be slightly convoluted, so bear with us. First of all, consider the expression (\ref{eq:Area_formula_regularized}) for the regularized areas on that surface.}
\label{fig:cut_WKB_plane_example}
\end{figure}

Now, let us think more carefully about the geometry of what we are doing. By recasting (\ref{eq:shG_linear_problem}) into the system (\ref{eq:WKB_triangular_system}) we have moved from an equation defined on $\mathbb C^2$ to a system living on the Riemann surface $\mathcal R_{\textrm{WKB}}$ defined by the algebraic equation $\zeta^2 = P\left(z\right)$. The quantities $S_k$ appearing in the expansion (\ref{eq:WKB_S_expansion_asymptotic}) are line integrals along curves on $\mathcal R_{\textrm{WKB}}$:
\beq
    S_k\left(z,\bar{z}\right) = \intop_{z_\ast}^{\left(z,\bar{z}\right)} s_k\;,\qquad \bar{S}_k = \intop_{z_\ast}^{(z,\bar{z})} \bar{s}_k\;,
\label{eq:WKB_one_forms}
\eeq
with $s_k$ and $\bar{s}_k$ being one-forms on $\mathcal R_{\textrm{WKB}}$\,, e.g.
\beq
    s_0 = \sqrt{P} dz\;,\qquad s_1 = \frac{dz}{8\sqrt{P}}  \left( \frac{P_{,zz}}{P} - \frac{5}{4}\left(\frac{P_{,z}}{P}\right)^2 + \varphi_{,z}^2 - 2\varphi_{,zz} \right)\;,\qquad \cdots \;.
\label{eq:one_forms_WKB}
\eeq
Figure \ref{fig:cut_WKB_plane_example} depicts the first sheet of the  Riemann surface in the case of a polynomial $P\left(z\right)$ having real roots.
In order to define the WKB solutions (\ref{eq:WKB_solution_large_lambda}, \ref{eq:WKB_solution_small_lambda}) correctly, on the one hand it is necessary to be careful in the choice of the base point $z_{\ast}$ and the integration contour. On the other hand, however, it is possible to pin down the specific solution correctly only in a certain sector of the complex plane; this is an example of the Stokes phenomenon 
and is a direct consequence of the presence of an irregular singularity at $\left(z,\bar{z}\right) \rightarrow\infty$.

To be more precise, consider the solution (\ref{eq:WKB_solution_large_lambda}) at large distances both from the origin and from any critical values of $P\left(z\right)$. Then $P(z)$ behaves as $P\left(z\right) \sim z^{2N}$ and we can compute the leading behaviour of the coefficients $S_0$ and $S_1$\,:
\beq
	S_0 \underset{\vert z\vert\rightarrow\infty}{\sim} \intop_{z_\ast}^{z} dz\, z^{N} = \frac{z^{N+1}-z_{\ast}^{N+1}}{N+1}\;,\qquad S_1 \underset{\vert z\vert\rightarrow\infty}{\sim} \frac{N}{8}\frac{N+2}{N+1}\left(z^{-N-1} - z_{\ast}^{-N-1}\right)\;.
\eeq
Similar expressions hold for $\bar{S}_0$ and $\bar{S}_1$. More generally, as shown in (\ref{eq:varphi_asymptotics}), solutions to the modified sinh-Gordon equation (\ref{eq:ADS3_GMC_simple}) behave at leading order in $\vert z\vert\rightarrow \infty$ as $\varphi \sim 2N \log\vert z\vert$; the only remaining terms in $S$ and $\bar{S}$ are then, respectively, $S_0$ and $\bar{S}_0$.  Hence one finds
\beq
	\boldsymbol{\Phi} \underset{\vert z\vert\rightarrow\infty}{\sim} \frac{\bar{z}^{N/4}}{z^{N/4}} \left(\begin{array}{c c}
		1 & 1 \\
		\frac{z^{N/2}}{\bar{z}^{N/2}} & -\frac{z^{N/2}}{\bar{z}^{N/2}}
	\end{array}
	\right) \cdot e^{-\frac{\lambda z^{N+1} + \frac{1}{\lambda} \bar{z}^{N+1}}{N+1} \sigma^3}\;.
\eeq

Let us denote by $\Phi^{(d)}$ and $\Phi^{(s)}$ the two column vectors 
comprising
the matrix $\boldsymbol{\Phi}$
\beq
	\boldsymbol{\Phi} = \left(\begin{array}{c c} \Phi^{(s)} & \Phi^{(d)} \end{array}\right)\;,
\eeq
so that for large $\vert z\vert$ and $\vert \vartheta \vert < \frac{\pi}{N+1}$ these vectors behave as
\begin{subequations}
\begin{align}
	 \Phi^{(s)} &\underset{\vert z\vert\rightarrow\infty}{\sim} \left(\begin{array}{c} e^{-\mathbbm i\frac{N}{4} \vartheta} \\ -e^{\mathbbm i\frac{N}{4} \vartheta} \end{array}  \right)\exp\left(-\frac{2\,\varrho^{N+1}}{N+1}\cos\left(\left(N+1\right)\vartheta-\mathbbm i\upsilon\right)\right)\;, \\
	 \Phi^{(d)} &\underset{\vert z\vert\rightarrow\infty}{\sim} \left(\begin{array}{c} e^{-\mathbbm i\frac{N}{4} \vartheta} \\ e^{\mathbbm i\frac{N}{4} \vartheta} \end{array}  \right)\exp\left(\frac{2\,\varrho^{N+1}}{N+1}\cos\left(\left(N+1\right)\vartheta-\mathbbm i\upsilon\right)\right) \label{subeq:dominant}\;,
\end{align}
\end{subequations}
where $z = \varrho e^{\mathbbm i\vartheta}$, $\bar{z} = \varrho e^{-\mathbbm i\vartheta}$ and $\lambda = e^{\upsilon}$. Much as before, we will call $\Phi^{(s)}$ the \emph{subdominant solution} and $\Phi^{(d)}$ the \emph{dominant solution}. It is clear from the above expressions that if we analytically continue from $\left(\varrho,\vartheta\right)$ to $\left(\varrho,\vartheta + \frac{\pi}{N+1}\right)$ the two asymptotics seem to swap r\^oles. However, while we can precisely pin down the asymptotic of $\Phi^{(s)}$, since no other term can be added to it without spoiling its asymptotic behaviour, the behaviour (\ref{subeq:dominant}) might be hiding a contribution coming from a decaying exponential, with a coefficient which in general will change as the Stokes line in the middle of this sector is crossed. Hence when we perform the analytic continuation, we will obtain the following asymptotics, valid for $\vert \vartheta \vert < \frac{\pi}{N+1}$  and 
$\vartheta^{(+)}=\vartheta+\frac{\pi}{N+1}$\,:
%#RT
\begin{subequations}
\begin{align}
	 \Phi^{(s)}\left(\varrho,\vartheta^{(+)}\right) &\underset{\vert z\vert\rightarrow\infty}{\sim} \left(\begin{array}{c} e^{-\mathbbm i\frac{N}{4} \vartheta^{(+)}} \\ -e^{\mathbbm i\frac{N}{4} \vartheta^{(+)}} \end{array}  \right)\exp\left(\frac{2\,\varrho^{N+1}}{N+1}\cos\left(\left(N+1\right)\vartheta-\mathbbm i\upsilon\right)\right)=\textrm{dominant}\;, \\
	 \Phi^{(d)}\left(\varrho,\vartheta^{(+)}\right)  &\underset{\vert z\vert\rightarrow\infty}{\sim} c_+\left(\lambda\right) \left(\begin{array}{c} e^{-\mathbbm i\frac{N}{4} \vartheta^{(+)}} \\ -e^{\mathbbm i\frac{N}{4} \vartheta^{(+)}} \end{array}  \right)\exp\left(\frac{2\,\varrho^{N+1}}{N+1}\cos\left(\left(N+1\right)\vartheta-\mathbbm i\upsilon\right)\right) \label{eq:schematic_stokes}\;.
\end{align}
\end{subequations}
%
%
%
%
%\beq
%	 \Phi^{(s)}\left(\varrho,\vartheta^{(+)}\right) \underset{\vert z\vert\rightarrow\infty}{\sim} \left(\begin{array}{c} e^{-\mathbbm i\frac{N}{4} \vartheta^{(+)}} \\ -e^{\mathbbm i\frac{N}{4} \vartheta^{(+)}} \end{array}  \right)\exp\left(\frac{2\,\varrho^{N+1}}{N+1}\cos\left(\left(N+1\right)\vartheta-\mathbbm i\upsilon\right)\right) = \textrm{dominant}\;, \label{eq:schematic_stokesA}
%\eeq	 
%and
%\beq	 
%	 \Phi^{(d)}\left(\varrho,\vartheta^{(+)}\right)   \underset{\vert z\vert\rightarrow\infty}{\sim} c_+\left(\lambda\right) \left(\begin{array}{c} e^{-\mathbbm i\frac{N}{4} \vartheta^{(+)}} \\ -e^{\mathbbm i\frac{N}{4} \vartheta^{(+)}} \end{array}  \right)\exp\left(\frac{2\, \varrho^{N+1}}{N+1}\cos\left(\left(N+1\right)\vartheta-\mathbbm i\upsilon\right)\right)\;. \label{eq:schematic_stokesB}
%\eeq
Therefore, for $\vartheta$ in the sector $\vert \vartheta \vert < \frac{\pi}{N+1}$, the continued solution $\Phi^{(d)}\left(\varrho,\vartheta^{(+)}\right)$      is in general  dominant but, exceptionally, it will be subdominant at zeros of the coefficient $c_+(\lambda)$. The story is similar to that of section \ref{sec:excited}, and the preliminary  discussion reported there will be formalised in the following sections.

\begin{figure}[t!]
 \centering
  \includegraphics[scale=1]{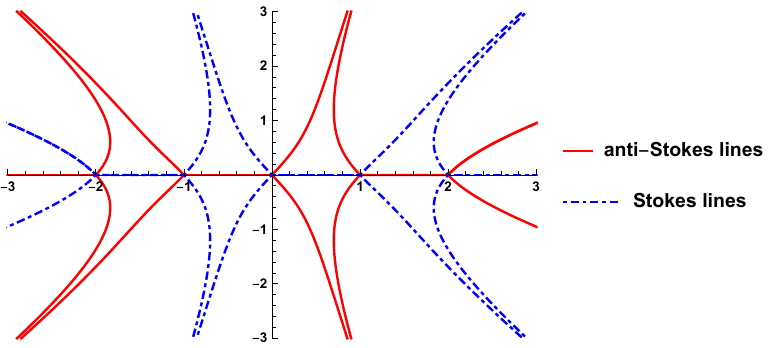}
\caption{\small Stokes and anti-Stokes lines for the function $P\left(z\right) = z\left(z^2-1\right)\left(z^2-4\right)$, with $\lambda\in\mathbb R$. Although not shown here, there are branch cuts connecting $-2$ with $-1$, $0$ with $+1$ and $+2$ with $\infty$.}
\label{fig:anti-Stokes_lines_example}
\end{figure}
\begin{figure}[t!]
\centering
\includegraphics[scale=0.5]{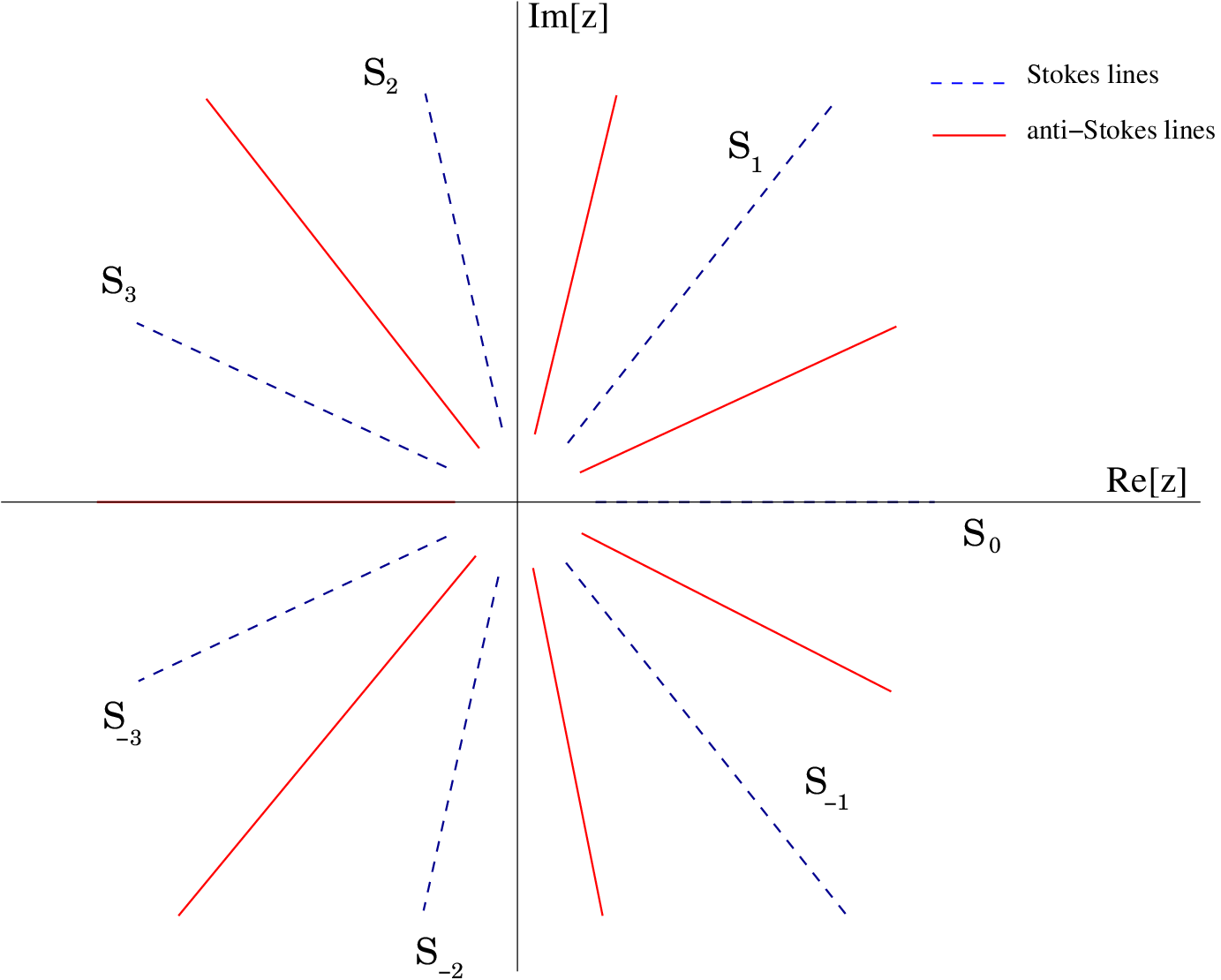}

\caption{\small Figure \ref{fig:anti-Stokes_lines_example} looked at from very large $\vert z\vert$. The fine details of the function $P\left(z\right)$ disappear and we only see the lines defined by $\textrm{Re}\left[z^{7/2}\right] = 0$ and $\textrm{Im}\left[z^{7/2}\right]=0$, that is $\vartheta^{\textrm{aS}}_k = \pi\frac{2k+1}{7}$ and $\vartheta^{\textrm{S}}_k = \pi\frac{2k}{7}$ with  $k=-3,-2,-1,0,1,2,3$. The Stokes sectors $\mathcal S_k$ are labeled by the index $k$ of the angles $\vartheta^{S}_k$.}
\label{fig:anti-Stokes_lines_example_large_z}
\end{figure}

\begin{figure}[t!]
 \centering
  \subfloat[$\textrm{arg}\left(\lambda\right) = \frac{\pi}{8}$]{\includegraphics[scale=0.5]{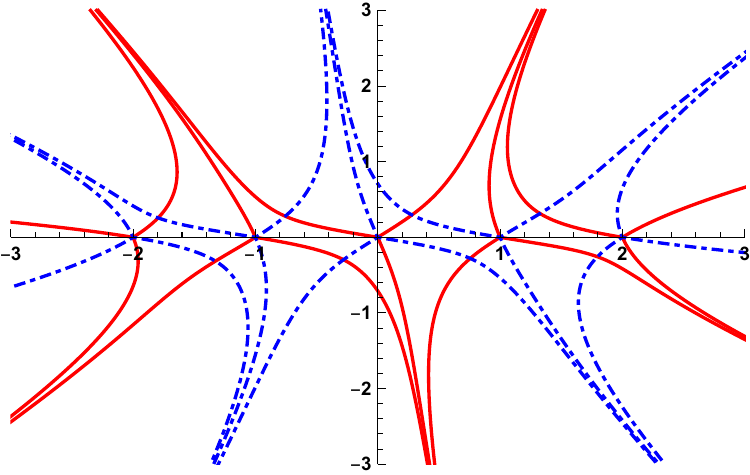}
  \label{fig:Stokes_pi_8_rotated}}
  \hspace{1.5cm}
  \subfloat[$\textrm{arg}\left(\lambda\right) = \frac{\pi}{4}$]{\includegraphics[scale=0.5]{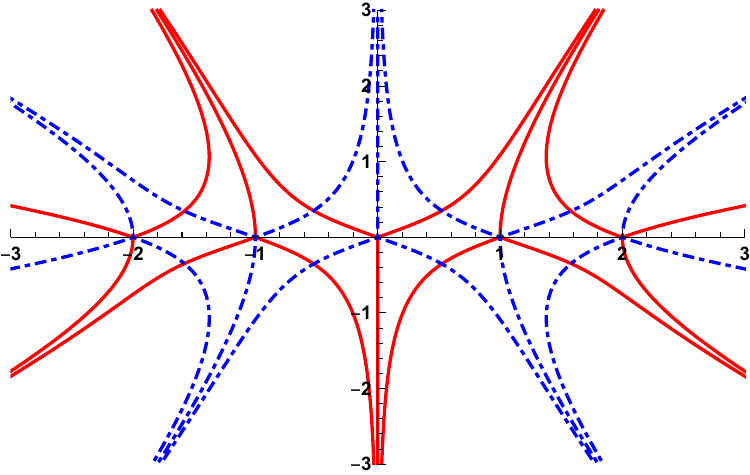}
  \label{fig:Stokes_pi_4_rotated}}
  \hspace{1.5cm}
  \subfloat[$\textrm{arg}\left(\lambda\right) = \frac{\pi}{2}$]{\includegraphics[scale=0.5]{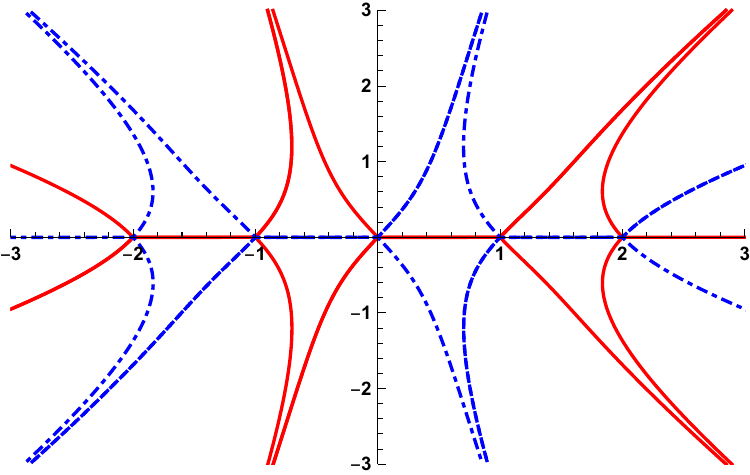}
  \label{fig:Stokes_pi_2_rotated}}
  \hspace{1.5cm}
  \subfloat[$\textrm{arg}\left(\lambda\right) = \frac{2\pi}{3}$]{\includegraphics[scale=0.5]{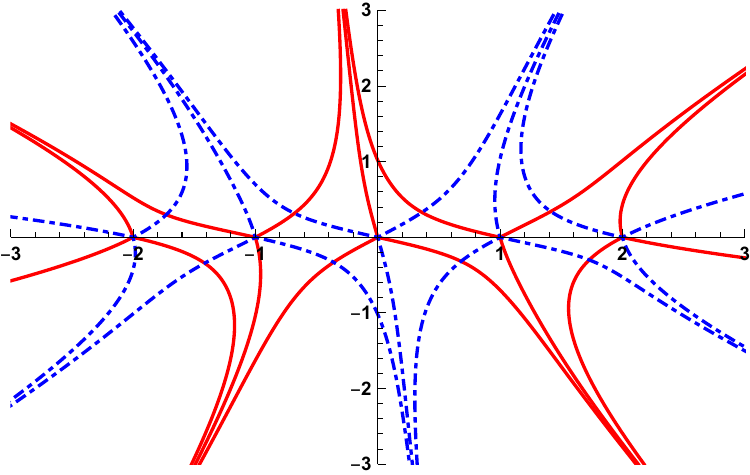}
  \label{fig:Stokes_pi_1.5_rotated}}
  \hspace{1.5cm}
  \subfloat[$\textrm{arg}\left(\lambda\right) = \frac{4\pi}{5}$]{\includegraphics[scale=0.5]{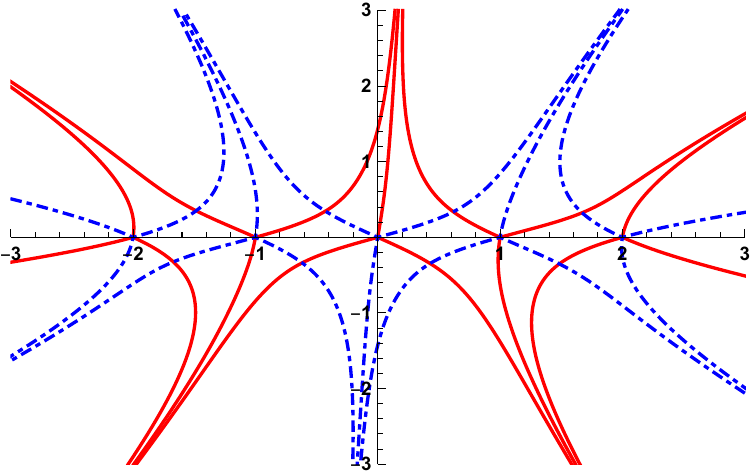}
  \label{fig:Stokes_pi_1.25_rotated}}
  \hspace{1.5cm}
  \subfloat[$\textrm{arg}\left(\lambda\right) = \pi$]{\includegraphics[scale=0.5]{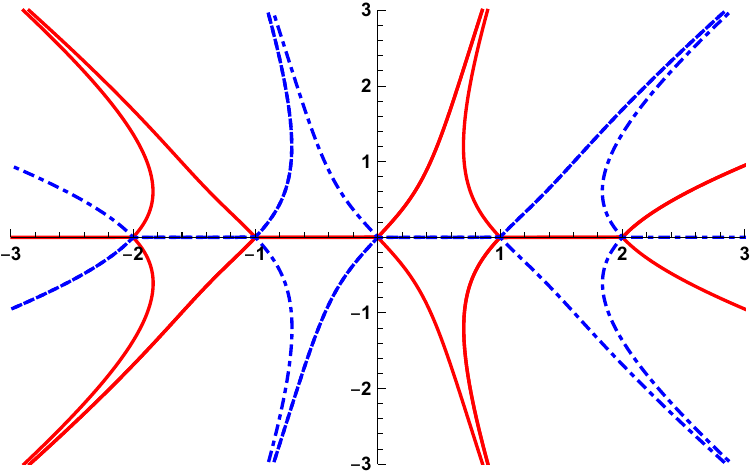}
  \label{fig:Stokes_pi_rotated}}
\caption{\small Plots of Stokes and anti-Stokes lines for the polynomial function $P\left(z\right) = z\left(z^2-1\right)\left(z^2-4\right)$ and various phases of the spectral parameter $\lambda$. One sees that a counter-clockwise rotation of $\lambda$ corresponds to a clockwise rotation of the sectors. For $\textrm{arg}\left(\lambda\right) = \pi$, the picture looks the same as figure \ref{fig:anti-Stokes_lines_example}, but the sectors have been exchanged in a clockwise fashion.}
\label{fig:Stokes_lines_rotation}
\end{figure}

Summarising, we see that the function $P\left(z\right)$ partitions the Riemann surface $\mathcal R_{\textrm{WKB}}$ into \emph{Stokes sectors} $\mathcal S_j$, bounded by \emph{anti-Stokes lines}, defined by $\textrm{Re}\left[\lambda S_0\right] = 0$. In each of these sectors we can define a matrix solution $\boldsymbol{\Phi}_j$ composed of a dominant and a subdominant solution
\beq
	\boldsymbol{\Phi}_j = \left(\begin{array}{c c} \Phi_j^{(s)} & \Phi_j^{(d)} \end{array}\right)\;.
\eeq
The decay (or growth) of this solution is largest whenever the solution lies on a \emph{Stokes line}, defined by $\textrm{Im}\left[\lambda S_0\right] = 0$.  Figure \ref{fig:anti-Stokes_lines_example}  depicts an example of the Stokes and anti-Stokes lines for a particular choice of $P\left(z\right)$, while  figure \ref{fig:anti-Stokes_lines_example_large_z} is a view of the same picture from very large $\vert z\vert$. The definition of Stokes and anti-Stokes lines depends on the phase of the spectral parameter $\lambda$ and, as displayed in figure \ref{fig:Stokes_lines_rotation}, a counter-clockwise rotation of $\lambda$ rotates the sectors in a clockwise direction. When $\textrm{arg}\left(\lambda\right)= \pi$, one returns to the same situation as for $\textrm{arg}\left(\lambda\right) = 0$, but with the sectors exchanged in a clockwise fashion. Consequently, exploiting the $\mathbb Z_2$ symmetry (\ref{eq:Z2_symmetry}), we can define the solutions $\boldsymbol{\Phi}_j$ as
\beq
	\boldsymbol{\Phi}_{j}\left(z,\bar{z}\vert\lambda\right) = \left(\sigma^3\right)^j \boldsymbol{\Phi}\left(z,\bar{z}\vert e^{j\mathbbm i\pi}\lambda\right)\;,
\eeq
where $\boldsymbol{\Phi}$, our starting solution, is defined in what we choose to be the $0$-th sector $\mathcal S_0$. In what follows we will label the sectors according to the index $k$ of the $\vartheta_k^{\textrm{S}} = \pi \frac{k}{N+1}$ solution of the Stokes line equation $\textrm{Im}\left[z^{N+1}\right]$ for large $\vert z \vert$.  Hence the sector $\mathcal S_0$ will be for $\lambda \in\mathbb R$ the one containing the positive real line at large enough $\vert z\vert$. See figure \ref{fig:anti-Stokes_lines_example_large_z} for an example.

\subsection{The connection matrices, the T-system and the Hirota equation}
\label{eq:Tsystem}
We can now make the relations (\ref{eq:schematic_stokes}) more precise as follows:\footnote{The $-1$ 
sign in the second equality is necessary to have $\det(\boldsymbol{\Phi}_{j-1}) = \det(\boldsymbol{\Phi}_j) = -2$.}
\begin{subequations}
\begin{align}
    \Phi^{(s)}_{j-1}\left(z,\bar{z}\vert\lambda\right) &= \Phi^{(d)}_{j}\left(z,\bar{z}\vert\lambda\right) \\ \Phi^{(d)}_{j-1}\left(z,\bar{z}\vert\lambda\right) &= -\Phi^{(s)}_{j}\left(z,\bar{z}\vert\lambda\right) + T\left( e^{j\mathbbm i\pi}  \lambda\right) \Phi^{(d)}_{j}\left(z,\bar{z}\vert\lambda\right) \;,
\end{align}
\end{subequations}
or, in matrix notation
\beq
	\boldsymbol{\Phi}_{j-1}\left(z,\bar{z}\vert\lambda\right) = \boldsymbol{\Phi}_{j}\left(z,\bar{z}\vert\lambda\right)\mathbf{T}\left(e^{j\mathbbm i\pi}\lambda\right)\;,\qquad \mathbf{T}\left(\lambda\right) = \left( \begin{array}{c c}
		0 & -1 \\
		1 & T\left(\lambda\right)
	\end{array}\right)\;.
\eeq
It is immediate to see that
\beq
	T\left(\lambda\right) = \frac{1}{\det(\boldsymbol{\Phi}_{0})}\det\left(\begin{array}{c c} \Phi_{0}^{(s)} & \Phi_{-1}^{(d)} \end{array}\right) =-\frac{1}{2}\det\left(\begin{array}{c c} \Phi_{0}^{(s)} & \Phi_{-2}^{(s)} \end{array}\right)\;,
\eeq
where we have used (\ref{eq:WKB_normalization}) and the fact that $\Phi_{j}^{(d)} = \Phi_{j-1}^{(s)}$. We can generalize this construction, introducing the \emph{lateral connection matrices} $\mathbf{T}_k\left(\lambda\right)$  which, as the name suggests, relate solutions living in $($next$)^k$-neighbouring Stokes sectors:
\beq
	\boldsymbol{\Phi}_j\left(z,\bar{z}\vert\lambda\right) = \boldsymbol{\Phi}_{j+k}\left(z,\bar{z}\vert\lambda\right)\mathbf{T}_k\left(\lambda e^{\left(j+\frac{k+1}{2}\right)\mathbbm i\pi}\right)\;.
\eeq
The form of these matrices is constrained by noticing that they need to satisfy the following consistency relation
\beq
	\mathbf{T}_{k}\left(\lambda\right) = \mathbf{T}_{k-j}\left(e^{\frac{j}{2}\mathbbm i\pi}\lambda\right)\mathbf{T}_{j}\left(e^{\frac{j-k}{2}\mathbbm i\pi}\lambda\right)\;,
\label{eq:T_system_matrix_form}
\eeq
which implies that we can parametrise the lateral connection matrices as follows
\beq
	\mathbf{T}_{k}\left(\lambda\right) = \left(\begin{array}{c c}
		-T_{k-2}\left(\lambda\right) & -T_{k-1}\left( e^{-\frac{1}{2}\mathbbm i\pi}\lambda\right) \\
		T_{k-1}\left( e^{\frac{1}{2}\mathbbm i\pi}\lambda\right) & T_k\left(\lambda\right)
	\end{array}
	\right)\;.
\eeq

Each function $T_k\left(\lambda\right)$, which we call a Stokes multiplier or \emph{lateral connection coefficient}, can be computed as a determinant of subdominant solutions defined in distinct Stokes sectors:
\begin{subequations}
\begin{align}
	&T_{2k-1}\left(\lambda\right) = \frac{1}{2}\det\left(\begin{array}{c c}\Phi_{-k-1}^{(s)} & \Phi_{k-1}^{(s)}\end{array}\right)\;,\\
	&T_{2k}\left(\lambda e^{\frac{1}{2}\mathbbm i\pi}\right) = \frac{1}{2}\det\left(\begin{array}{c c}\Phi_{-k-1}^{(s)} & \Phi_{k}^{(s)}\end{array}\right)\;.
\end{align}
\label{eq:T_as_wronskians}
\end{subequations}
One must clearly have $\mathbf{T}_0\left(\lambda\right) = \mathbf{1}$, implying that
\beq
    T_{-2}\left(\lambda\right) = -1\;,\qquad T_{-1}\left(\lambda\right) = 0\;,\qquad T_{0}\left(\lambda\right) = 1\;,
\eeq
which agree with the determinant expressions (\ref{eq:T_as_wronskians}).

The relation (\ref{eq:T_system_matrix_form}) can be used to extract a series of additional constraints on the functions $T_k\left(\lambda\right)$. First of all one has the unimodularity condition
\beq
    \det(\mathbf{T}_k\left(\lambda\right)) = 1\;,
\eeq
to which we will return momentarily. Another obvious relation is the following
\beq
    \mathbf{T}_0\left(e^{-\frac{j}{2}\mathbbm{i}\pi}\lambda\right) = \mathbf{1} = \mathbf{T}_{-j}\left(\lambda\right)\mathbf{T}_j\left(\lambda\right)\quad \Longrightarrow \quad T_{-k-1}\left(\lambda\right) = -T_{k-1}\left(\lambda\right)\;.
\eeq
We also require that a rotation of $2\left(N+1\right)$ Stokes sectors brings us back to the same solution (modulo a $\pm1$ factor), from which we deduce that
\beq
    \mathbf{T}_{j+2N+2}\left(\lambda\right) = \pm \mathbf{T}_j\left(e^{\left(N+1\right)\mathbbm i\pi}\lambda\right)\quad \Longrightarrow\quad T_{2N+1}\left(\lambda\right) = 0\;.
\eeq
Finally, we obtain a recursive definition for $T_k\left(\lambda\right)$ by looking at the components of (\ref{eq:T_system_matrix_form})
\beq
	T_k\left(\lambda\right) = T_{j}\left(e^{\frac{j-k}{2}\mathbbm i\pi}\lambda\right) T_{k-j}\left(e^{\frac{j}{2}\mathbbm i\pi}\lambda\right) - T_{j-1}\left(e^{\frac{j-k-1}{2}\mathbbm i\pi}\lambda\right)T_{k-j-1}\left(e^{\frac{j+1}{2}\mathbbm i\pi}\lambda\right)\,,
\label{eq:T_system}
\eeq
which is called the T-system. An equivalent, more elegant, form is obtained by the simple unimodularity requirement mentioned above
\beq
	\det\left(\mathbf{T}_{k+1}\left(\lambda\right) \right)= 1 \quad \Longrightarrow \quad T_{k}\left(e^{\frac{1}{2}\mathbbm i\pi}\lambda\right)T_{k}\left(e^{-\frac{1}{2}\mathbbm i\pi}\lambda\right) = 1 + T_{k+1}\left(\lambda\right)T_{k-1}\left(\lambda\right) \;.
\label{eq:Hirota_equation}
\eeq
This equation needs to be supported by the boundary conditions found above,  $T_0\left(\lambda\right) = 1$ and $T_{2N+1}\left(\lambda\right) = 0$, and is known in the literature as \emph{Hirota bilinear equation} \cite{Hiro_1981,Krichever:1996qd,Kuniba:1993cn}. One can check that the T-system is obtained by iteration from the Hirota equation.

There are various manipulations one can perform on the Hirota equation. For example, one can formally solve it by parametrizing the functions $T_k\left(\lambda\right)$ by a pair of Q functions $\left\lbrace Q_a\left(\lambda\right)\right\rbrace_{a=1,2}$ as follows
\beq
    T_k\left(\lambda\right) = \det\left( \begin{array}{c c}
        Q_1\left(e^{\frac{k+1}{2}\mathbbm i\pi}\lambda\right) & Q_1\left(e^{-\frac{k+1}{2}\mathbbm i\pi}\lambda\right)\\
        Q_2\left(e^{\frac{k+1}{2}\mathbbm i\pi}\lambda\right)  & Q_2\left(e^{-\frac{k+1}{2}\mathbbm i\pi}\lambda\right)
    \end{array}\right)\;.
\label{eq:T_as_Q_det}
\eeq
Then it is easy to see that the Hirota equation is equivalent to the following one
\beq
    \det\left( \begin{array}{c c}
        Q_1\left(e^{\frac{1}{2}\mathbbm i\pi}\lambda\right) & Q_1\left(e^{-\frac{1}{2}\mathbbm i\pi}\lambda\right) \\
        Q_2\left(e^{\frac{1}{2}\mathbbm i\pi}\lambda\right) & Q_2\left(e^{-\frac{1}{2}\mathbbm i\pi}\lambda\right)
    \end{array}\right) = 1\;,
\label{eq:QQ-system}
\eeq
which, in the literature, is known as a  \emph{quantum Wronskian} \cite{Bazhanov:1994ft,Lukyanov:2010rn}. The relation (\ref{eq:QQ-system}) is the off-critical version of the constraint (\ref{eq:QQ-system0}), obtained within the quantum KdV context. From  (\ref{eq:T_as_Q_det}) and (\ref{eq:QQ-system}) we obtain Baxter's TQ equation
\beq
    T_1\left(\lambda\right)Q_a\left(\lambda\right) = Q_a\left(e^{\mathbbm i\pi}\lambda\right) + Q_a\left(e^{-\mathbbm i\pi}\lambda\right)\;,\qquad (a=1,2)\;,
\label{eq:Baxter_TQ}
\eeq
by simply expanding the trivial identity
\beq
    \det\left(\begin{array}{c c c}
        Q_1\left(e^{\mathbbm i\pi}\lambda\right) & Q_1\left(\lambda\right)  & Q_1\left(e^{-\mathbbm i\pi}\lambda\right)  \\
        Q_2\left(e^{\mathbbm i\pi}\lambda\right) & Q_2\left(\lambda\right) & Q_2\left(e^{-\mathbbm i\pi}\lambda\right) \\
        Q_a\left(e^{\mathbbm i\pi}\lambda\right) & Q_a\left(\lambda\right) & Q_a\left(e^{-\mathbbm i\pi}\lambda\right)
    \end{array}\right) = 0\;,\qquad (a=1,2)\;.
\eeq
The matrix
\beq
    \mathbf{Q}\left(\lambda\right) = \left( \begin{array}{c c}
        Q_1\left(e^{\mathbbm i\pi}\lambda\right) & Q_1\left(\lambda\right) \\
        Q_2\left(e^{\mathbbm i\pi}\lambda\right) & Q_2\left(\lambda\right)
    \end{array}\right)
\eeq
has a geometrical interpretation: it is the \emph{central connection matrix} of the central problem for our linear system. In other words, it relates the solutions $\boldsymbol{\Phi}_j$ to another fundamental solution $\boldsymbol{\Xi}$, defined via local analysis at a point where no Stokes phenomenon is present\footnote{In the first incarnations of the ODE/IM correspondence \cite{Dorey:1998pt,Bazhanov:1998wj} this point was the origin $z = 0$, which represents a regular singularity of the differential equation. Consequently the solution obtained by local analysis around $z=0$ does not exhibit any Stokes phenomena. The term ``central'' also descends from these first examples, in which the eigenvalue problem associated to the central connection matrix concerned functions with  behaviour defined at $z = 0$ and $\vert z\vert\rightarrow\infty$. In our case the linear system possesses no singularity at finite $z$, however we can still define an eigenvalue problem for functions with given behaviour as $\vert z\vert \rightarrow \infty$ and at an arbitrary point $z$ which, being regular, will not give rise to a Stokes phenomenon. We stick to the tradition and call such an eigenvalue problem ``central''.
}. Then $\boldsymbol{\Xi}$ is insensitive to the rotation of $\lambda$ by integer multiple of $\mathbbm i\pi$ and one has the relation
\beq
    \boldsymbol{\Phi}_j\left(z,\bar{z}\vert\lambda\right) = \boldsymbol{\Xi}\left(z,\bar{z}\vert\lambda\right) \mathbf{Q}\left(e^{j\mathbbm i\pi}\lambda\right)\;.
\eeq
Playing with this relation and  (\ref{eq:T_system_matrix_form}), we obtain the following identity
\beq
    \mathbf{Q}\left(\lambda\right) = \mathbf{Q}\left(e^{k\mathbbm i\pi} \lambda\right)\mathbf{T}_k\left(e^{\frac{k+1}{2}\mathbbm i\pi}\lambda\right)\;,
\eeq
from which it is possible to derive both the Baxter TQ equation (\ref{eq:Baxter_TQ}) (by simply setting $k=1$) and the parametrization (\ref{eq:T_as_Q_det}) of the functions $T_k$ (by Cramer's rule). The QQ-system (\ref{eq:QQ-system}) corresponds  to the unimodularity requirement $\det(\mathbf{Q}(\lambda)) = 1$.

Although Q-functions are interesting objects, we find it more convenient to introduce a new set of functions: the Y-functions. These are defined as follows
\beq
	Y_k\left(\lambda\right) = T_{k-1}\left(\lambda\right)T_{k+1}\left(\lambda\right)\;,\quad (k = 1,\ldots , 2N-1)\;,
\eeq
or, in a more invariant form, and using the fact that $\det\left(\begin{array}{c c} \Phi^{(s)}_{k} & \Phi^{(s)}_{k+1} \end{array}\right) = - \det\boldsymbol{\Phi}_0$,
\begin{subequations}
\begin{align}
	&Y_{2k}\left(\lambda\right) = \frac{\det\left(\begin{array}{c c} \Phi^{(s)}_{-k-2} & \Phi^{(s)}_{k} \end{array}\right) \det\left(\begin{array}{c c} \Phi^{(s)}_{-k-1} & \Phi^{(s)}_{k-1} \end{array}\right)}{\det\left(\begin{array}{c c} \Phi^{(s)}_{-k-1} & \Phi^{(s)}_{-k-2} \end{array}\right) \det\left(\begin{array}{c c} \Phi^{(s)}_{k} & \Phi^{(s)}_{k-1} \end{array}\right)} \;, \\
	&Y_{2k+1}\left(\lambda e^{\frac{1}{2}\mathbbm i\pi}\right) = \frac{\det\left(\begin{array}{c c} \Phi^{(s)}_{-k-2} & \Phi^{(s)}_{k+1} \end{array}\right) \det\left(\begin{array}{c c} \Phi^{(s)}_{-k-1} & \Phi^{(s)}_{k} \end{array}\right)}{\det\left(\begin{array}{c c} \Phi^{(s)}_{-k-1} & \Phi^{(s)}_{-k-2} \end{array}\right) \det\left(\begin{array}{c c} \Phi^{(s)}_{k+1} & \Phi^{(s)}_{k} \end{array}\right)} \;.
\end{align}
\end{subequations}
In term of the functions $Y$, the Hirota equation (\ref{eq:Hirota_equation}) becomes
\beq
	Y_k\left(\lambda e^{\frac{1}{2}\mathbbm i\pi}\right)Y_k\left(\lambda e^{-\frac{1}{2}\mathbbm i\pi}\right) = \left(1+Y_{k+1}\left(\lambda\right)\right) \left(1+Y_{k-1}\left(\lambda\right)\right)\;.
\label{eq:Y_system}
\eeq
This set of equations is known in the literature as a \emph{Y-system}; see for example  \cite{Zamolodchikov:1991et,Ravanini:1992fi,Kuniba:1993cn, Kuniba:2010ir}.
\subsection{Properties of the Y-functions and the TBA equation}
\label{sec:TBA}
Although the rewriting (\ref{eq:Y_system}) of the Hirota equation does not seem to change  the situation much, it actually allows us to derive an integral equation for the logarithm of the Y functions. Let us briefly review how this is done.

Using the definition (\ref{eq:WKB_solution_ansatz}) of the WKB solution, we easily see that
\beq
	Y_{2k}\left(\lambda\right) = \exp\left(-\lambda \ointop_{\gamma_{2k}} s\right)\;,\qquad Y_{2k+1}\left(\lambda e^{\frac{1}{2}\mathbbm i\pi}\right) = \exp\left(-\lambda \ointop_{\gamma_{2k+1}} s\right)\;,
\label{eq:Y_funct_WKB_expr}
\eeq
where $s = \sum_{k=0}^{\infty} \lambda^{-2k} s_k$ and the one-forms $s_k$ were introduced in (\ref{eq:WKB_one_forms}). The $\gamma_k$ are closed contours, elements of a basis of the first homology group $\textrm{H}_1\left(\mathcal R_{\textrm{WKB}},\mathbb Z\right)$. Since our branch cuts can all be taken to lie on the real axis (remember, we chose the polynomial $P\left(z\right)$ to only have real roots), we can arrange them as shown in figure \ref{fig:cut_WKB_plane_example}. 
It is evident that the $Y_k\left(\lambda\right)$ functions are analytic in $\lambda$ with essential singularities sitting at $\lambda = 0$ and $\lambda = \infty$. In particular, a perturbative analysis of the WKB solutions tells us that
\beq
	\log Y_{2k} = -\lambda \ointop_{\gamma_{2k}} dz \, \sqrt{P}  + \mathcal O\left(\lambda^{-1}\right)\;,\qquad \log Y_{2k+1} = \mathbbm i\lambda \ointop_{\gamma_{2k+1}}  dz \, \sqrt{P}+ \mathcal O\left(\lambda^{-1}\right)\;.
\eeq
A similar result holds for the expansion around $\lambda = 0$, with $\sqrt{P}dz$ replaced by $\sqrt{\bar{P}}d\bar{z}$. Hence we find that the Y functions have the following asymptotic for large $\vert \upsilon\vert$, with $\upsilon = \log\lambda$, 
\beq
	\log Y_k\left(\upsilon\right) \underset{\vert\upsilon\vert\rightarrow\infty}{\sim} - m_k \cosh(\upsilon)\;\qquad \left\lbrace\begin{array}{l}
	m_{2k} = 2\ointop_{\gamma_{2k}} dz\,\sqrt{P} = 2\ointop_{\gamma_{2k}} d\bar{z} \,\sqrt{\bar{P}} \\
	\\
	m_{2k+1} = -\mathbbm{i} \, 2  \ointop_{\gamma_{2k+1}} dz\, \sqrt{P} = -\mathbbm{i} \, 2\ointop_{\gamma_{2k+1}} d\bar{z} \,\sqrt{\bar{P}}
	\end{array}
	\right.\;.
\label{eq:TBA_masses}
\eeq
Note that this behaviour is valid for $\textrm{Im}\left[\upsilon\right]\in\left(-\pi,\pi\right)$, since beyond this range, the WKB approximation we have used may no longer be reliable.\footnote{Actually the WKB approximation can be shown to be valid in the range $\textrm{Im}\left[\upsilon\right]\in\left(-\frac{3}{2}\pi,\frac{3}{2}\pi\right)$.} The quantities $m_k$ can be shown to be real when all the zeroes of $P\left(z\right)$ are real.\footnote{Consider a polynomial with $2N$ roots
\beq
    P\left(z\right) = \left(z-z_1\right)\left(z-z_2\right)\cdots \left(z-z_{2N}\right)\;, \nonumber
\eeq
and suppose that $z_1,z_2\in\mathbb R$. We wish to compute the integral
\beq
    I = \ointop_{\gamma_{1,2}} dz\sqrt{P\left(z\right)} \;, \nonumber
\eeq
where $\gamma_{1,2}$ is a cycle encircling in a  counter-clockwise sense the cut running from $z_1$ to $z_2$. Moreover, without loss of generality, suppose $z_1=0$, $z_2>0$ and $z_j\notin \left[0,z_1\right]\,,\;\forall j=3,\ldots,2N$. Then our integral becomes
\beq
    I = -2\intop_{0}^{z_2}  dz \, \sqrt{z\left(z-z_2\right)\cdots \left(z-z_{2N}\right)}\;, \nonumber
\eeq
since the integrals on infinitesimal circles around $z=1$ and $z=z_2$ vanish. The integral $I$ is explicitely a real number, as long as $z_j\in\mathbb R\,,\;\forall j=3,\ldots 2N$.}

Now, from the properties just mentioned, we deduce that the auxiliary function
\beq
	y_k\left(\upsilon\right) = \log\left(Y_k\left(\upsilon\right) e^{m_k\cosh(\upsilon)}\right)\;,
\eeq
is analytic in the strip $\mathsf S_{\upsilon} = \vert\textrm{Im}\left[\upsilon\right]\vert<\frac{\pi}{2}$ and decays at large $\vert\textrm{Re}\left[\upsilon\right]\vert$ therein. Moreover it obeys the logarithmic form of (\ref{eq:Y_system})
\beq
	y_k\left(\upsilon+\frac{1}{2}\mathbbm i\pi\right) + y_k\left(\upsilon-\frac{1}{2}\mathbbm i\pi\right) = \log\left(1+Y_{k+1}\left(\upsilon\right)\right) + \log\left(1+Y_{k-1}\left(\upsilon\right)\right)\;.
\eeq
This form is very useful, because the operator effecting the shift in the right-hand side above is inverse to the convolution kernel $\mathcal K\left(\upsilon\right) = \frac{1}{2\pi\cosh(\upsilon)}$. In mathematical terms
\begin{align}
	&\left[\mathcal K\ast y_k\right]\left(\upsilon+\frac{1}{2}\mathbbm i\pi\right) + \left[\mathcal K\ast y_k\right]\left(\upsilon-\frac{1}{2}\mathbbm i\pi\right)  \notag \\
	&= \intop_\RR \frac{d\upsilon'}{2\pi} \frac{y_k\left(\upsilon'+\frac{1}{2}\mathbbm i\pi\right) + y_k\left(\upsilon'-\frac{1}{2}\mathbbm i\pi\right) }{\cosh\left(\upsilon-\upsilon'\right)} = \ointop_{\partial \mathsf S_{\upsilon}} \frac{d\upsilon'}{2\pi\mathbbm i} \frac{y_k\left(\upsilon'\right)}{\sinh\left(\upsilon-\upsilon'\right)} = y_k\left(\upsilon\right)\;,
\end{align}
where $\partial \mathsf S_{\upsilon}$ is the boundary of the strip $\mathsf S_{\upsilon} = \vert\textrm{Im}\left[\upsilon\right]\vert<\frac{\pi}{2}$ and we used, in turn, that $y_k$ decays in $\mathsf S_{\upsilon}$ for $\textrm{Re}\left[\upsilon\right]\rightarrow \pm\infty$, and that it has no singularities in $\mathsf S_{\upsilon}$. Thus we have arrived at the following integral TBA-like equation \cite{ZamolodchikovTBA}
\beq
	\varepsilon_{k}\left(\upsilon\right) = m_k \cosh(\upsilon) - \intop_\RR \frac{d\upsilon'}{2\pi} \frac{\log\left(1+ e^{-\varepsilon_{k-1}\left(\upsilon'\right)}\right) + \log\left(1+ e^{-\varepsilon_{k+1}\left(\upsilon'\right)}\right)}{\cosh\left(\upsilon - \upsilon'\right)}\;,
\label{eq:TBA_equation}
\eeq
where we introduced the \emph{pseudo-energies} (borrowing the language of the TBA)
\beq
	Y_k\left(\upsilon\right) = e^{-\varepsilon_{k}\left(\upsilon\right)}\;.
\eeq

If we were to choose a polynomial $P\left(z\right)$ with complex roots, then everything that has been said and shown above will essentially remain the same, with the exception of the assertion $m_k\in\mathbb R$. What will now happen is that the `masses' $m_k$ will be complex numbers and the TBA equation (\ref{eq:TBA_equation}) will need to be adjusted to the following, more general, form
\beq
    \varepsilon_k\left(\upsilon\right) = \frac{m_k}{2} e^{\upsilon} + \frac{m_k^{\ast}}{2} e^{-\upsilon} - \intop_{\RR} \frac{d\upsilon'}{2\pi} \frac{\log\left(1+ e^{-\varepsilon_{k-1}\left(\upsilon'\right)}\right) + \log\left(1+ e^{-\varepsilon_{k+1}\left(\upsilon'\right)}\right)}{\cosh\left(\upsilon - \upsilon'\right)}\;.
\label{eq:complex_mass_TBA}
\eeq
Note that as long as $\left\vert\textrm{arg}\left(m_k\right) - \textrm{arg}\left(m_{k+1}\right)\right\vert<\pi/2\,,\;\forall k$, the above equation is perfectly well defined. However, as soon as we go beyond this regime, it is necessary to pick out the appropriate pole contribution from the kernel.\footnote{In fact, the equations (\ref{eq:complex_mass_TBA}) can be rewritten in the form (\ref{eq:TBA_equation}), by shifting $\upsilon \rightarrow \upsilon - \textrm{arg}\left(m_k\right)$. These equations will involve kernels $1/\cosh\left(\upsilon - \upsilon' - \mathbbm{i}\,\textrm{arg}\left(m_k\right) + \mathbbm{i}\,\textrm{arg}\left(m_{k\pm 1}\right)\right)$, which present singularities on the real $\upsilon'$-line whenever $\left\vert\textrm{arg}\left(m_k\right) - \textrm{arg}\left(m_{k+1}\right)\right\vert= \left(2n+1\right)\pi/2\,,\;n\in\mathbb Z_\geq$.} Although the integral equation changes form, the functions $Y$ turn out to be continuous; this phenomenon is known as wall-crossing and has been discussed in \cite{Gaiotto:2008cd,Gaiotto:2009hg}.

We have arrived at an integral equation whose only inputs are the `masses' $m_k$, i.e.\ the integrals of the WKB one-form $s_0$ along the basis cycles of $\textrm{H}_1\left(\mathcal R_{\textrm{WKB}},\mathbb Z\right)$, and whose outputs are some functions $\varepsilon_k$ of the spectral parameter $\lambda$. As we will now explain, the knowledge of these functions will allow us to compute the regularized area (\ref{eq:Area_formula_regularized}) of the minimal surface in AdS$_3$, the boundary of which is a polygonal light-like Wilson loop determined by the function $P\left(z\right)$, as explained in section \ref{subsec:boundary_p}.

\subsection{The area as the free energy}
\label{sec:Area}
Now we wish to show that the regularized area is really the \emph{Free Energy} associated to the TBA equation (\ref{eq:TBA_equation}) -- or, more generally, (\ref{eq:complex_mass_TBA}). In order to do so we will take a route which might appear to be slightly convoluted, so bear with us. First of all, consider the expression (\ref{eq:Area_formula_regularized}) for the regularized area
\beq
    \mathcal A_{\textrm{reg}} = 2\alpha^2 \intop_{\Sigma} dz \, d\bar{z}\left(P\bar{P} e^{-\varphi} - \sqrt{P\bar{P}}\right)\;.
\label{eq:reg_area_2}
\eeq
We notice that it is possible to write this in terms of the one-forms $s_0$ and $\bar{s}_0$ (\ref{eq:one_forms_WKB}) and a one-form $u$ 
\beq
    s_0 = \sqrt{P}dz\;,\qquad \bar{s}_0 = \sqrt{\bar{P}}d\bar{z}\;,\qquad u = u_z dz + u_{\bar{z}}d\bar{z}\;,
\eeq
as
\beq
    \mathcal A_{\textrm{reg}} = 2\,\alpha^2 \intop_{\mathcal R_{\textrm{WKB}}} \left(s_0 \wedge u - s_0\wedge \bar{s}_0\right)\;,
\label{eq:reg_area_as_wedge}
\eeq
where, in order to reproduce (\ref{eq:reg_area_2}), we are forced to fix the anti-holomorphic part of $u$ as
\beq
    u_{\bar{z}} = \sqrt{P}\bar{P}e^{-\varphi}\;.
\eeq

It is evident that both $s_0$ and $\bar{s}_{0}$ are exact, since their components are, respectively, holomorphic and anti-holomorphic. In general the form $u$ is not exact, but it can be made so by precisely  choosing the $z$ component $u_z$, which does not contribute to the integral (\ref{eq:reg_area_as_wedge}). One easily verifies that the following choice
\beq
    u = \left(\frac{\varphi_{,z}^2 - 2 \varphi_{,zz}}{8\sqrt{P}} + f\left(z\right)\right)dz + \sqrt{P}\bar{P} e^{-\varphi}d \bar{z}\;,
\eeq
where $f\left(z\right)$ is an arbitrary function of $z$, fits the bill since
\beq
    du = \frac{e^{\varphi}}{2\sqrt{P}}\frac{\partial}{\partial z}\left(P\bar{P}e^{-2\varphi} + \frac{1}{2}\varphi_{,z\bz}e^{-\varphi}\right)dz\wedge d\bz = 0\;,
\eeq
due to the modified sinh-Gordon equation (\ref{eq:ADS3_GMC_simple}). We still have the freedom to choose the function $f(z)$ at will, and in the following we take 
\beq
	f\left(z\right) = \frac{1}{8\sqrt{P}}\left(\frac{P_{,zz}}{P} - \frac{5}{4}\left(\frac{P_{,z}}{P}\right)^2\right)\;,
\eeq
so that we can express the form $u$ in terms of $s_1$ (\ref{eq:one_forms_WKB}) as
\beq
	u = s_1 + \sqrt{P}\bar{P}e^{-\varphi} d\bar{z}\;.
\eeq
We are then able to rewrite the regularized area as an integral (\ref{eq:reg_area_as_wedge}) over the Riemann surface $\mathcal R_{\textrm{WKB}}$ of the external product of two exact one-forms: $s_0$ and $u-\bar{s}_0$. Why would we want to do this? The answer comes from the following neat property of integration on Riemann surfaces:

\begin{theorem}\cite{spivak_1}
Consider a Riemann surface $\Sigma_g$ of genus $g$ and let $\left\lbrace a_i,b_i \right\rbrace_{i=1}^g$ be a standard basis of cycles, i.e. a standard basis of $\textrm{H}_1\left(\Sigma_g,\mathbb Z\right)$. Take two exact one-forms $\omega$ and $\omega'$ and define $$ \alpha_i = \ointop_{a_i}\omega\;,\qquad \beta_i = \ointop_{b_i} \omega \;,$$ $$ \alpha_i' = \ointop_{a_i}\omega'\;,\qquad \beta_i' = \ointop_{b_i} \omega'\;. $$
    Then the integral of the two-form $\omega\wedge\omega'$ over the Riemann surface can be decomposed as
    \beq
        \intop_{\Sigma_g} \omega \wedge \omega' = \sum_{i=1}^g \left(\alpha_i\beta_i' - \beta_i \alpha_i'\right)\;.
    \eeq
\end{theorem}
Thanks to this result we can write the expression (\ref{eq:reg_area_as_wedge}) for the area as
\beq
    \mathcal A_{\textrm{reg}} = 2\alpha^2 \sum_{i,j} w_{i,j} \left(\ointop_{\gamma_i} s_{0}\right)\left(\ointop_{\gamma_j} s_1-\hat{\bar{s}}_{0}\right) \;,
\label{eq:reg_area_as_forms_product}
\eeq
where
\beq
	\hat{\bar{s}}_0 = \sqrt{P}\bar{P}e^{-\varphi} d\bar{z} - \bar{s}_0 = \sqrt{\bar{P}}\left(\sqrt{P\bar{P}}e^{-\varphi}-1\right)d\bar{z}\;,
\eeq
$\left\lbrace \gamma_i \right\rbrace$ is a basis of $\textrm{H}_1\left(\mathcal R_{\textrm{WKB}},\mathbb Z\right)$ and $w_{i,j}$ are the intersection numbers of these cycles.\footnote{The cycles $\gamma_i$ depicted in figure \ref{fig:cut_WKB_plane_example} do form a basis but not a normalized one. Hence the need to insert the intersection numbers.}

Now we need to identify the contour integrals in (\ref{eq:reg_area_as_forms_product}). To this end, let us introduce the functions $\hat{\varepsilon}_k$ defined as
\beq
    \hat{\varepsilon}_{2k}\left(\upsilon\right) = \varepsilon_{2k}\left(\upsilon\right)\;,\qquad \hat{\varepsilon}_{2k+1}\left(\upsilon\right) = \varepsilon_{2k+1}\left(\upsilon+\mathbbm i\frac{\pi}{2}\right)\;.
\eeq
We can describe their large $\lambda$ behaviour in two equivalent ways:
\begin{itemize}
    \item using the expression (\ref{eq:Y_funct_WKB_expr}) in terms of WKB integrals
\beq
	\hat{\varepsilon}_k = \lambda \ointop_{\gamma_k} s_0 + \frac{1}{\lambda} \ointop_{\gamma_k} s_1 +\mathcal O\left(\lambda^{-2}\right)\;, 
\eeq
    \item using the TBA equation (\ref{eq:complex_mass_TBA})
\beq
    \hat{\varepsilon}_k = \lambda\ointop_{\gamma_k}s_0  + \frac{1}{\lambda}\left ( \ointop_{\gamma_k}\bar{s}_0  - \frac{1}{\pi}\intop_{-\infty}^{\infty} d\upsilon'e^{\upsilon'} \sum_{j} w^{k,j}\log\left(1+e^{-\hat{\varepsilon}_j\left(\upsilon'\right)}\right) \right) +\mathcal O\left(\lambda^{-2}\right)\;,
\eeq
where we have used the definition (\ref{eq:TBA_masses}) of the dimensionless mass parameters $m_k$ and their complex conjugates $m^{\ast}_k$. 

In the case in which the parameters $m_k$ satisfy $\left\vert \textrm{arg}\left(m_k\right) - \textrm{arg}\left(m_{k+1}\right) \right\vert < \pi/2$, $w^{j,k}$ has the simple expression $w^{j,k} = \delta^{j+1,k} + \delta^{j-1,k}$, and if $2N\in 2\,\mathbb Z_{\geq} +1$ it is invertible with inverse given by the cycle intersection number $w_{i,j}$ introduced above.
\end{itemize}

Since the above two large-$\lambda$ expansions must agree term by term, we find the \emph{exact} expression for the integral of the $1$-form $s_1$ on the contours $\gamma_k$:  
\beq
    \ointop_{\gamma_k} s_1 = \ointop_{\gamma_k}\bar{s}_0  - \frac{1}{\pi}\intop_{-\infty}^{\infty} d\upsilon' e^{\upsilon'} \sum_{j} w^{k,j}\log\left(1+e^{-\hat{\varepsilon}_j\left(\upsilon'\right)}\right) \;.
\eeq
The expression for the area (\ref{eq:reg_area_as_forms_product}) then takes the following form:
\begin{subequations}
\begin{align}
    &\mathcal A_{\textrm{reg}} =  2\frac{\alpha^2}{\pi} \sum_{i,j} w_{i,j} Z_i\left(\,\intop_{-\infty}^{\infty} d\upsilon' e^{\upsilon'} \sum_{j} w^{j,k}\log\left(1+e^{-\varepsilon_k\left(\upsilon'\right)}\right)\right) \;, \\
    &Z_i = - \ointop_{\gamma_i} s_0\;.
\end{align}
\end{subequations}
The exact same reasoning as above can be repeated for the small $\lambda$ limit; this yields
\beq
    \mathcal A_{\textrm{reg}} =  2\frac{\alpha^2}{\pi} \sum_{i,j} w_{i,j} \bar{Z}_i\left(\,\intop_{-\infty}^{\infty} d\upsilon' e^{-\upsilon'} \sum_{j} w^{j,k}\log\left(1+e^{-\varepsilon_k\left(\upsilon'\right)}\right)\right) \;.
\eeq
Finally, as these two expressions must give the same result,\footnote{This statement is equivalent to the requirement that the total momentum of the TBA vanishes identically, or, in other words, that the \emph{pseudo-energies} $\varepsilon_k$ are even functions of $\upsilon$.} we can take  their mean value to find 
\beq
    \mathcal A_{\textrm{reg}} = \frac{\alpha^2}{\pi}\sum_i \left\vert m_i\right\vert \intop_{\RR} d\upsilon \cosh(\upsilon)\, \log\left(1+e^{-\varepsilon_i\left(\upsilon - \mathbbm{i}\, \textrm{arg}\left(m_i\right)\right)}\right)\;,
\label{eq:area_free_energy}
\eeq
which coincides with the free energy expression for the TBA equation (\ref{eq:complex_mass_TBA}). Note that we made the implicit assumptions that $\sum_j w_{i,j} w^{j,k} = \delta_i^k$, which is true only if $2N\in 2\,\mathbb Z_{\geq} +1$, and $\left\vert \textrm{arg}\left(m_k\right) - \textrm{arg}\left(m_{k+1}\right) \right\vert < \pi/2$. 
If instead we have $N\in\mathbb Z_\geq$ with the constraint on the phases of the masses still in place, the area keeps the form (\ref{eq:area_free_energy}), though acquiring an extra term as  studied in detail in \cite{Alday:2009yn}.
% If this last hypothesis holds when we choose $N\in\mathbb Z_\geq$, the area keeps the form (\ref{eq:area_free_energy}) and acquires an extra term as  studied in detail in \cite{Alday:2009dv}.
On the other hand, if this constraint
% on the phases of the masses 
is relaxed and we cross a wall, new cycles enter the game and one needs to track their contributions with care. 
However by adapting the derivation we followed it is possible to show that an expression of the form (\ref{eq:area_free_energy}) continues to hold. See \cite{Alday:2010vh}, appendix B, for more details.

\subsection{The IM side of ODE/IM correspondence and the conformal limit}

We conclude this excursion in the realm of minimal surfaces by briefly making contact with the IM side of the ODE/IM correspondence. In fact what we have done so far in this section pertains to the ODE part of the correspondence: we investigated the classical linear problem (\ref{eq:shG_linear_problem}) and showed how its monodromy data can be used to compute the area of a minimal surface in AdS$_3$ sitting on a light-like polygonal loop on the boundary $\partial$AdS$_3$. Through some non-trivial manipulations of the monodromy data, we arrived at the expression (\ref{eq:area_free_energy}) in terms of a set of auxiliary functions $\varepsilon_k\left(\upsilon\right)$ which satisfy the system of non-linear integral equations (\ref{eq:TBA_equation}). As mentioned above, these equations have the flavour of TBA equations for quantum integrable field theories and, as a matter of fact, have appeared earlier in the literature as the equations describing the finite-size ground state spectrum of the $SU\left(2N\right)_2/U\left(1\right)^{2N-1}$ Homogeneous sine-Gordon model\footnote{Actually, the equations (\ref{eq:TBA_equation}) are associated to a particular instance of the $SU\left(2N\right)_2/U\left(1\right)^{2N-1}$ HsG model, in which the so-called \emph{resonance parameters} are chosen to vanish, see \cite{CastroAlvaredo:2001ih,Miramontes:1999hx}.
} \cite{FernandezPousa:1996hi,FernandezPousa:1997iu,FernandezPousa:1997zb,CastroAlvaredo:1999em,CastroAlvaredo:2000nr,CastroAlvaredo:2001ih,Dorey:2004qc}. Hence we conclude that the linear system (\ref{eq:shG_linear_problem}) works as a bridge, connecting the geometry of minimal surfaces in AdS$_3$ -- and, consequently, the properties of light-like Wilson loops in $\partial$AdS$_3$ -- to the properties of the quantum $SU\left(2N\right)_2/U\left(1\right)^{2N-1}$ HsG model in finite-size geometry.

It is known \cite{CastroAlvaredo:1999em,Dorey:2004qc} that the CFT limit of the $G_k/U\left(1\right)^{r_G}$, with $G$ a compact simple Lie group, $r_G$ the rank of the group $G$ and $k$ its level, is described by the parafermionic $G_k/U\left(1\right)^{r_G}$ coset CFT with central charge
\beq
    c = \frac{k-1}{k+h_G}r_G \, h_G\;,
\eeq
where $h_G$ is the Coxeter number of the group $G$. In the case considered in this section, that is $G=SU\left(2N\right)$, one has $r_G=2N-1$ and $h_G=2N$ and choosing $k=2$ one obtains the central charge
\beq
    c = N\frac{2N-1}{N+1}\;.
\eeq

As mentioned in section \ref{sec:gen}, the integrable structure of these CFTs is conjectured to be described by a Sturm-Liouville problem for (\ref{eq:sh0}) with the particular choice (\ref{eq:polPot}) of the potential $P\left(x\right)$. In order to verify this fact, we need to perform the \emph{conformal limit} on the linear system (\ref{eq:shG_linear_problem}). We thus first pick a generic point $\left(z_0,\bar{z}_0\right)$, such that $P\left(z_0\right) = p_0 \neq 0,\infty$ and $\bar{P}\left(\bar{z}_0\right) = \bar{p}_0 \neq 0,\infty$. Without loss of generality we will suppose that $\left(z_0,\bar{z}_0\right) = \left(0,0\right)$. As the point $\left(0,0\right)$ need to be generic, we require the Gauss curvature (\ref{eq:Gauss_in_term_of_potential}) to be a finite constant at that point
\beq
    e^{-2\varphi}P\bar{P} \underset{\left(z,\bar{z}\right)\rightarrow \left(0,0\right)}{\sim} \mathcal O\left(z^0,\bar{z}^0\right)\;,
\eeq
which means that the sinh-Gordon field $\varphi$ will have the following simple, regular behaviour
\beq
    \varphi\left(z,\bar{z}\right) \underset{\left(z,\bar{z}\right)\rightarrow \left(0,0\right)}{\sim} \frac{1}{2}\log\left(P_0\bar{P}_0\right) + \sum_{k=1}^{\infty} \left(\varphi_k z^k + \bar{\varphi}_k \bar{z}^k\right)\;.
\eeq
The coefficients $\varphi_k$ and $\bar{\varphi}_k$ are fixed by inserting the above ansatz into the modified sinh-Gordon equation (\ref{eq:ADS3_GMC_simple});
their explicit form is of no relevance, but we list here the first few
\begin{subequations}
\begin{align}
    \varphi_1 &= \frac{P_1}{2P_0}\;,\quad \varphi_2 = \frac{P_2}{2P_0} - \frac{P_1^2}{4P_0^2}\;,\quad \varphi_3 = \frac{P_3}{2P_0} -\frac{P_1P_2}{2P_0^2} + \frac{P_1^3}{6P_0^3}\;, \\
    \varphi_4 &= \frac{P_4}{2P_0} - \frac{P_2^2+2P_1P_3}{4P_0^2} + \frac{P_1^2P_2}{2P_0^3}-\frac{P_1^4}{8P_0^4} \;,
\end{align}
\end{subequations}
with
\beq
P\left(z\right) = P_0 + \sum_{k=1}^{2N} P_k z^k = \prod_{k=1}^{2N}\left(z-z_k\right)\,.
\eeq
Similar expressions hold for $\bar{\varphi}_k$ and $\bar{p}\left(\bar{z}\right)$. We see that when taking the \emph{light-cone} limit $\bar{z}\rightarrow 0$,   the field assumes the following form
\beq
    \varphi\left(z,\bar{z}\right) \underset{\bar{z}\rightarrow0}{\sim} \frac{1}{2}\log\left(P_0\bar{P}_0\right) + \sum_{k=1}^\infty \varphi_k z^{k}\;.
\eeq

Let us look back at the linear system (\ref{eq:shG_linear_problem})
\beq
    \Phi_{,z} = \mathcal L \Phi \;,\qquad \Phi_{,\bar{z}} = \bar{\mathcal L} \Phi\;,
\eeq
with
\beq
    \mathcal L\left(\lambda\right) = \left(\begin{array}{c c}
        -\frac{1}{4}\varphi_{,z} & \lambda e^{\frac{\varphi}{2}} \\
        \lambda P e^{-\frac{\varphi}{2}} & \frac{1}{4}\varphi_{,z}
    \end{array}\right)\;,\qquad \bar{\mathcal L}\left(\lambda\right) = \left(\begin{array}{c c}
        \frac{1}{4}\varphi_{,\bar{z}} & \frac{1}{\lambda} \bar{P} e^{-\frac{\varphi}{2}} \\
        \frac{1}{\lambda} e^{\frac{\varphi}{2}} & -\frac{1}{4}\varphi_{,\bar{z}}
    \end{array}\right)\;.
\eeq
We now consider the unknown $\Phi$ as a vector, i.e. an arbitrary column of a generic matrix solution of (\ref{eq:shG_linear_problem}), which we can parametrise in the two following ways  
\beq
    \Phi = \left(\begin{array}{c}
        \lambda e^{\frac{\varphi}{4}}\chi \\
        e^{-3\frac{\varphi}{4}} \partial\left(e^{\frac{\varphi}{2}}\chi\right)
    \end{array}\right) = \left(\begin{array}{c}
        e^{-3\frac{\varphi}{4}} \bar{\partial}\left(e^{\frac{\varphi}{2}}\bar{\chi}\right) \\
        \frac{1}{\lambda}e^{\frac{\varphi}{4}}\bar{\chi}
    \end{array}\right)\;.
\label{eq:vector_parametrization_for_chiral_limit}
\eeq
One then easily checks that the linear problem reduces to the following pair of second order differential equations
\beqa
   \chi_{,zz}\left(z,\bar{z}\right) + \left(\frac{1}{2}v\left(z,\bar{z}\right)-\lambda^2 P\left(z\right)\right)\chi\left(z,\bar{z}\right) &=& 0\;, \label{eq:chiral_PDE} \\
   \bar{\chi}_{,\bar{z}\bar{z}}\left(z,\bar{z}\right) + \left(\frac{1}{2}\bar{v}\left(z,\bar{z}\right)-\frac{1}{\lambda^{2}} \bar{P}\left(\bar{z}\right)\right)\bar{\chi}\left(z,\bar{z}\right) &=& 0\;, \label{eq:anti_chiral_PDE}
\eeqa
where
\beq
    v\left(z,\bar{z}\right) = \varphi_{,zz}\left(z,\bar{z}\right) - \frac{1}{2}\varphi_{,z}\left(z,\bar{z}\right)^2\;,\qquad \bar{v}\left(z,\bar{z}\right) = \varphi_{,\bar{z}\bar{z}}\left(z,\bar{z}\right) - \frac{1}{2}\varphi_{,\bar{z}}\left(z,\bar{z}\right)^2\;,
\eeq
are the \emph{Miura transforms} of the field $\varphi$.

Now we will consider the conformal limit in the form of a double limit: we first take the light cone limit $\bar{z}\rightarrow 0$, which will `freeze' the anti-holomorphic dependence, and subsequently consider the regime $z\sim 0$. In order to consistently perform this last limit, we first rescale all the quantities in play by the appropriate power of $\lambda$ as follows
\beq
    z = \lambda^{-\frac{1}{N+1}}x\;,\qquad \bar{z} = \lambda^{\frac{1}{N+1}}\bar{x}\;,
\eeq
and scale the zeroes $z_k$ of the potential $P\left(z\right)$ as $z \to 0 $ so that
\beq
    P\left(z\right) = \prod_{k=1}^{2N}\left(z-z_k\right) = \lambda^{-\frac{2N}{N+1}}\prod_{k=1}^{2N}\left(x-x_k\right) = \lambda^{-\frac{2N}{N+1}}P\left(x\right)\;,
\eeq
then consider the limit $\lambda\rightarrow\infty$. Let us first concentrate on what happens to equation (\ref{eq:chiral_PDE}) when we send $\bar{z}\rightarrow 0$. The Miura transform $v$ becomes
\beq
    v\left(z,\bar{z}\right) = \mathcal O\left(z^{0}\right) = \lambda^{\frac{2}{N+1}}\mathcal O\left(\lambda^{-\frac{2}{N+1}}\right)\;,
\eeq
while the differential equation itself now reads
\beq
   \chi_{,xx}\left(x,\bar{x}\right) - \left(\mathcal O\left(\lambda^{-\frac{2N}{N+1}}\right) + P\left(x\right) \right)\chi\left(x,\bar{x}\right) = 0\;.
\eeq
Then we take the limit $\lambda\rightarrow\infty$ while keeping the scaling variables $x$ and $x_k$ finite, so that we arrive at the following equation
\beq
    \chi_{,xx}\left(x\right) =  P\left(x\right)\chi\left(x\right)\;, \label{eq:chiral_x_PDE}
\eeq
which is clearly holomorphic in form and the reason why we dropped the $\bar{x}$ dependence of $\phi$.

What is the fate of the equation (\ref{eq:anti_chiral_PDE})? Let us look at what happens to the potential $\bar{P}$ in the light-cone limit
\beq
    \bar{P}\left(\bar{z}\right) = \prod_{k=1}^{2N}\left(\bar{z} - z_k\right) \underset{\bar{z}\rightarrow 0}{\sim} \prod_{k=1}^{2N} z_k = \lambda^{-\frac{2N}{N+1}} \prod_{k=1}^{2N}x_k = \lambda^{-\frac{2N}{N+1}} X_N\;.
\eeq
On the other hand, in the light-cone limit we have $\bar{v} \rightarrow 0$. Consequently the equation (\ref{eq:anti_chiral_PDE}) reduces to
\beq
    \bar{\chi}_{,\bar{x}\bar{x}}\left(x,\bar{x}\right) - \lambda^{-\frac{4N}{N+1}}X_N\bar{\chi}\left(x,\bar{x}\right) = 0\;,
\eeq
which in the limit $\lambda\rightarrow\infty$ becomes
\beq
    \bar{\chi}_{,\bar{x}\bar{x}}\left(x,\bar{x}\right) = 0\;.
\eeq
We easily check that this equation is consistent with the relation imposed by the two parametrizations (\ref{eq:vector_parametrization_for_chiral_limit}) of the vector $\Phi$, since considering the identity
\beq
    \chi = \frac{1}{\lambda}e^{-\varphi}\bar{\partial}\left(e^{\frac{\varphi}{2}}\bar{\chi}\right)\;,
\eeq
and taking a derivative with respect to $\bar{z}$, we obtain 
\beq
    \chi_{,\bar{z}} = \frac{e^{-\frac{\varphi}{2}}}{\lambda}\left(\bar{\chi}_{,\bar{z}\bar{z}}\left(z,\bar{z}\right) + \frac{1}{2} \bar{v}\left(z,\bar{z}\right)\bar{\chi}\left(z,\bar{z}\right)\right)\;\longrightarrow\; 0\;.
\eeq
This proves that  in the double scaling limit, the function $\phi$ is  indeed holomorphic. Hence, as expected, we have recovered the ODE (\ref{eq:sh0}) with a potential 
\beq
P\left(x\right) = \prod_{k=1}^{2N}\left(x-x_k\right)\,,\quad (2 N \in \mathbb Z_>)\,,
\eeq
of the same form as (\ref{eq:polPot}).

\section{Conclusions}
\label{sec:conclusions}
The discovery of a connection between the theory of ordinary differential equations and  2D quantum field theories was a completely unexpected surprise for the integrable model community.  It has allowed the investigation of problems in pure mathematics, in statistical mechanics and condensed matter physics,  strings and supersymmetric gauge theories.  However, most of the mathematical structures and connections that have emerged over the past 20 years in the  ODE/IM context have only been superficially explored.  
Among the many mysterious facts concerning the ODE /IM correspondence, perhaps one of the most fascinating is that it provides a compelling alternative way to quantise classical integrable systems.  In this respect, it will be essential to put more effort toward the implementation of this novel quantisation scheme in the context of non-linear sigma models, as initiated in \cite{Bazhanov:2014joa}.

The ODE/IM correspondence might also provide a way to extend fundamental concepts related to the renormalisation group to the Hamiltonian picture \cite{Bazhanov:2019xvy} and to implement the quantisation of effective quantum field theories. 
 
Concerning the last topic,  the so-called $\TbT$-perturbation, where $\TbT$ is the composite operator defined as the determinant of the stress-energy tensor \cite{Zamolodchikov:2004ce}, is known to be integrable at both classical and quantum level \cite{Dubovsky:2012wk, Caselle:2013dra, Smirnov:2016lqw, Cavaglia:2016oda,Bonelli:2018kik}.  On the classical side,   deformed EoMs and  Lax operators coincide with the undeformed quantities up to a field-dependent local change of the space-time coordinates \cite{Dubovsky:2012wk, Conti:2018tca, Coleman:2019dvf}.  The effect of this deformation on the finite-size quantum TBA spectrum is also well understood; however, what is still missing are the ODE/IM steps connecting the classical to the quantum TBA answer.    
For instance,  it would interesting to know the fate of the polygonal Wilson loops, in particular of the area/
free-energy equivalence described in this review, under the $\TbT$ perturbation or the Lorentz-breaking generalisations studied in \cite{Guica:2017lia, Chakraborty:2018vja, Conti:2019dxg}. 
\\  

\medskip
\noindent{\bf -- Acknowledgments --}\\
SN and RT  thank the organisers of the ``Young Researchers Integrability School 2017'' in Dublin, for the invitation and opportunity to talk about the ODE/IM correspondence.  We would also like to thank Andrea Cavagli\`a, Riccardo Conti, Davide Fioravanti, Alba Grassi, Sylvain Lacroix,  Davide Masoero, Sergei Lukyanov and Hongfei Shu, for useful discussions.
This project was partially supported by the INFN project SFT, the EU networks SAGEX (grant agreement no.\ 764850) and GATIS+, NSF Award  PHY-1620628, STFC consolidated grant ST/P000371/1, and by the FCT Project PTDC/MAT-PUR/30234/2017 ``Irregular connections on algebraic curves and Quantum Field Theory".\\
%
%\appendix
%
\medskip

\bibliography{Biblio3} 

\end{document}